\documentclass[11pt,a4paper]{article}

\usepackage{jheppub}
\usepackage{amsmath}
\usepackage{amssymb}
\usepackage{amsthm}
\usepackage[mathscr]{eucal}
\usepackage{graphicx}
\usepackage{slashed}
\usepackage{setspace}
\usepackage{hyperref}

\newcommand{\bZ}{\mathbb{Z}}
\newcommand{\bC}{\mathbb{C}}
\newcommand{\bP}{\mathbb{P}}
\newcommand{\bR}{\mathbb{R}}
\newcommand{\bF}{\mathbb{F}}

\newcommand{\cE}{\mathcal{E}}
\newcommand{\cN}{\mathcal{N}}
\newcommand{\cM}{\mathcal{M}}

\newcommand{\cR}{\mathcal{R}}

\newcommand{\cO}{\mathcal{O}}

\newcommand{\cL}{\mathcal{L}}

\newcommand{\cS}{\mathcal{S}}

\newcommand{\eA}{\mathscr{A}}
\newcommand{\eB}{\mathscr{B}}
\newcommand{\eC}{\mathscr{C}}

\newcommand{\ov}{\overline}
\newcommand{\un}{\underline}

\newcommand*{\longhookrightarrow}{\ensuremath{\lhook\joinrel\relbar\joinrel\rightarrow}}
\newcommand*{\longtwoheadedrightarrow}{\ensuremath{\relbar\joinrel\twoheadrightarrow}}
\newcommand{\Pfaff}{\text{\it pfaff}}

\DeclareMathOperator{\ch}{ch}

\title{Three Looks at Instantons in F-theory\\
  {\large -- New Insights from Anomaly Inflow, String Junctions and
    Heterotic Duality --}}

\author{Mirjam Cveti\v c,}
\author{I\~ naki Garc\' ia-Etxebarria and}
\author{James Halverson}
\affiliation{Department of Physics and Astronomy \\ University of Pennsylvania, Philadelphia, PA 19104-6396, USA}

\abstract{We discuss the physics of zero modes of ED3/M5 instantons at
  strong coupling from three different viewpoints. Motivated by an
  anomaly inflow argument, we give a prescription for describing
  neutral instanton modes in terms of string junctions, unifying the
  language with that used for charged modes. We proceed to discuss the
  physics of charged modes as we move between different points in the
  moduli space of F-theory compactified on $K3$. In particular, we
  show how, in going from the $E_6^3$ point to the $SO(8)^4$ point,
  the structure of $SO(8)$ zero modes arises from a non-trivial mixing
  of massless {\bf 27}'s of $E_6$ with massive modes stretching
  between different $E_6$ stacks. We observe a similar mixing in going
  from $SO(8)^4$ to $E_6^3$. Finally, we see how the zeroes of some
  exact worldsheet instanton superpotentials in heterotic backgrounds
  preserving $E_6$ symmetry admit a physical interpretation in terms
  of low energy physics. We also discuss the behavior of the dual
  F-theory compactification as the superpotential approaches a
  zero. An interesting observation is that in the examples we study
  some of the zeroes of the superpotential correspond to points of
  $E_8$ enhancement in the worldvolume of the dual $M5$ instanton, and
  more generally from enhancements of the singularity over the
  worldvolume of the instanton.}

\begin{document}

\begin{flushright}
{\small \tt
  \tt UPR-1230-T
}
\end{flushright}
\maketitle
\section{Introduction}
Over the last few years there has been much focus on euclidean D3/D2
instantons\footnote{For a comprehensive review of D-instantons, see
  \cite{Blumenhagen:2009qh}. For reviews of intersecting D-branes in
  type II, see \cite{Blumenhagen:2005mu,Blumenhagen:2006ci}, and for
  an introduction to both D-instantons and intersecting branes, see
  \cite{Cvetic:2011vz}.} in weakly coupled type IIB/IIA string theory,
which can generate superpotential corrections involving chiral matter
fields that are forbidden by global U(1) selection rules in string
perturbation theory
\cite{Blumenhagen:2006xt,Ibanez:2006da,Florea:2006si}. Phenomenologically
this is very important, as instantons can generate the $10\,10\,5_H$
top-quark Yukawa coupling in Georgi-Glashow GUTs
\cite{Blumenhagen:2007zk} as well as a Majorana neutrino mass
\cite{Blumenhagen:2006xt,Ibanez:2006da}, both of which are
\emph{always} forbidden in perturbation theory. It is also well known
that D-instanton effects can play a role in moduli stabilization, as
used in \cite{Kachru:2003aw} for the stabilization of K\"ahler
moduli, for example.

Given the recent progress in understanding non-perturbative effects in
weakly coupled type II, it is natural to investigate their relation to
similar effects in F-theory \cite{Vafa:1996xn}, which contains type
IIB as a weak coupling limit \cite{Sen:1997gv}. F-theory gives a nice
framework for constructing semi-realistic GUT models
\cite{Donagi:2008ca,Beasley:2008dc} in a way that takes advantage of
both the local constructions of particle physics offered by
intersecting braneworlds and the exceptional gauge symmetries common
in the heterotic string. There has been much follow up work, both on
important subtleties and model building
\cite{Beasley:2008kw,Donagi:2008kj,Hayashi:2008ba,Collinucci:2008zs,Donagi:2009ra,Marsano:2009gv,Marsano:2009wr,Collinucci:2009uh,Blumenhagen:2009up,Blumenhagen:2009yv}.
An advantage of this framework is that one can study important aspects
of particle physics such as the appearance of chiral matter and Yukawa
couplings by studying codimension two and three singularities on the
GUT 7-brane.

That one can learn much about the particle physics of an F-theory
compactification by studying local geometry around the GUT brane does
not mean that global effects are unimportant, however.  The GUT
7-brane is in fact only one divisor in the Calabi-Yau fourfold base
$B_3$, and effects which influence the physics of supersymmetry
breaking and moduli stabilization (for example) can occur away from
the GUT brane in $B_3$. Moreover, many three generation local models
do not even admit a global embedding \cite{Donagi:2009ra}. For these
and other reasons, there have been a number of works studying the
physics of global F-theory GUTs
\cite{Blumenhagen:2009yv,Grimm:2009yu,Cvetic:2010rq,Chen:2010tp,Chen:2010ts,Grimm:2010ez,Chung:2010bn,Chen:2010tg,Knapp:2011wk},
including some explicit global models where the elliptic Calabi-Yau
fourfold is a CICY in a toric variety.

The physics of instantons in F-theory is one example which depends
heavily on the global geometry, as they wrap divisors which are
generically different from the GUT brane divisor and sometimes do not
even intersect it. Though instantons in F-theory are not ``needed" to
generate the $10\, 10\, 5_H$ Yukawa coupling as they are in type IIB,
due to the existence of that Yukawa coupling at a point of $E_6$
enhancement, they nevertheless still generate superpotential
corrections, play a crucial role in the stabilization of K\"ahler
moduli, and can account for large hierarchies due to exponential
suppression. From a more pragmatic point of view, one must account for
them simply because they are there and will affect the physics.

In studying instanton effects in F-theory, it often proves useful to
understand them via duality to M-theory and the heterotic string, and
also in the type IIB weak coupling limit.  For example, via duality
with M-theory on a elliptic Calabi-Yau fourfold in the limit of
vanishing fiber, it was shown in \cite{Witten:1996bn} that in an
F-theory compactification to four dimensions a necessary condition for
M5-brane instantons to correct the superpotential is that the divisor
$D$ of the M5 instanton satisfies $h^0(D,\cO_D)=1$ and all other
$h^i(D,\cO_D)=0$. In addition, the instanton must be a ``vertical" M5
brane, meaning that $D$ wraps the total space of the fibration over a
divisor $\tilde D$ of the base $B_3$. One can relate the description
in terms of the M5 brane in the four-fold (the M-theory description of
the system) with a description purely in terms of an euclidean D3 on
the base $B_3$ (the IIB description). In particular the relation
between the counting of neutral zero modes was given in
\cite{Blumenhagen:2010ja}. For this reason, one often refers to ED3/M5
instantons in F-theory.

Under heterotic / F-theory duality, some ED3/M5 instantons in F-theory
dualize to heterotic worldsheet instantons, while others dualize to
NS5-brane instantons. More precisely, given a compactification of
F-theory on an elliptically fibered $K3$, dual to a $T^2$
compactification of the heterotic string, M5 branes wrapping the whole
$K3$ dualize to worldsheet instantons, while those wrapping the fiber
but not the base of the $K3$ dualize to NS5 branes on the
heterotic side. This extends naturally to the $K3$ fibered backgrounds we
will be considering in this paper, accordingly we will be focusing on
M5 branes wrapping the whole $K3$ fiber. Some aspects of ED3/M5
instantons in F-theory were studied from the point of view of
heterotic worldsheet instantons in \cite{Donagi:2010pd}. Other works
on instantons in F-theory include the lift of an ED3 instanton which
generates the $10\,10\,5_H$ coupling to a global F-theory model
\cite{Cvetic:2010rq}, instantons in local F-theory models
\cite{Heckman:2008es,Marsano:2008py}, and the use of instanton flux
\cite{Grimm:2011dj} to alleviate the generic tension between moduli
stabilization and chirality \cite{Blumenhagen:2007sm}.

Despite this progress, there are still a number of issues regarding
instantons in F-theory which must be addressed. One issue is that a
complete understanding of the chirality inducing G-flux, inherently in
F-theory without reference to a heterotic dual, is still lacking,
though a proposal has been made in
\cite{Marsano:2010ix,Marsano:2011nn}. This issue, which faces
F-theory compactifications generically, also has strong implications
for instantons. Aside from these issue, though, there are very basic
questions which still need to be addressed. For example, given a
globally defined F-theory compactification with G-flux
\begin{itemize}
\item[$\bullet$] What are the ``charged" zero modes stretching between
  the instanton and 7-branes, and how do they relate to the
  well-understood charged modes in type IIB?
\item[$\bullet$] How does one compute the superpotential correction
  due to an ED3/M5 instanton?
\end{itemize}	
In weakly coupled type II compactifications with a CFT description
(such as models compactified on a toroidal orbifold) both of these
questions have concrete answers.  The charged zero modes are
represented by vertex operators obtained by the quantization of open
strings stretching between and ED2(ED3) instanton and a D6(D7) brane
in type IIA(IIB). Their superpotential corrections, if any, can be
calculated according to the instanton calculus of
\cite{Blumenhagen:2006xt}, and couplings between charged instanton
zero modes and chiral matter fields can be computed via disc diagrams
in the CFT.  Of course, F-theory compactifications generically involve
highly curved manifolds and strong coupling, so that they do not admit
at CFT description. Therefore, one must address and answer these
questions using a different formalism.

\medskip

In this paper we aim to shed some light on this questions by using a
multi-pronged approach. In section~\ref{sec:anomaly inflow} we will
reinterpret an observation of \cite{Harvey:2007ab} regarding anomaly
inflow towards orientifolds in the context of instantons in
F-theory. This will give us a way of understanding neutral zero modes
in terms of string junctions, and in particular we give a strong
coupling re-interpretation of the familiar $\theta$ mode as a
particular junction with prongs on both orientifold components. We
proceed to study charged zero modes in section~\ref{sec:string
  junctions} using string junctions techniques. We will see that modes
that are massless in certain regions of moduli space can come from
rather complicated multi-pronged strings in other regions. These
multi-pronged strings often stretch between distant 7-branes, and therefore
can correspond to massive BPS states. We
illustrate this discussion in a particularly simple family of
configurations that interpolates between $E_6^3$ gauge symmetry and
$SO(8)^4$. Finally, in section~\ref{sec:heterotic duality} we analyze
the non-perturbative physics of certain heterotic backgrounds with
$E_6$ symmetry at low energies. The exact dependence on the vector
bundle moduli of the superpotential due to a particular worldsheet
instanton was computed in \cite{Buchbinder:2002ic,Buchbinder:2002pr},
and we give a physical understanding of this dependence in terms of
low energy effective field theory. By analyzing the behavior of the
dual F-theory compactification as we reach the zeroes of the
superpotential, we observe some interesting features of the physics of
M5 instantons in the non-perturbative regime. Certain points in moduli
space corroborate what is expect from weakly coupled type II, where
a zero of the superpotential correction due to an ED3 instanton
corresponds to the appearance of extra light matter in the spectrum.

\medskip

{\bf Note added:} As we were readying this paper for publication, we
received a draft copy of \cite{Marsano:2011nn}, which has some overlap
with the discussion in section~\ref{sec:heterotic duality}. We thank
the authors of \cite{Marsano:2011nn} for letting us know of their work
prior to publication.

\section{Anomaly inflow and a new description of neutral zero modes}
\label{sec:anomaly inflow}

We will start our discussion by giving in this section a formulation
of neutral zero modes of rigid $O(1)$ instantons (that is, the modes
commonly known in the recent instanton literature as $\theta$ modes)
valid in regions of moduli space arbitrarily far away from weak
coupling.

In the conformal field theory approach to instantons, neutral and
charged zero modes arise from fundamentally different objects: neutral
zero modes arise from strings going from the instanton to itself,
while charged modes arise from strings going from the instanton to
space-time filling D7 branes intersecting the instanton. As we will
see in section~\ref{sec:theta-split}, analyzing the system in detail
reveals that $\theta$ modes localize in the intersections of the
instanton with the background orientifolds. This fact partially blurs
the strong split one observes in the perturbative approach between
both kinds of zero modes, since now both localize around defects in
the worldvolume theory on the instanton.

In going to strong coupling the orientifold decomposes into $(p,q)$-7
branes, and close to each of the components of the orientifold the
system is a $SL(2,\bZ)$ transform of the one giving a charged mode. So
our next task is to explain why this mutually non-local pair of
charged modes gives rise to something that at large distances looks
like an uncharged mode. As we will see in
section~\ref{sec:zero-mode-carriers}, doing this properly blurs even
further the distinction between both kinds of modes.

Even if non-perturbatively both neutral and charged modes admit a
unified description, they are still fundamentally different at low
energies (they transform differently under the Lorentz group of
spacetime, for one). In section~\ref{sec:theta-as-hypers} we provide a
criterion for determining microscopically which kind of zero mode one
is dealing with.

We will come back to study the behavior of charged modes from a
non-perturbative perspective in section~\ref{sec:string junctions},
but for the rest of this section we will focus mainly on neutral
modes.

\medskip

Let us consider F-theory compactified on $K3$, starting from
situations close to weak coupling. We will take the euclidean D3 to
wrap the $\bP^1$ base of the $K3$, and two extra directions transverse
to the K3.\footnote{We take F-theory to mean type IIB with a varying
  axio-dilaton $\tau$, the elliptic fiber is just encoding the varying
  of $\tau$. The natural non-perturbative object is thus an euclidean
  $D3$ wrapping 4 directions in the base of the fibration.} In this
setting the worldvolume theory of the instanton is four dimensional
topologically twisted $\cN=4$ $U(1)$ SYM, compactified on an $\bP^1$,
and in the presence of 24 string-like defects which are point-like on
the $\bP^1$. Close to weak coupling the 24 defects split naturally
into 16 mutually local defects (the D7 branes), and 4 pairs of
mutually non-local defects (the 4 $O7$-planes).

In order to keep the discussion as simple as possible, we will take a
decompactification limit of the $\bP^1$ in which the curvature
vanishes, while keeping the 24 branes at a finite distance. Our
discussion will only involve topological quantities, which are robust
under this deformation. In this flat limit the twisting becomes
unimportant, and the worldvolume theory of the instanton becomes 4d
$\cN=4$ $U(1)$ SYM in the presence of string-like defects.

\subsection{$1/4$ of a $\theta$ mode per orientifold}
\label{sec:theta-split}

We will like to understand how to describe instanton zero modes in
such a background in a way valid away from weak coupling. In order to
better motivate our later results we will need to take a small detour
and study anomaly inflow in our configuration. This analysis has been
done with a different motivation (AdS/CFT) in \cite{Harvey:2007ab}; in
this section we review their discussion and present some numeric
results that support and illustrate their result. We will reinterpret
the discussion of \cite{Harvey:2007ab} in
sections~\ref{sec:zero-mode-carriers} and \ref{sec:theta-as-hypers}.

\medskip

The basic puzzle that \cite{Harvey:2007ab} addresses can be summarized
as follows: take the euclidean D3 brane to be wrapping a 4-manifold
$X$ (in our case $X$ is a copy of flat space $\bR^4$). There is also a
$O7^-$ plane wrapping a divisor of $X$ (i.e. a $\bR^2$ subspace of
$\bR^2$ in the flat case), and 6 extra dimensions. We denote by $Y$
the total space wrapped by the $O7^-$.\footnote{This configuration was
  studied previously from the point of view of $O(1)$ instanton
  physics in F-theory in \cite{Cvetic:2009ah}.}

The Chern-Simons coupling on the $O7^-$ plane is given by \cite{Dasgupta:1997cd,Dasgupta:1997wd,Morales:1998ux,Stefanski:1998yx,Scrucca:1999uz}:
\begin{align}
  \label{eq:O7-CS-coupling}
  S_{cs}^{O7^-} = \int_Y C \wedge \sqrt{\frac{\hat L(TY/4)}{\hat
      L(NY/4)}}\, ,
\end{align}
where $C$ denotes the formal sum of $RR$ forms, and $TY$ and $NY$
denote respectively the tangent bundle to $Y$ and the normal bundle to
$Y$ inside the ambient Calabi-Yau. Given a vector bundle $E$, $\hat
L(E)$ denotes the Hirzebruch genus of $E$, defined by:
\begin{align}
  \hat L(E) = \prod \frac{x}{\tanh x} = 1 + \frac{1}{3}(c_1^2(E) -
  2 c_2(E)) + \ldots\, ,
\end{align}
and as it is conventional we have denoted by $x$ the components of $E$
under splitting. The Chern-Simons coupling on the D3 is also well
known, and it is given by:
\begin{align}
  \label{eq:D3-CS-coupling}
  S_{cs}^{D3} = \int_X C \wedge \ch(V)\wedge \sqrt{\frac{\hat
      A(TX)}{\hat A(NX)}}\, ,
\end{align}
with the same conventions as before for $C$, $TX$ and $NX$; and $\hat
A$ being the $A$-roof (or Dirac) genus, given to the first few orders
by:
\begin{align}
  \hat A(E) = \prod \frac{x_i/2}{\sinh(x_i/2)} = 1 - \frac{1}{24}(c_1^2(E) - 2 c_2(E)) + \ldots
\end{align}
In order to be general we have also included a factor of $\ch(V)$,
where $V$ is a possible vector bundle on the stack of D3 branes. In
our case we have a single D3 with gauge group $\bZ_2$, so we take
$V=1$ in the following.

Given \eqref{eq:O7-CS-coupling} and \eqref{eq:D3-CS-coupling}, a
standard anomaly inflow argument (see \cite{Harvey:2005it} for a
review) shows that there is an anomaly localized on the 2d
intersection between the $O7^-$ and the $D3$. The associated 4-form
anomaly polynomial is given by:
\begin{align}
  \label{eq:ED3-O7-inflow}
  I_{inflow} \sim p_1(TX) - p_1(NX) - \frac{1}{2}(p_1(TY) - p_1(NY))\, ,
\end{align}
where we have omitted some numerical factors which are irrelevant at
this level of the argument, and for conciseness we have introduced the
first Pontryagin class $p_1(E) = c_1^2(E) - 2 c_2(E)$.

The puzzle is now evident: the fact that~\eqref{eq:ED3-O7-inflow} is
non-vanishing indicates that consistency of the background requires
some chiral degrees of freedom to live in the intersection in order to
cancel the anomaly. Nevertheless, there are no obvious candidates in
the form of strings between the D3 and the $O7^-$ plane. By this we
mean simply that there is no twisted sector for orientifolds. Strings
going from the instanton to the $O7^-$ are interpreted as unoriented
D3-D3 strings, and if they are the source for the chiral modes, the
analysis is necessarily somewhat subtle: 3-3 states are associated
with states in the 4d theory on the instanton which survive the
orientifold projection, and one would think that states in the 4d
theory look non-chiral from the 2d point of view.

Indeed, the answer that \cite{Harvey:2007ab} suggests, and
convincingly argues in a beautiful analysis, is that the required
chiral matter comes from a zero mode of the gaugino on the euclidean
D3. In our particular context, this means that the $\theta$ mode
localizes on the intersections of the instanton with the background
orientifolds. To our knowledge, this is a new (and to us, surprising)
observation in the context of instanton physics, and as we will see in
the rest of this section it provides the key to understanding
instanton zero modes at arbitrarily strong coupling.

\medskip

Before going into the consequences of this observation for instanton
physics, and since the claim may be a bit surprising, let us present
some simple numerical results that clearly illustrate the phenomenon.

Let us consider F-theory compactified on an elliptically fibered $K3$
with section. We wrap the euclidean D3 on the $\bP^1$ base of the
$K3$, and thus the D7 and $O7^-$ planes appear as 24 point-like
defects on the $\bP^1$. We want to stay close to weak coupling for
ease of interpretation, so let us parameterize the Weierstrass
parameters of the elliptic fibration in Sen's form \cite{Sen:1996vd,
  Sen:1997gv}:
\begin{align}
  f & = -3 h(z)^2 + \epsilon \eta(z)\\
  g & = -2 h(z)^3 + \epsilon h(z) \eta(z) - \frac{1}{12}\epsilon^2 \chi(z) 
\end{align}
with $z$ the complex coordinate on $\bP^1$, $\epsilon$ is an arbitrary
number parameterizing how close we are to weak coupling (we take
$\epsilon=10^{-3}$), and $h(z)$, $\eta(z)$ and $\chi(z)$ arbitrary
functions. In terms of this parameterization, the discriminant is
given by
\begin{align}
  \label{eq:Sen-discriminant}
  \begin{split}
    \Delta & = 4 f^3 + 27 g^2 \\ &= 9 \epsilon^2 h^2(h\chi - \eta^2) -
    \frac{1}{2}\epsilon^3\eta(9h\chi - 8\eta^2)\, ,
  \end{split}
\end{align}
where we have dropped the explicit dependence on $z$ for
readability. The weak coupling limit is given by $\epsilon\to 0$, in
which case we have $\Delta\sim h^2(h\chi - \eta^2)$. The 7 branes are
given by the roots of $\Delta$, are they are thus located at $h(z)=0$
and $(h\chi - \eta^2)=0$. A monodromy analysis shows that the first
set of roots corresponds to the location of $O7^-$ planes, and the
second set to the location of D7 branes.

We take these arbitrary functions to be:
\begin{align}
  \label{eq:Sen-parameterization}
  \begin{split}
    h(z) & = \prod_{n=1}^4 (z - h_n) \\
    \chi(z) & = 0\\
    \eta(z) & = \prod_{n=1}^8 (z-\eta_n)
  \end{split}
\end{align}
with
\begin{align}
  \label{eq:eta-roots}
  \eta_n = 1.3\, e^{2\pi i n/8} \qquad ; \qquad h_n = \{0.6+0.35i,
    0.35-i/2, -0.25-0.45i, -1/2+i/2\}
\end{align}

\begin{figure}[ht]
  \includegraphics[width=\textwidth]{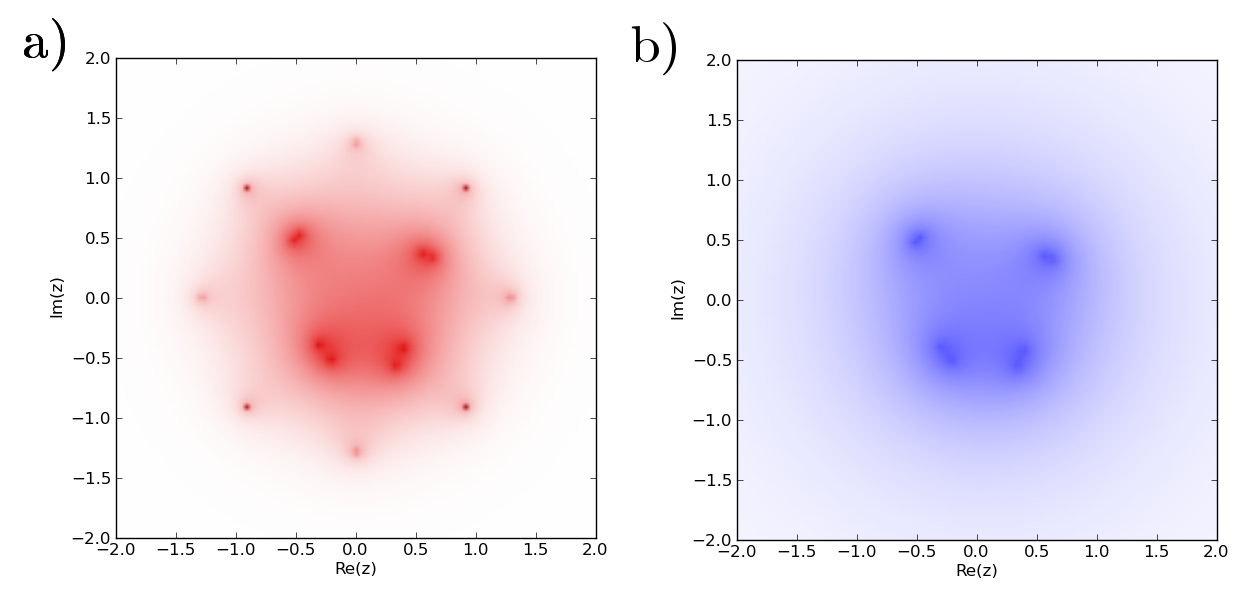}

  \caption{a) Discriminant $\Delta$ for the choice of $h,\eta,\chi$ in
    \eqref{eq:Sen-parameterization}, \eqref{eq:eta-roots}. b)
    Corresponding $b(z)$ for $\theta$, we are plotting $|b(z)|$.}

  \label{fig:delta-and-zero-mode}
\end{figure}

The exact numeric values are inessential, the basic feature of this
choice being that the orientifold planes get distributed in the
corners of a (slightly deformed) square, with positions given roughly
by the zeroes of $h(z)$.\footnote{We say ``roughly'' since the
  identification of $h(z)=0$ with orientifolds is only valid if we
  disregard the second term in~\eqref{eq:Sen-discriminant} --- which
  given our small value for $\epsilon$ is a good approximation --- but
  the plots in the text are exact in $\epsilon$. Similar remarks apply
  to the position of the D7 branes.} The 16 D7 branes split into 8
pairs of branes, each pair located at a zero of $\eta(z)$, which we
choose in \eqref{eq:eta-roots} to be arranged concentrically. We show
a plot of the resulting discriminant in
figure~\ref{fig:delta-and-zero-mode}a.

We now want to see how the $\theta$ zero mode localizes near the
orientifold planes. The solution for the wavefunction of $\theta$ on
$\bP^1$ is given in \cite{Harvey:2007ab} in terms of a function $b(z)$
given by:
\begin{align}
  \label{eq:theta-wavefunction}
  b(z) = \frac{\eta(\tau(z))}{\Delta^{1/24}}\, ,
\end{align}
with $\Delta$ the discriminant of the elliptic fibration. The coupling
$\tau(z)$ can be determined in terms of hypergeometric functions using
the expression for the inverse of Klein's $J$-invariant:
\begin{align}
  \tau(z) = J^{-1}(j(z)/1728)
\end{align}
where
\begin{align}
  J^{-1}(\lambda) = \frac{i(r(\lambda)-s(\lambda))}{r(\lambda)+s(\lambda)}
\end{align}
and we have introduced
\begin{align}
  \begin{split}
    r(\lambda) & = \Gamma\left(\frac{5}{12}\right)^2
    \,_2F_1\left(\frac{1}{12},\frac{1}{12};\frac{1}{2};1-\lambda\right)\\
    s(\lambda) & = 2(\sqrt{3}-2)\Gamma\left(\frac{11}{12}\right)^2
    \sqrt{\lambda -1 }
    \,_2F_1\left(\frac{7}{12},\frac{7}{12};\frac{3}{2};1-\lambda\right)
  \end{split}
\end{align}
and $\,_2F_1$ denotes the ordinary or Gaussian hypergeometric
function. We have plotted the absolute value of $b(z)$ in
figure~\ref{fig:delta-and-zero-mode}b, where it is clear that the
resulting non-trivial behavior of the wavefunction localizes around
the orientifold planes.

While figure~\ref{fig:delta-and-zero-mode}b shows the localization of
the zero mode structure in regions close to the orientifolds, the full
story is slightly more complicated. The wavefunction obtained by
\cite{Harvey:2007ab} can be written as:\footnote{The wavefunction
  \eqref{eq:Psi-wavefunction} was derived using techniques valid only
  at weak coupling, and as such we can only trust it far away from the
  orientifolds. It will nevertheless suffice for showing
  localization.}
\begin{align}
  \label{eq:Psi-wavefunction}
  \Psi(z) = e^{-i \arg(b(z))}
\end{align}
where $\arg(e^{i\alpha})=\alpha$. Let us analyze first the monodromy
around a D7. As we go around a D7 brane we have that $\tau \to
\tau+1$, which sends $\eta(\tau)\to \eta(\tau+1) = e^{\frac{\pi
    i}{12}}\eta(\tau)$. We are also going around a single zero of the
discriminant, which thus sends $\Delta^{\frac{1}{24}}\to e^{\frac{\pi
    i}{12}}\Delta^{\frac{1}{24}}$. From the form of $b(z)$ in
\eqref{eq:theta-wavefunction}, we then immediately see that $b(z)$,
and thus $\Psi(z)$ is invariant as we go around a D7 brane. On the
other hand, as we go around an $O7^-$ plane, we have that $\tau \to
\tau-4$. Correspondingly, we have that
$\eta(\tau)\to\eta(\tau-4)=e^{-\frac{\pi i}{3}}\eta(\tau)$. Since in
going around an $O7^-$ we pick a double zero of $\Delta$, we have that
$\Delta^{\frac{1}{24}}\to e^{\frac{\pi
    i}{6}}\Delta^{\frac{1}{24}}$. In terms of the wavefunction we thus
have that $\Psi(z)\to -i\Psi(z)$. (As a side remark, notice that this
is the behavior one should expect from the action of $SL(2,\bZ)$ on
the fermions of the $\cN=4$ theory, as derived in
\cite{Kapustin:2006pk} and used in this same context in
\cite{Cvetic:2009ah}.) So we see that the non-trivial variation of the
$\theta$ mode concentrates on the orientifold planes.

\subsection{What flows}
\label{sec:zero-mode-carriers}

The results in the previous section can be given a very useful
interpretation in the following way. Notice that in the case of
ordinary D7 branes intersecting the instanton the anomaly inflow is
canceled due to the zero modes localized at the D7 defect, i.e. the
massless modes associated to the fundamental strings between the
instanton and the D7 brane. The analogous candidate string at weak
coupling for canceling the inflow to the orientifold defect (which,
as we saw, is also where the $\theta$ mode localizes) are strings
going from the instanton to the orientifold.

This is perhaps a little bit surprising to people familiar with
instanton dynamics: usually we think of the $\theta$ mode as the
gaugino of the $\cN=4$ theory on the instanton that survives the
orientifold projection. As such, we are not used to thinking of it as
a localized mode. Nevertheless, it is important to realize that zero
modes of the gaugino on the worldvolume of the instanton, by their
very nature, are extended objects, and can appear or disappear in the
presence of defects. In the case of $\theta$ the counting does not
change, but the profile of the zero mode does.

Strings from the instanton to the orientifold can be alternatively
interpreted as the 3-3 strings mapped to themselves under the
orientifold action, the statement above being in this language that
$\theta$ modes come from invariant 3-3 strings.

\medskip

\begin{figure}[htp]
  \begin{center}
    \includegraphics[width=0.7\textwidth]{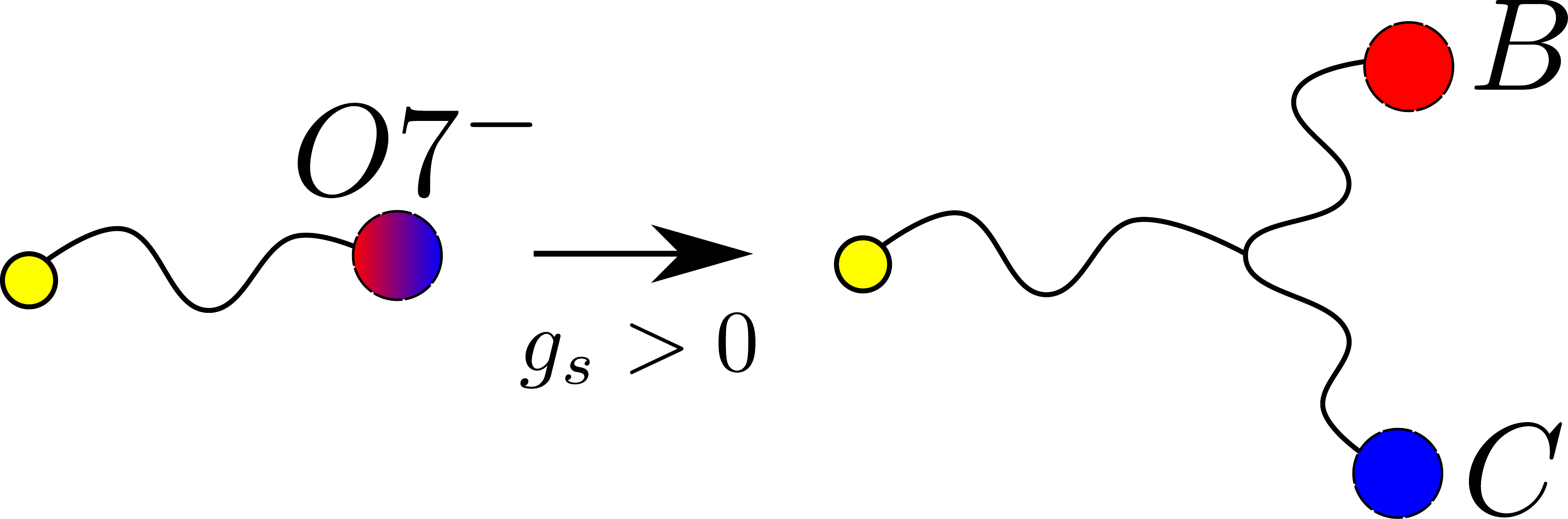}
  \end{center}

  \caption{As the orientifold splits into its $B,C$ components due to
    the effect of $D(-1)$ instantons, the orientifold invariant 3-3
    string corresponding to the $\theta$ mode becomes a string
    junction. Note that the junction is formally the same as the one
    describing $W$ bosons in the string realization of
    Seiberg-Witten.}
  \label{fig:orientifold-split}
\end{figure}

Having identified the $\theta$ mode in the perturbative language, we
would now like to turn on $g_s$. We encounter a small puzzle here: as
we turn on the coupling the orientifold splits into two components,
and there is no massless fundamental string anymore. This is analogous
to what happens in Seiberg-Witten theory for the $W$ boson
\cite{Seiberg:1994rs,Seiberg:1994aj}, and we propose that the
resolution is precisely the same in our case: the fundamental string
associated to the $\theta$ mode turns into a string junction. See
figure~\ref{fig:orientifold-split}.

The way to understand the fact that a massive string junction is
associated to a zero mode is the following. Let us turn on an electric
field on the instanton worldvolume in the directions parallel to the
defect. This is slightly ill-defined in the strongly coupled region
inside the orientifold, so let us postpone discussing what happens
there momentarily, and work far away from the orientifold. Due to the
monodromy of the axion around the orientifold, the electric field
produces an electric current \emph{towards} the defect. The derivation
is elementary, and we review it here for convenience: in the presence
of an axion $a(x)$, the $U(1)$ Yang-Mills Lagrangian looks like:
\begin{align}
  S_{U(1)} = \frac{-1}{g^2}\int F\wedge \star F + a F \wedge F\, +
  \ldots ,
\end{align}
where we have ignored terms irrelevant for our analysis. Integrating
by parts, and taking the variation with respect to the $U(1)$
connection, we have:
\begin{align}
  d\star F = g^2 d(a F)\, .
\end{align}
This implies that in the presence of a non-constant axion and a
non-trivial background $F$ we have a current vector $j = \star g^2
d(aF)$ for the electromagnetic field. In particular, choosing
$x^0,x^1$ as the directions along the defect, and $\theta,r$ the
normal directions (in polar coordinates), it is clear that choosing $F
= E\, dx^0\wedge x^1$, $a=\theta$ gives a current $j \sim dr$.

This current towards the string will violate charge conservation
unless something happens at the string that takes away the charge. The
``something'' that happens is that generically there are chiral zero
modes on the string which are charged under the bulk $U(1)$, so when
the electric field is applied they flow along the string taking away
the charge. These modes are localized in the sense that they have
exponentially decaying wavefunctions, and we will refer to them as
\emph{carriers} in what follows. (See \cite{Naculich:1987ci} for a
careful study of the dynamics of the system using the same viewpoint
that we are taking here.)

In our orientifold case, we also have axion monodromy, and thus
current flowing into the orientifold. The modes that take away the
current are the $\theta$ junctions.

We come now to an important point: the $\theta$ junctions exist as
stable carriers when we are far way from the orientifold. As we enter
the region of strong coupling the junction becomes unstable, and
decays into a couple of elementary constituents, going to each of the
components of the orientifold. The analysis is identical to the
familiar decay of the $W$ boson at strong coupling in the
Seiberg-Witten theory, and we omit repeating the well-known details of
this process (see
\cite{Fayyazuddin:1995pk,Ferrari:1996sv,Bilal:1996sk} for in-depth
discussions from the field theory point of view, and
\cite{Bergman:1998br,Mikhailov:1998bx} for the string junction point
of view). Seeing the same decay process inversely, this gives a
natural explanation to the question of what happens to the putative
charged zero modes associated to each component of the orientifold:
locally there are indeed dyonic zero modes of the instanton, carried
by the corresponding strings, but globally the carriers for the local
zero modes form bound states which are purely electric, the $\theta$
junctions.

\medskip

Let us give a couple of further arguments in favor of this
proposal. First of all, a somewhat counter-intuitive fact is that the
$\theta$ mode has charge under the $U(1)$ that exists locally on the
instanton. More precisely, its charge is 2, in the conventions where
an ordinary 3-7 string has charge 1. (The analogous statement in the
Seiberg-Witten case is that the $W$ boson has charge 2 in the
conventions where the charge of the elementary flavor hypermultiplet
is 1.) This is in fact in agreement with the fact that the resulting
zero mode is uncharged under the $\bZ_2$ symmetry remaining on the
instanton after the orientifold action.

Also notice that the $\theta$ modes form a doublet under the
$SU(2)_\cR$ symmetry (in analogy with the Seiberg-Witten
case). Together with the fact that the junction has double the $U(1)$
charge as an ordinary string, we see that the $\theta$ junctions carry
precisely 4 times the charge of an elementary string carrier (an
ordinary 3-7 string). This is in agreement with the fact that the
monodromy of the axion around an orientifold is 4 units, as opposed to
1 unit for an ordinary D7 brane.

\subsection{Neutral zero modes as vector multiplets}
\label{sec:theta-as-hypers}

The previous proposal gave a very nice unified description in the
language of string junctions of the origin of both neutral and charged
zero modes of the instanton. One may wonder how to distinguish both
cases in practice. The answer has already been implicitly given in the
previous section, but let us be explicit about it here.

Take the string junction one is interested in analyzing. There will be
an analogous junction in Seiberg-Witten theory. The resulting state
will be either a hypermultiplet or a vector multiplet, this can be
determined using standard techniques. The discussion in the previous
section shows that we should understand hypers as giving rise to
charged zero modes, while vectors give rise to neutral zero modes. The
basic distinction is their behavior under $SU(2)_\cR$, which we can
identify as rotations in spacetime: vectors, just as the $\theta$,
transform as spinors, while hypers, such as the 3-7 strings, are
singlets.

\section{Charged modes as string junctions}
\label{sec:string junctions}

In the previous section we argued that neutral zero modes of the
instanton admit a very similar description to their charged
counterparts, once we formulate the physics using string junctions.
In this section we continue our study of instanton zero modes and
address the question of how do charged modes in strongly coupled
limits of F-theory relate to their perturbative type IIB counterparts.

We would like, in particular, to clarify those aspects of instantons
in F-theory which are qualitatively different from the physics of
instantons in type IIB, such as those arising in the presence of
exceptional 7-brane singularities in the instanton worldvolume.
Whatever the correct formalism is for charged instanton corrections in
F-theory, it should recover the known type IIB results in Sen's limit,
so one might hope to see this directly by studying instanton effects
as one moves in the complex structure moduli space of the F-theory
compactification.

Our approach is to study the structure of charged instanton zero
modes, realized as string junctions, as one moves in the complex
structure moduli space of an F-theory compactification. We will
present a particular path in the moduli space of F-theory on $K3$ and
explicitly study the fate of $\textbf{27}$'s of $E_6$ upon movement in
moduli space to the type IIB limit. We will also study an
\textbf{8}$_v$ of $SO(8)$ in type IIB and will show that, while some
of the junctions in the \textbf{8}$_v$ contribute to $\textbf{27}$'s
of $E_6$ elsewhere in moduli space, others are massive BPS states with
prongs stretching between distant $E_6$ singularities. It will become
clear that this type of behavior is not uncommon upon movement in
moduli space.

\subsection{A review of 7-branes and junctions}
\label{sec:7 brane review}
Let us review those aspects of 7-branes and junctions necessary for
our discussion of charged instanton zero modes. We will make some
definitions and present some constant coupling results relevant for
our study. Our conventions are mostly taken from
\cite{DeWolfe:1998zf}.

Much of the interesting physics arising from F-theory
compactifications descend from the fact that F-theory allows for more
generic 7-branes than type IIB, with a notable feature being that the
$7$-branes have non-trivial monodromy. The monodromy upon
counterclockwise crossing of a (downwards-directed) branch
cut\footnote{For an in depth discussion of 7-branes, branch cuts, and
  junctions, see \cite{Gaberdiel:1998mv}.} is given by
\begin{equation}
	\cM = \begin{pmatrix} a & b\\ c & d 
  \end{pmatrix}\, \in SL(2,\bZ)
\end{equation} 
and acts on the axio-dilaton $\tau=C_0 + \frac{i}{g_s}$ as 
\begin{equation}
	\tau \mapsto \frac{a\tau + b}{c \tau + d}\, .
\end{equation}
A $(p,q)$ 7-brane is such a $7$-brane, 
where the monodromy matrix
is given by:
\begin{align}
  \cM_{(p,q)} = \begin{pmatrix} 1+pq & -p^2\\ q^2 & 1-pq
  \end{pmatrix}\, .
\end{align}
An important fact is that 7-branes with exceptional gauge symmetry (as
well as others) can not be represented by a single $(p,q)$ 7-brane,
but instead can be represented by a number of $(p,q)$ 7-branes. A
convenient choice of basis for $(p,q)$ 7-branes that give rise to the
ADE groups is:
\begin{align}
\eA = \cM_{(1,0)} = \begin{pmatrix} 1 & -1 \\ 0 & 1 \end{pmatrix} \qquad
\eB = \cM_{(1,-1)} = \begin{pmatrix} 0 & -1 \\ 1 & 2 \end{pmatrix} \qquad
\eC = \cM_{(1,1)} = \begin{pmatrix} 2 & -1 \\ 1 & 0 \end{pmatrix}\,.
\end{align}
In the type IIB picture, an $A$-brane is a $D7$ brane and a $B$ and
$C$ brane together give rise to an $O7$-plane. This is the famous
splitting of an $O7$-plane in F-theory. It splits into two branes, a
$B$ and a $C$, with separation set by $e^{-1/g_s}$.

As an example, a 7-brane with $SO(8)$ singularity can be represented as 
\begin{equation*}
\includegraphics[scale=.5]{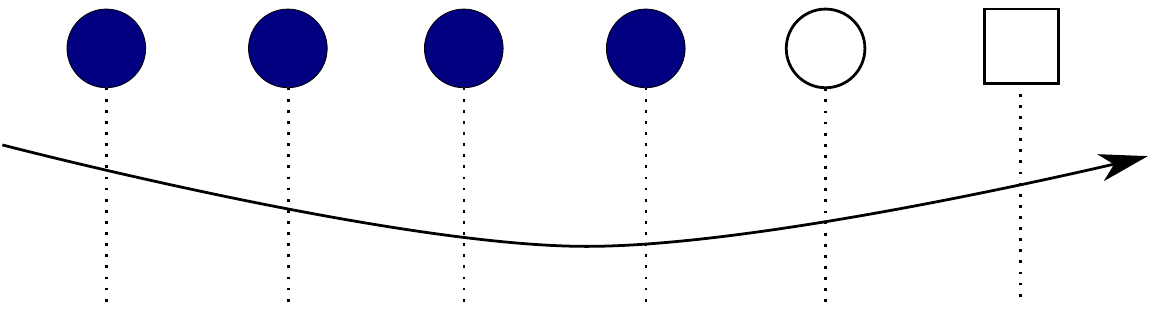}
\end{equation*}
where solid circles, hollow circles, and hollow squares represent
A branes, B branes, and C branes, respectively. The dotted lines represent
the downward directed brane cuts, and the arrow represents the ``direction" of
the monodromy in the conventions we are using. The monodromy matrix is
\begin{align}
  \cM_{SO(8)} = \eC\eB\eA^4 = \begin{pmatrix}-1 & 0 \\ 0 & -1\end{pmatrix}\, .
\end{align}
The monodromy matrices act to the right, so the order of matrices is
the opposite of the left-right ordering of branes. We recognize this
convention is somewhat confusing, but it is fairly standard in the
literature and so we adopt it.\footnote{We will attempt to minimize
  the resulting confusion by denoting the branes themselves in italic
  script: $A,B,C,\ldots$, while the corresponding monodromy matrices
  will be written in calligraphic script: $\eA,\eB,\eC,\ldots$}

Similarly, the $E_6$ singularity can be obtained from a stack with left-right
ordering $AAAAABCC$, with monodromy
\begin{equation} 
\cM_{E_6} = \eC^2\eB\eA^5= \begin{pmatrix} -2 & 3 \\ -1 & 1 \end{pmatrix}.
\end{equation}
This singularity leaves constant the coupling satisfying:
\begin{align}
  \tau = \frac{-2\tau + 3}{-\tau + 1}\, ,
\end{align}
which has the solutions $\tau = e^{\pm\frac{\pi i}{3}}+1$. Only the
plus sign is physical, since $Im(\tau)=\frac{1}{g_s}$ needs to be
positive. Thus, using these conventions, the monodromy associated with
an $E_6$ singularity leaves $\tau = e^{\frac{\pi i}{3}} + 1$
invariant. We will be studying constant coupling configurations with
$E_6$ singularities, which were introduced in
\cite{Dasgupta:1996ij}. The conventions of \cite{Dasgupta:1996ij} are
such that $\tau = e^{\frac{\pi i}{3}}$, which differs from our case by
$1$, i.e. a $T$-monodromy. However, we stick with these conventions
throughout, as they agree with \cite{DeWolfe:1998zf} , the methods of
which we will use extensively.
 
It will be of importance four our analysis that the monodromy associated with
an $E6$ 7-brane can be written as
\begin{equation} 
  \cM_{E_6} =\eC^2\eB\eA^5 = \eC\,\cM_{D_4} \eA,
\end{equation}
corresponding to a left-right ordering $AAAAABCC$, where the monodromy of an
$SO(8)$ 7-brane sits in the middle. The leftmost A brane can
be brought past the $SO(8)$ stack with an orientation reversal, so that the
order becomes $AAAABCAC$. This structure will be important in what follows, because the
movement in moduli space that we will use to study instanton zero
modes involves ``popping off" the $AC$ associated to each $E_6$
singularity, leaving behind an $SO(8)$ singularity. It is also
important to note that
\begin{equation}
	\cM_{AC} = \eC\eA = \begin{pmatrix} 2 & -3 \\ 1 & -1 \end{pmatrix}\, ,
\end{equation}
from which it can be seen that $AC$ also leaves $\tau = e^\frac{\pi
  i}{3}+1$ invariant.

In addition to acting on the axio-dilaton $\tau$, the monodromy associated with a generic
$7$-brane also acts on the field strengths $H_3 \equiv dB_2$ and $F_3 \equiv dC_2$ 
of the Neveu-Schwarz B-field and the Ramond-Ramond two-form.
This means, for example, that by crossing the branch cut associated
with a generic $7$-brane, the $B_2$ charge associated with the
fundamental string becomes a combination of NS and RR charge, so that
the fundamental string becomes a string with both F-string and
D-string charge. A string with $p$ units of F-string charge and $q$
units of D-string charge is known as a $(p,q)$-string, which are the
types of strings that can end on $(p,q)$ 7-brane.

\begin{figure}[thbp]
	\centering
	\includegraphics[scale=1]{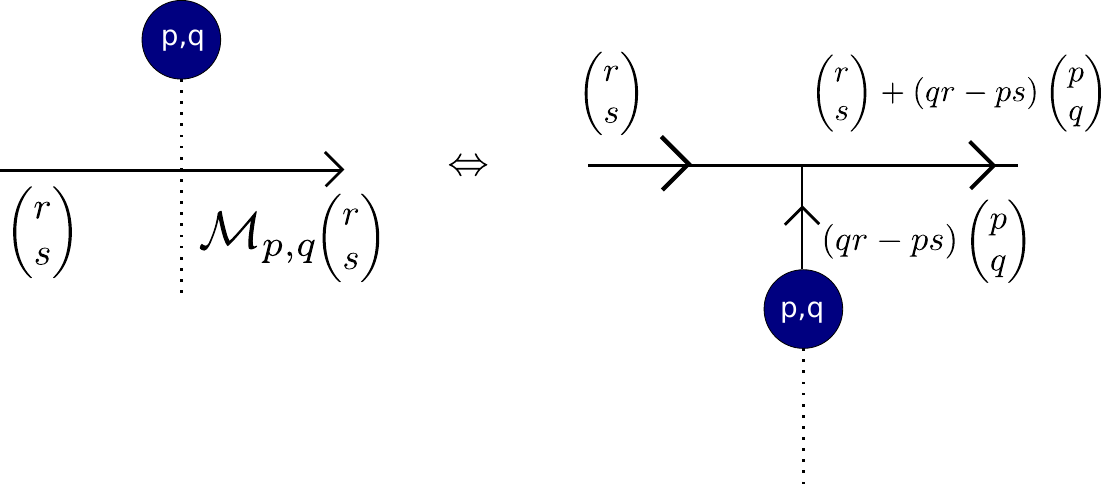}
	\caption{The transformation of an $(r,s)$-string upon crossing
          the branch cut associated with a $(p,q)$ 7-brane. On the
          right is a junction configuration, which is equivalent to
          the original configuration on the left via the Hanany-Witten
          effect.}
	\label{fig:junction example}
\end{figure}

It is known \cite{Gaberdiel:1997ud} that a string which crosses the
branch cut of a $(p,q)$ 7-brane is equivalent to a configuration where
the string has passed through the brane and grown a prong by a process
that is the U-dual of the Hanany-Witten effect
\cite{Hanany:1996ie}. The resulting configuration is a string junction
of the kind that we have encountered in the previous section. Unlike
fundamental strings, which only can give two-index representations of
the classical Lie algebras, string junctions can also fill out
representations of exceptional algebras \cite{DeWolfe:1998zf}. We have
depicted the equivalence between crossing a branch cut and growing a
prong, together with the type and multiplicity of prongs in the
junction, in figure~\ref{fig:junction example}.  Junctions between a
$D3$ brane probe and a set of 7 branes are also known to reproduce the
BPS states of $\cN=2$ $d=4$ theories with ADE flavor symmetry
\cite{DeWolfe:1998bi}, where the $7$ branes define the flavor
algebra. Such junctions can also be used to study charged instanton
zero modes if the junction is connected to a $D3$ brane transverse to
the $7$-branes, rather than to a $D3$ probe. We will proceed to do so
in the rest of this section.

\subsection{F-theory on K3: from strongly coupled $E_6^3$ to $SO(8)^4$
  at weak coupling}

We present an illustrative example using F-theory on an elliptically
fibered $K3$. We choose this example because it is simple, while still
allowing for the type of physics we would like to study, since the
moduli space of $K3$ contains regions of weak coupling and regions of
strong coupling with exceptional singularities. In particular, there
is a path in the moduli space of $K3$ which interpolates between three
$E_6$ singularities at strong coupling and four $SO(8)$ singularities
at weak coupling, which then allows for the appearance of sixteen
D7-branes and four O7-planes, as usual. Wrapping a euclidean D3 on the
$\bP^1$ base of $K3$, we show that some of the standard sixteen 3-7
strings of weakly coupled type IIB ($\lambda$ modes) are states which
contribute to $\textbf{27}$'s of $E_6$, while others are massive string
junctions which become massless in certain regions of moduli space.

Our study of $\lambda$ modes as string junctions will involve following the
movement of the junctions on a particular path in complex
structure moduli space. This path includes both strong coupling and weak
coupling points, but always maintains \emph{constant} coupling.
It is important to note that the constant coupling limit
is different from Sen's limit, which is a weak coupling limit
with possibly varying (though only slightly) coupling. On some regions of moduli
space they will overlap. For example, the region of the moduli space at
weak coupling with $SO(8)^4$ gauge symmetry is also at constant coupling, since
the D7-branes locally cancel the tadpoles of the O7-planes.

Let us review the constant coupling limit \cite{Sen:1996vd} of F-theory on $K3$. We write elliptically fibered $K3$ in Weierstrass form as 
\begin{equation}
y^2 = x^3 + fx + g,
\end{equation}
where $f\in H^0(\bP^1, K_{\bP^1}^{-4})$ and $g\in H^0(\bP^1,K_{\bP^1}^{-6})$.
Since $\bP^1$ is $\bC$ with a point at infinity, we can go to a patch while still
being fairly general, which turns $f$ and $g$ into (generically) inhomogeneous
polynomials of degree $8$ and $12$ in a complex coordinate $z$. The modular
parameter $\tau$ of the fiber determines the $j$-function as
\begin{equation}
j(\tau) = \frac{4(24f)^3}{4f^3+27g^2}\, .
\end{equation}
Let us go to a region of moduli space where $f(z) = \alpha(\phi(z))^2$
and $g(z) = \phi(z)^3$, with $\phi(z) = \prod_{i=1}^4(z-z_i)$ an
arbitrary polynomial of degree four. In this case the $j$-function
becomes
\begin{equation}
  \label{eq:j-constant-coupling}
j(\tau) = \frac{55296 \alpha^3}{4\alpha^3 + 27}\, .
\end{equation}
It is clear that the $j$-function, and therefore $\tau$, is constant
on the base, and therefore this region of $K3$ moduli space
corresponds to F-theory at constant coupling. If one restricts even
further where we are in moduli space by tuning $\alpha$ such that
$4\alpha^3 + 27$ vanishes, the $j$-function blows up, which
corresponds to $g_s \to 0$. This is precisely the specialization of
Sen's limit of F-theory on K3 to constant coupling. In general,
however, the constant coupling parameterization giving
\eqref{eq:j-constant-coupling} is \emph{not} at weak coupling.

\begin{figure}[thbp]
	\includegraphics[scale=.7]{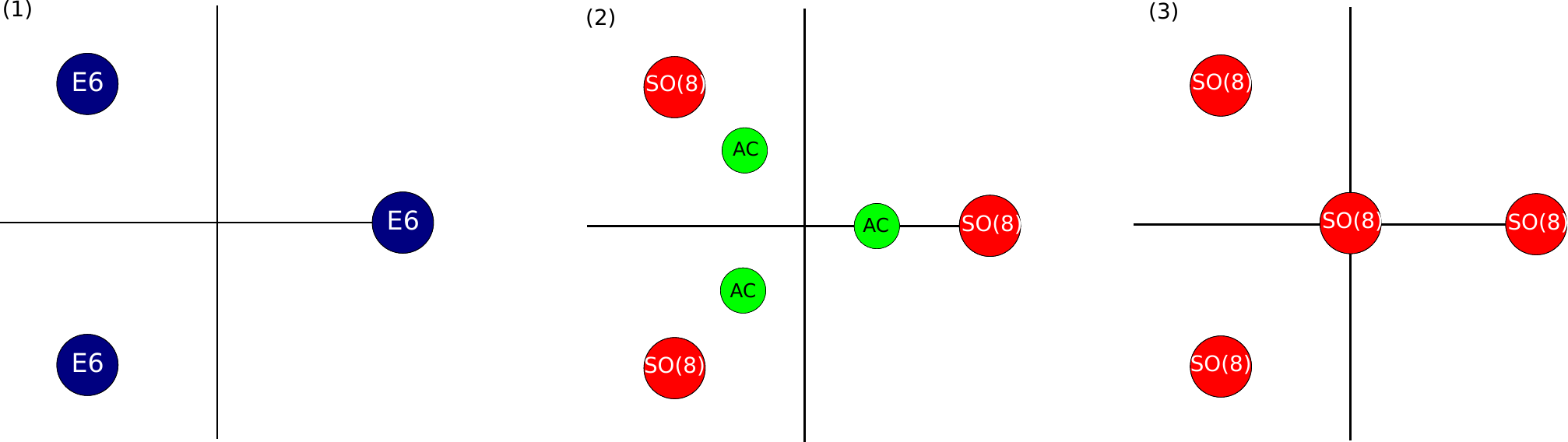}
	\caption{Depicted above is the gauge symmetry and position of
          $7$-branes as one moves from $\beta = 1$ to $\beta=0$ in the
          moduli space of $K3$. This $\alpha=0$ branch of the moduli
          space has $\tau=e^\frac{\pi i}{3}$ and all branes depicted
          have monodromy such that $\tau$ is constant under taking a
          circle around them.}
	\label{fig:betachange}
\end{figure}

It was pointed out in \cite{Dasgupta:1996ij} that there are other
branches of the constant coupling moduli space. One such branch,
useful for our purposes, comes about in the limit $\alpha \to 0$, so
that $f$ is identically zero and thus $j=0$. This corresponds to $\tau
= e^{\frac{i\pi}{3}} + 1$. This branch of moduli space is fundamentally
different than the $\alpha\ne 0$ branch, since $f=0$ allows $g$ to be
an arbitrary polynomial of degree $12$ in $z$ while maintaining
constant coupling. The two branches meet at $\alpha = 0$ when $g$ is
the cube of a quartic polynomial. Out on the $\alpha=0$ branch, one
could write (for example)
\begin{equation}
f = 0 \qquad \qquad \qquad g = \prod_{n\in\{0,1,2\}}(z-e^{\frac{2\pi n i}{3}})^3(z-\beta \, e^\frac{2\pi n i}{3})
\end{equation}
where $\beta$ is a parameter that can be tuned to move on this
branch. The intersection with the $\alpha\ne 0$ branch occurs at
$\beta = 0$. The singularity type and position of the branes as one
moves from $\beta=1$ to $\beta=0$ is depicted in
figure~\ref{fig:betachange}.  Physically, $\beta=1$ corresponds to a
point with three stacks of $7$-branes with $E_6$ gauge symmetry, which
can be represented in terms of $A$, $B$, and $C$ branes as $AAAAABC^2$,
as discussed in section \ref{sec:7 brane review}.

As one tunes $\beta$ to be less than one, one $A$ brane and one $C$
brane are pulled off of each $E_6$ stack, leaving behind three copies
of $A^4BC=SO(8)$. When $\beta=0$, the three sets of $AC$ branes have
come together to form an $(AC)^3$ stack,
which has the monodromy of an $SO(8)$ (we will present an explicit
brane motion that takes a $(AC)^3$ stack to the conventional $A^4BC$
presentation of $SO(8)$ below).  At $\beta=0$, it is important to note
that $g$ is the cube of a quartic polynomial, as in the original
constant coupling parameterization. One can then turn on $\alpha$ and
tune it to $\alpha=-(\frac{27}{4})^\frac{1}{3}$, which is the weak
coupling limit.

To summarize, the path in moduli space which we study is as
follows. Starting from the $E_6^3$ point at $\tau=e^\frac{\pi i}{3} + 1$,
move to the $SO(8)^4$ point at the same $\tau$. At these points
$g_s\sim \cO(1)$, and we can then send $g_s\mapsto 0$ by
sending $\alpha$ from $0$ to $-(\frac{27}{4})^\frac{1}{3}$ while
maintaining $SO(8)^4$ gauge symmetry. We are then in weakly coupled
type IIB.

\subsection{Exceptional charged modes, massive BPS states, and relating to IIB}
To this point, we have demonstrated that there is a path in the moduli
space of $K3$ which interpolates between $E_6^3$ and $SO(8)^4$ at $\tau
= e^{\frac{i\pi}{3}} + 1$, and then goes to $SO(8)^4$ at weak
coupling. From that region of moduli space, the $D7$-branes and
$O7$-planes can spread out over the $\bP^1$, at which point it is
clear that the 3-7 strings are the standard $\lambda$ modes of weakly
coupled type IIB. What becomes of the $\lambda$ modes as one moves in
moduli space between the $SO(8)^4$ point at weak coupling and the
$E_6^3$ point at $\tau = e^{\frac{i\pi}{3}} + 1$?

Recall that the instanton wraps the $\bP^1$ base of $K3$ and two
directions of $8d$ spacetime. At generic points in moduli space, there
is some set of 7 branes intersecting the instanton at a complex
codimension one defect in its worldvolume. Massless charged modes are
given by 3-7 junction of zero length.\footnote{We will use the
  terminology ``zero modes'' for zero length junctions. In view of the
  discussion in section~\ref{sec:anomaly inflow}, we should more
  properly be talking about massless carriers, since there we saw that
  the map between junctions and zero modes is subtle. We have not
  performed the analogous anomaly inflow analysis for charged modes,
  so we will stick to this slightly imprecise nomenclature. We hope to
  come back to this question in future work.} That is, a zero mode
represented by a string junction can attach to one set of localized
$(p,q)$ 7-branes, but not to multiple sets which are separated in the
$\bP^1$. Given a set of localized 7-branes, it was shown in
\cite{DeWolfe:1998zf} how to classify the junctions ending on that set
and determine the representation of those junctions.

Of importance is the notion of asymptotic charge, which is nothing but
the $(p,q)$ type of a free prong. For example, if the junction in
figure~\ref{fig:junction example} had its prong with charge $(r,s)$
attached to an $(r,s)$ 7-brane, then there would be one free end with
asymptotic charge $(r + (qr-ps)p, q + (qr-ps)q)$. Since the free ends
of the junctions we study will attach to a $D3$ brane and we want to
relate those junctions to the standard charged instanton zero modes in
the IIB limit, we will focus on junctions with asymptotic charge
$(1,0)$. That is, we will study junctions which are asymptotically
fundamental strings.

\subsubsection{The fate of junctions in the {\bf{27}} of $E_6$}
\label{sec:fate 27}
At the $E_6^3$ point in moduli space, the charged instanton zero modes
we study are those of asymptotic charge $(1,0)$.\footnote{There is a
  further condition on the junction: it should have self-intersection
  equal to 1 \cite{DeWolfe:1998zf}. The states that we study below all
  have this property.} To remain massless, the junctions cannot
stretch between $E_6$ stacks, and thus the study of zero modes at the
$E_6^3$ point amounts to the study of string junctions of asymptotic
charge $(1,0)$ connecting to a single $E_6$ stack, picking up a
multiplicity of $3$ for the three stacks.

The set of junctions of asymptotic charge $(1,0)$ and
self-intersection $-1$ attached to an $E_6$ stack represented by
left-right ordering $AAAAABCC$ were classified in
\cite{DeWolfe:1998zf}.  There are four
types of such junctions, given by
\begin{align}
\begin{split}
  \text{Type I:}& \qquad \qquad \,\,\,\,\,\,\, a_i \\
  \text{Type II:}& \qquad \qquad -a_i + b + c_j \\
  \text{Type III:}& \qquad \qquad -\sum_{p=1}^3 a_{i_p} + 2b+c_1+c_2 \\
  \text{Type IV:}& \qquad \qquad  -\sum_{p=1}^5 a_p + 3b + c_k + c_1 +
c_2,
\end{split}
\end{align}
where the appearance of a term $a_i$ denotes that the junction has a
prong on the $i^{th}$ $A$ brane, with a $+$ sign denoting outgoing and a
$-$ sign denoting incoming, with the terms $c_i$ similarly
defined. The $A$ brane indices run from $1\dots 5$ and the $C$ brane
indices from $1\dots 2$, as one would expect. Junctions of Type III
end on three of the five $A$ branes, with $i_1 < i_2 < i_3$. A simple
counting shows that there are $5$ junctions of Type I, $10$ junctions
of Type II, $10$ junctions of Type III, and $2$ junctions of Type
$IV$, for a total of $27$ junctions. Moreover, they have the Lie
algebraic structure of a \textbf{27} of $E_6$.

\begin{figure}[ht]
\centering
\begin{minipage}[b]{0.3\linewidth}
(1)\includegraphics[scale=.3]{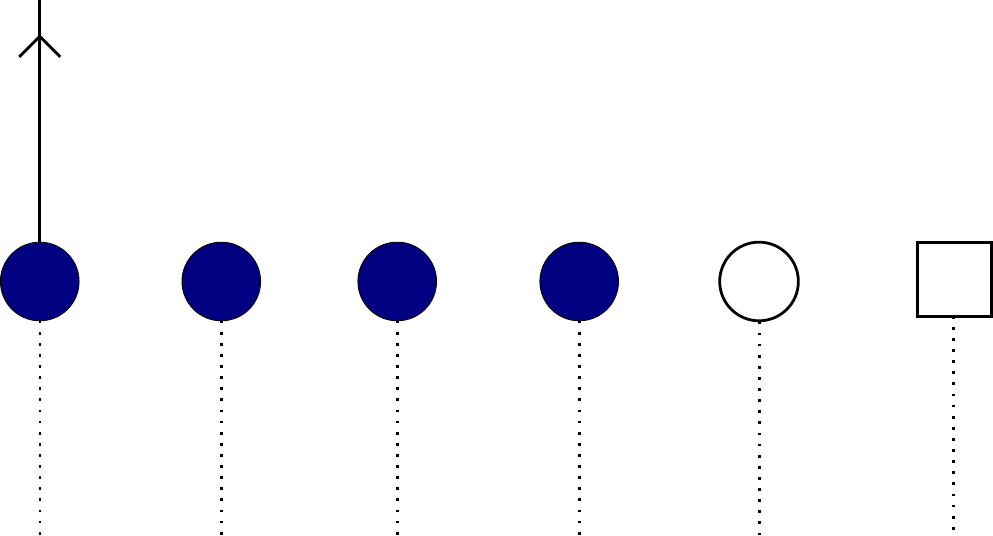}
\end{minipage}
\hspace{1.5cm}
\begin{minipage}[b]{0.3\linewidth}
(2)\includegraphics[scale=.3]{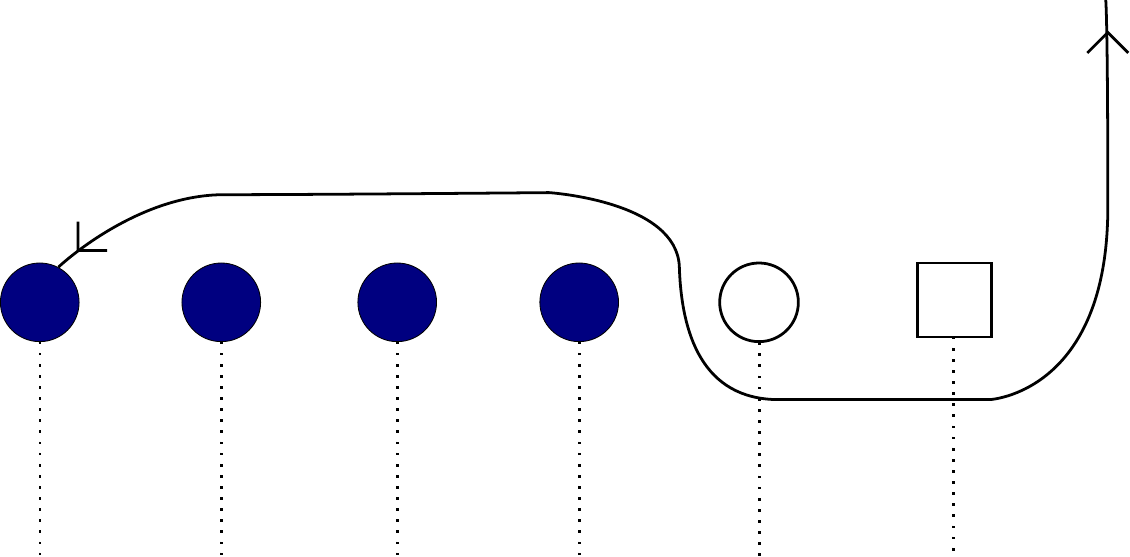}
\end{minipage}
\vspace{1cm}\\
\begin{minipage}[b]{0.3\linewidth}
(3)\includegraphics[scale=.3]{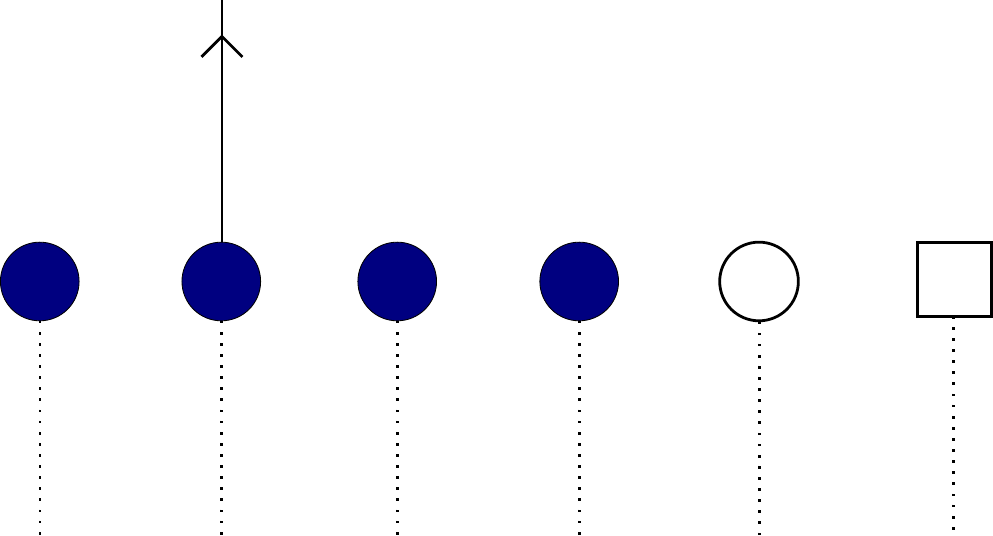}
\end{minipage}
\hspace{1.5cm}
\begin{minipage}[b]{0.3\linewidth}
(4)\includegraphics[scale=.3]{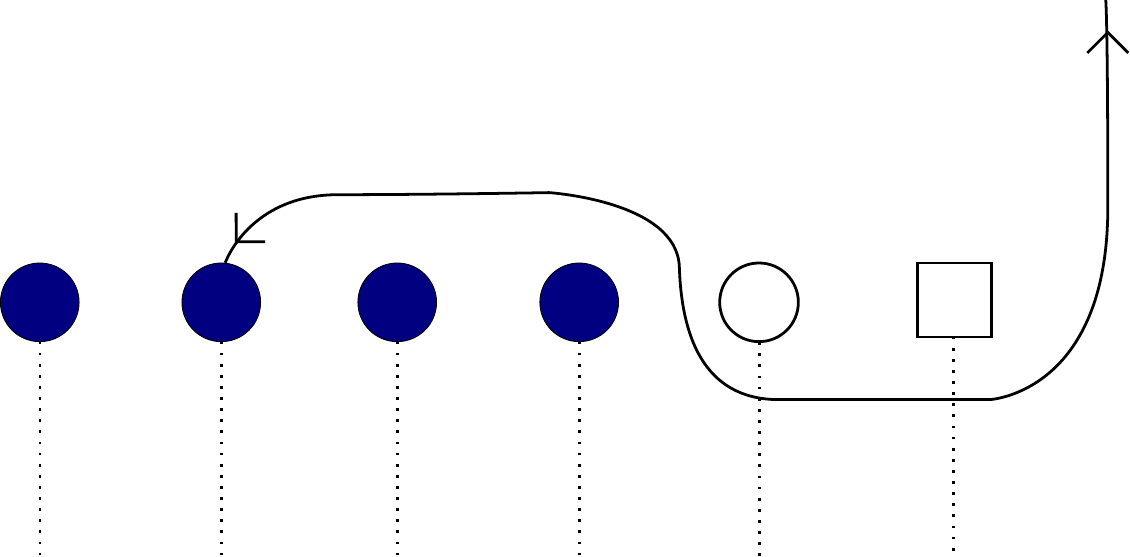}
\end{minipage}
\vspace{1cm}\\
\begin{minipage}[b]{0.3\linewidth}
(5)\includegraphics[scale=.3]{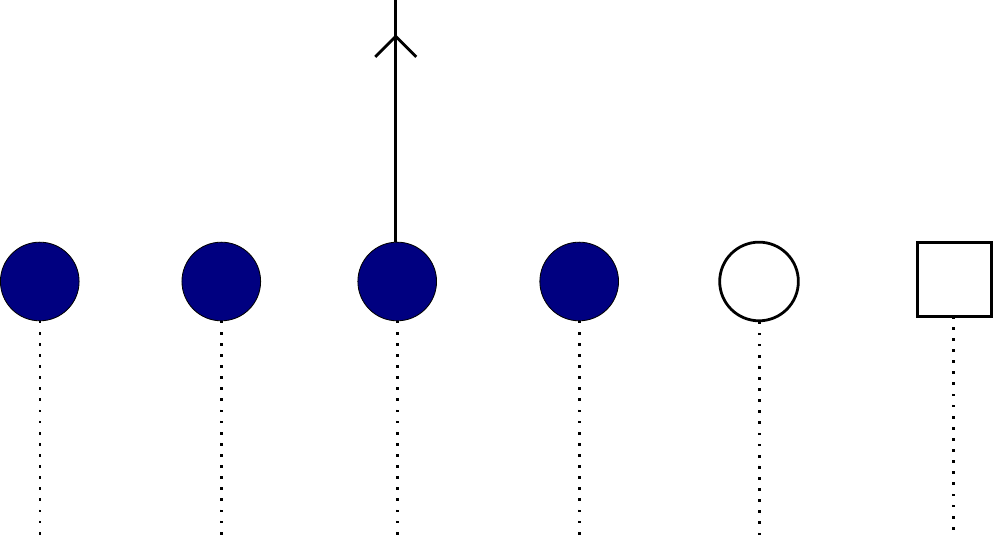}
\end{minipage}
\hspace{1.5cm}
\begin{minipage}[b]{0.3\linewidth}
(6)\includegraphics[scale=.3]{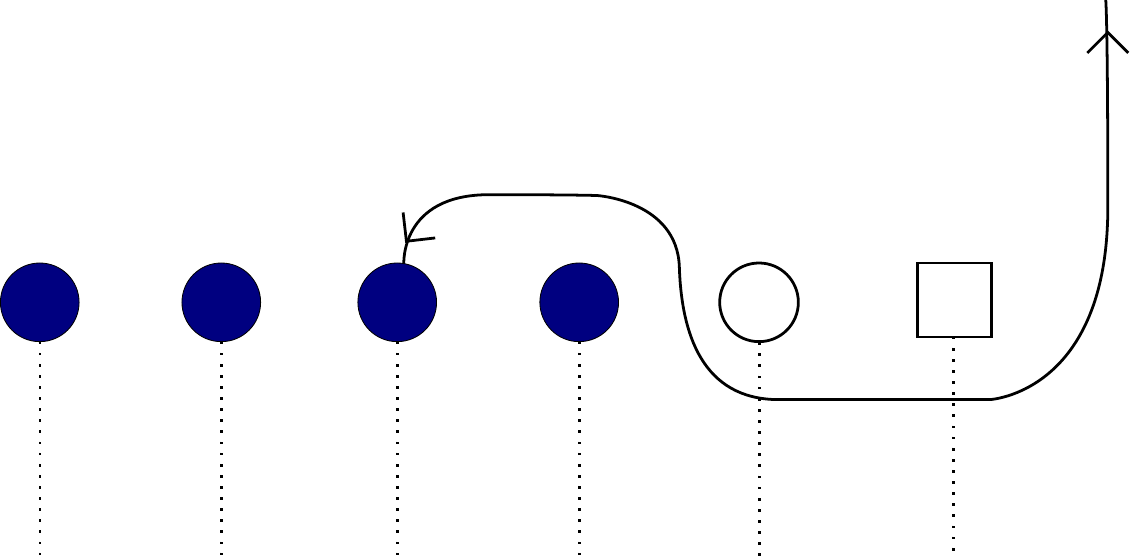}
\end{minipage}
\vspace{1cm}\\
\begin{minipage}[b]{0.3\linewidth}
(7)\includegraphics[scale=.3]{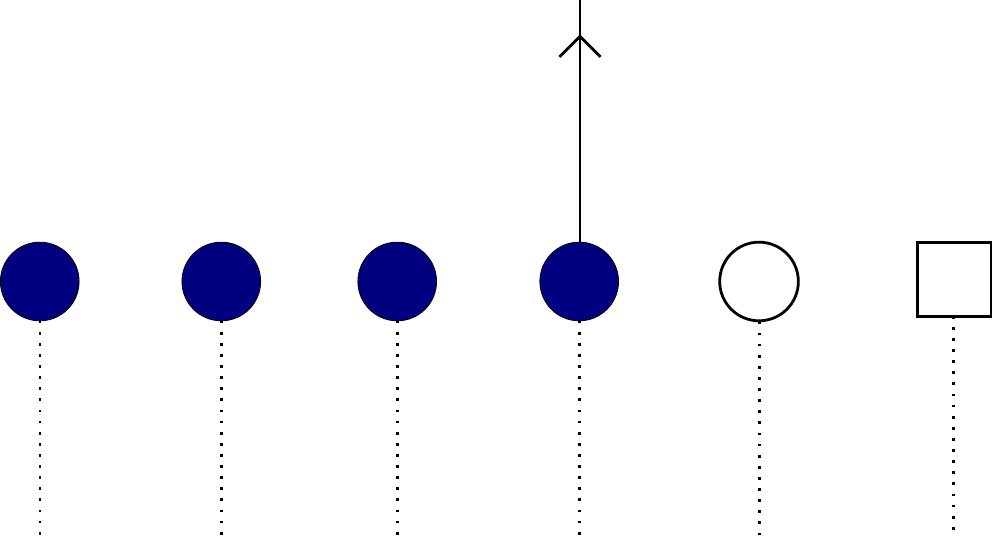}
\end{minipage}
\hspace{1.5cm}
\begin{minipage}[b]{0.3\linewidth}
(8)\includegraphics[scale=.3]{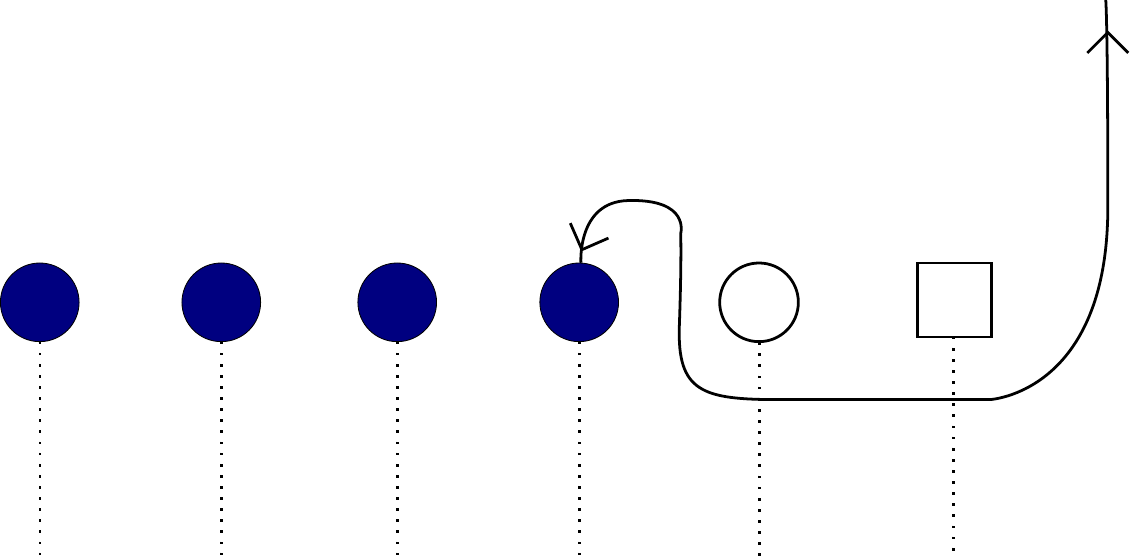}
\end{minipage}
\caption{Upon tuning $\beta < 1$, an $A$ and a $C$ brane are pulled
  away from the $E_6$ stack, leaving an $SO(8)$ stack. Depicted are the
  $8$ junctions out of the \textbf{27} which are localized at the
  $SO(8)$ after tuning $\beta < 1$. These junctions fill out an
  \textbf{8}$_v$ of $SO(8)$.}
\label{fig:8v}
\end{figure}

What happens to these junctions as one moves in moduli space? Recall
that the path we have defined interpolating between $E_6^3$ and
$SO(8)^4$ involves removing an A brane and a C brane from each $E_6$,
leaving an $SO(8)$ behind. Any junction in the $\textbf{27}$ with a
prong on the removed $A$ or $C$ brane becomes of finite length upon
moving in the defined path on moduli space. Thus, all junctions of
Type III and Type IV have finite length. Six junctions of Type II have
finite length while the other four are left behind, localized at the
$SO(8)$. All junctions of Type I remain of zero length, though one is
localized on the A brane that has been removed, while the other four
are localized at the $SO(8)$.  This gives a total of $8$ junctions
that are still of zero length and localized at the $SO(8)$. These
junctions, depicted in figure~\ref{fig:8v} are known to fill out an
\textbf{8}$_v$ of $SO(8)$.

In summary, of the $27$ junctions filling out a \textbf{27} of $E_6$,
$18$ become massive states, $1$ is a massless junction attached to the
removed $A$ brane, and $8$ are left behind as massless junctions that
fill out an \textbf{8}$_v$ of $SO(8)$. Since there are three $E_6$
stacks initially, there is an overall multiplicity of three for all of
the mentioned junctions. This leaves a puzzle, however: in type IIB at
the $SO(8)^4$ point, there should be four sets of charged instanton
zero modes transforming in the \textbf{8}$_v$ of $SO(8)$, whereas in
the present analysis we have seen that only three copies of
\textbf{8}$_v$ arise from Higgsing $\textbf{27}$'s of $E_6$. More
specifically, there are $3$ massless junctions at the fourth $SO(8)$
coming from the Type I junctions associated with the removed $A$ brane
of each $E_6$, but there are still $5$ junctions in the fourth
\textbf{8}$_v$ which are not accounted for.

\subsubsection{The fourth \textbf{8}$_v$ of $SO(8)$ and massive states}
There is one complication that we must discuss before addressing the
appearance of the fourth \textbf{8}$_v$ as a continuous process upon
movement in moduli space.  The junctions that fill out an
\textbf{8}$_v$ are given in figure~\ref{fig:8v}, but those junctions
are for the case where the $SO(8)$ is realized as $AAAABC$. In the
case of moving from $E_6^3$ to $SO(8)^4$, the fourth $SO(8)$ stack
located at $z=0$ has left-right ordering $ACACAC$. Before we can
discuss the appearance of the \textbf{8}$_v$ we must perform a
technical exercise to determine how the junctions forming an
$\textbf{8}_v$ of $SO(8)$ as $AAAABC$ are transformed under the brane
movement $AAAABC \mapsto ACACAC$. In appendix~\ref{sec:untangle} we
explicitly untangle each junction that contributes to the
$\textbf{8}_v$ of $AAAABC$ so that we can discuss what becomes of the
fourth $\textbf{8}_v$ upon movement in moduli space. Here we present
one of the more difficult examples in detail, so that our method is
clear.

Consider the example of junction $(3)$ of figure~\ref{fig:untangle2}
in appendix~\ref{sec:untangle}, which we will show becomes junction
$(4)$ after untangling. This involves determining a way to take the
branes around the nearby branch cuts to transform $AAAABC$ into
$ACACAC$. It also requires keeping track of the string as one moves
the branes.  The starting configuration is junction $(3)$
\begin{equation*}
\includegraphics[scale=.5]{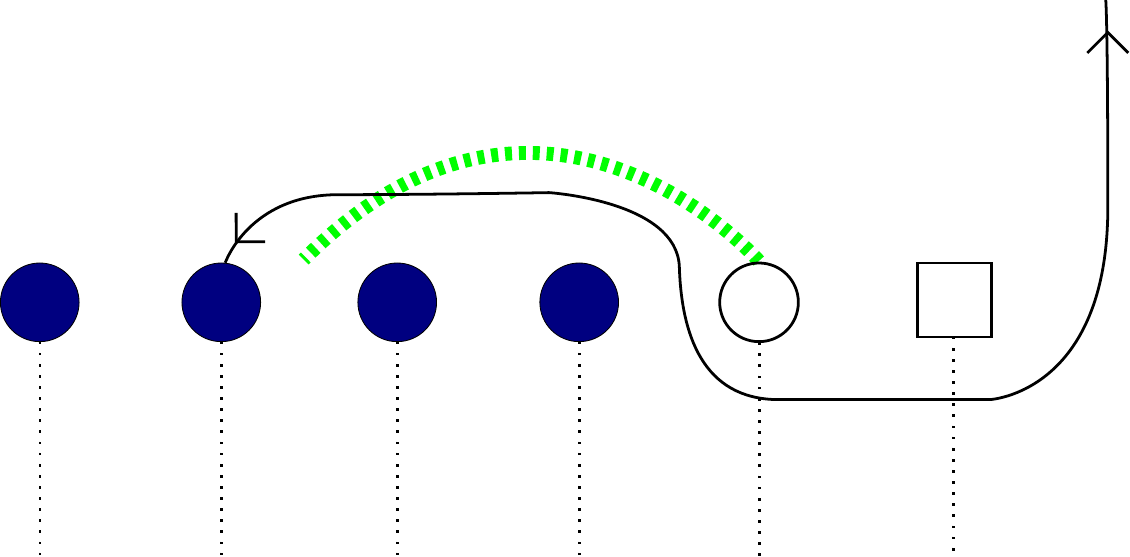}
\end{equation*}
where we have represented $A$-branes, $B$-branes, and $C$-branes as
filled circles, hollow circles, and hollow squares, respectively. The
$(p,q)$ string is represented by a solid line, and the arrow into the
A-brane indicates a $(1,0)$ string going in, or equivalently a
$(-1,0)$ string coming out. It has asymptotic charge $(1,0)$. The
dotted green line represents the path taken for the next step in the
untangling process. Taking that path transforms the configuration to
\begin{equation*}
\includegraphics[scale=.5]{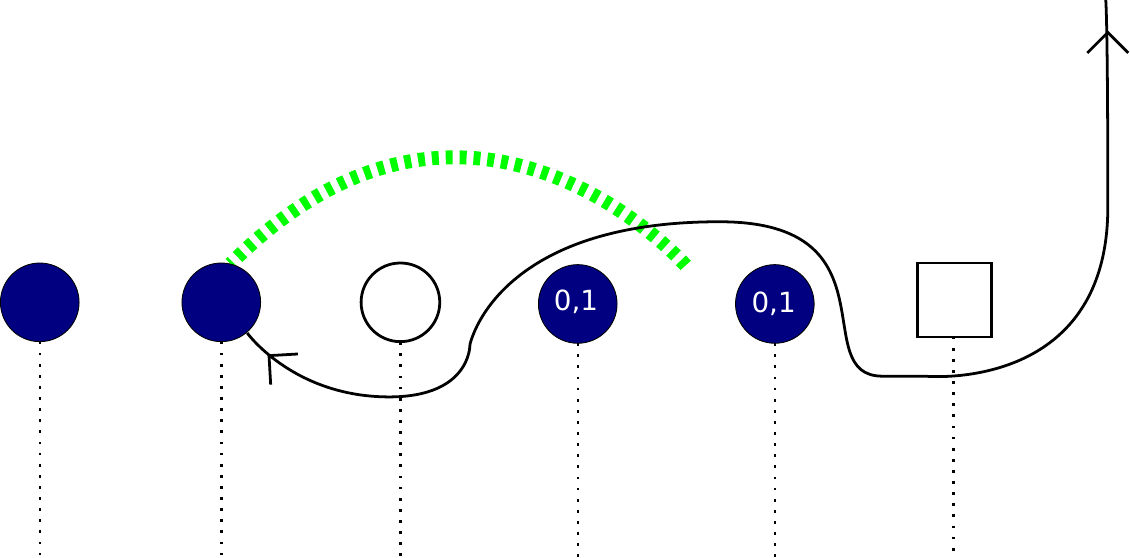}
\end{equation*}
where branes labeled with white numbers $(p,q)$ denote a brane of that
type, and the green path again denotes the next brane movement in the
untangling process. The two branes are $(0,1)$ type because they have
passed the branch cut of the $B$ brane that was moved. Taking the next
movement, we arrive at
\begin{equation*}
\includegraphics[scale=.5]{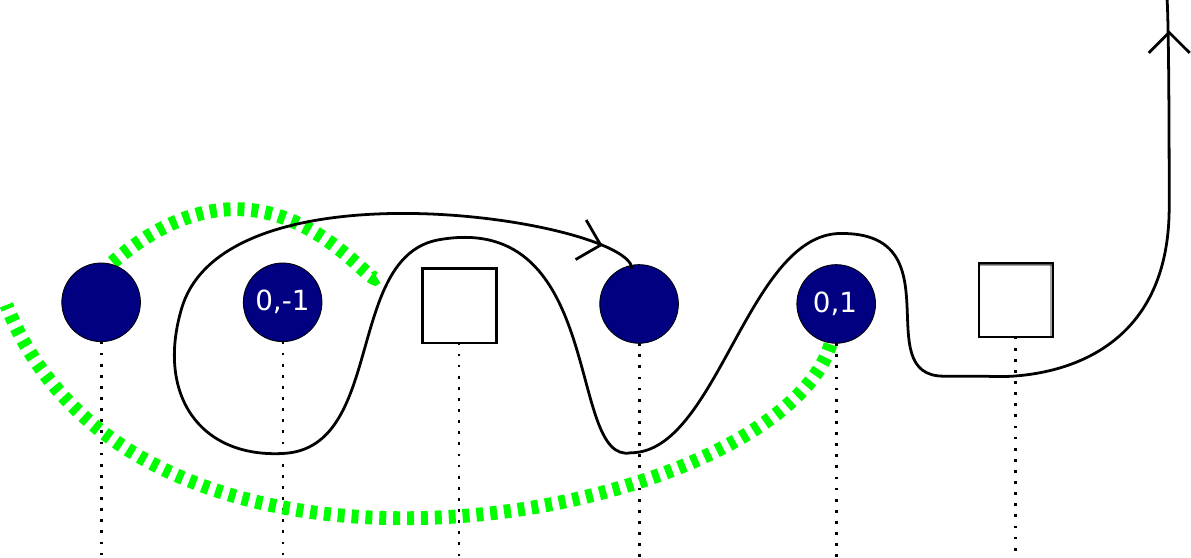}
\end{equation*}
and then takes the final two movements to arrive at
\begin{equation*}
\includegraphics[scale=.5]{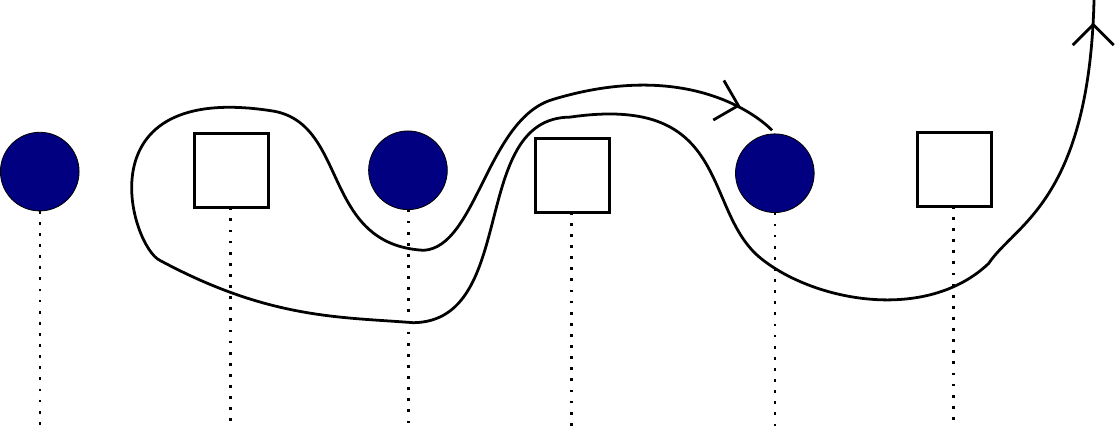}
\end{equation*}
As a double check, if one starts with the $(-1,0)$ string attached to
the rightmost $A$ brane and applies the appropriate $SL(2,\bZ)$
transforms, one sees that this configuration does indeed have
asymptotic charge $(1,0)$, as it must.

In untangling the branes and strings, here and in appendix
\ref{sec:untangle}, we have not used the Hanany-Witten effect to
change strings crossing branch cuts into junctions. The main reason
for this is that fewer steps offer more simplicity, and we have found
that it also makes comparing states a bit easier.
Nevertheless, as an example let us convert the final $(p,q)$ string
configuration just derived into a junction.  Applying the rules for
converting $(p,q)$ strings crossing branch cuts into junctions, as
exemplified in figure~\ref{fig:junction example}, one arrives at
\begin{equation*}
\includegraphics[scale=.4]{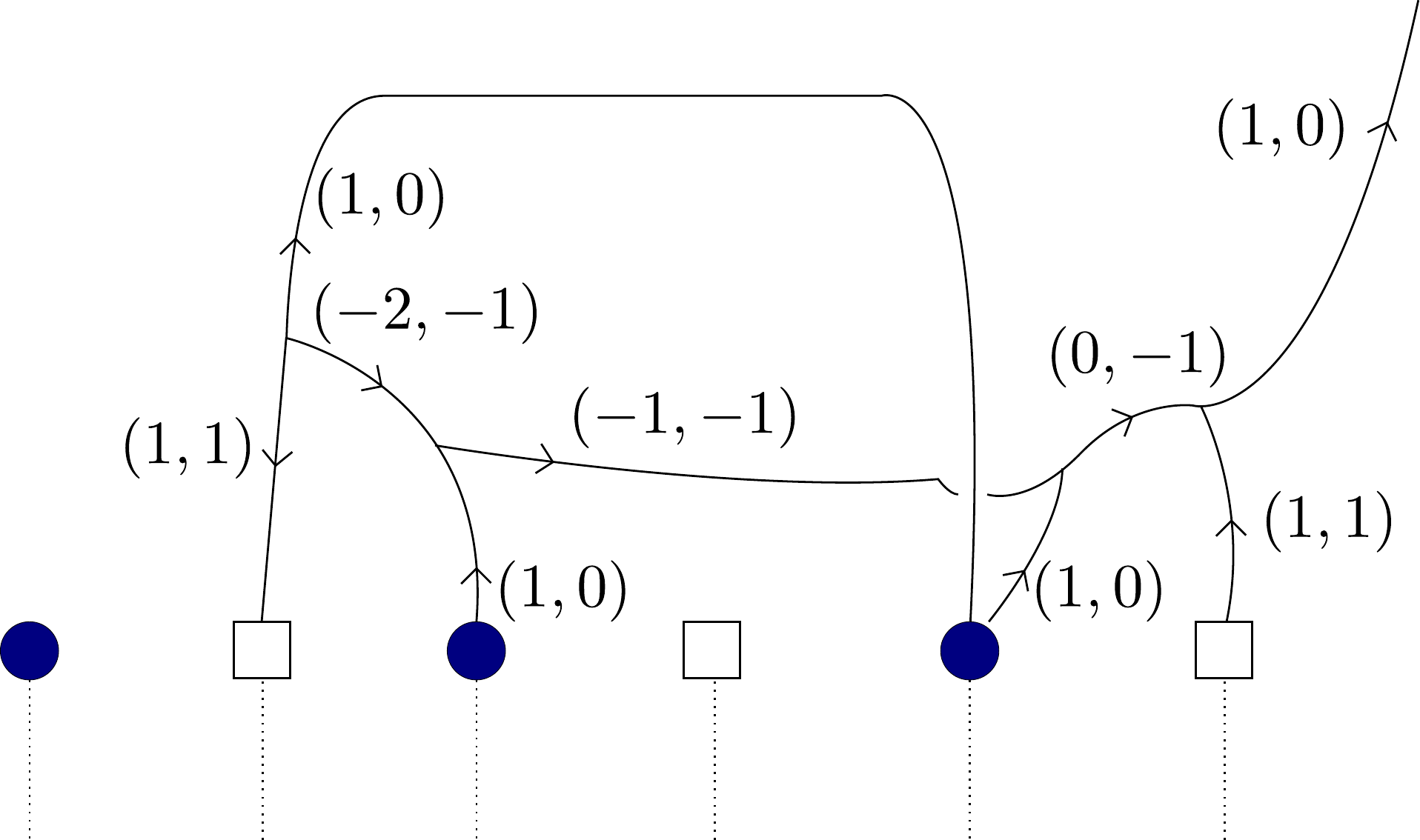}
\end{equation*}
which looks quite complicated. This junction is massless when the
three $AC$ combinations from the $E_6$'s coalesce at $\beta=0$ and
contributes to an \textbf{8}$_v$. If one considering the opposite path
in moduli space, though, moving from $\beta = 0$ to $\beta = 1$, the
junction is such that there will be prongs stretching between
different $E_6$ stacks. That is, this component of the fourth
\textbf{8}$_v$ is a massive BPS states at the $E_6^3$ point in moduli
space.

In fact, upon moving to the $E_6^3$ point, five of the junctions in the
fourth \textbf{8}$_v$ stretch between distinct $E_6$ stacks and are
therefore massive. Junctions (2) and (4) of figure~\ref{fig:untangle1}
and junction $(8)$ of figure~\ref{fig:untangle2} are the three which
do not become massive upon moving to the $E_6^3$ point. They are
precisely the three junctions of Type I that contribute to the
$\textbf{27}$'s, one for each $E_6$, that were discussed in section
\ref{sec:fate 27}.

\subsection{A brief summary and outlook}
We have seen that string junctions are useful for studying 3-7
instanton zero modes in F-theory, even upon movement in moduli
space. We defined a very specific path in the complex structure moduli
space of F-theory on $K3$ which interpolated between $E_6^3$ symmetry
at strong, constant coupling and $SO(8)^4$ symmetry in weakly coupled
type IIB.

At the $E_6^3$ point in moduli space, the 3-7 instanton zero modes with
charge (1,0) ending on the D3 instanton are three \textbf{27}'s of
$E_6$. In terms of $A$ branes, $B$ branes, and $C$ branes $E_6$ is
represented as $AAAAABCC$. Moving from the $E_6^3$ point to the
$SO(8)^4$ point, an $A$ brane and a $C$ brane are pulled off of each
$E_6$ stack, which coalesce to form the fourth $SO(8)$.  Of the 27
junctions filling out each \textbf{27} at the $E_6^3$ point, $18$
become massive BPS states with prongs between separated branes, $1$ is
a fundamental string attached to the $A$ brane which was removed, and
$8$ form the \textbf{8}$_v$ of the $SO(8)$ that was left behind when
the $AC$ combination was removed. Those $3$ fundamental strings
associated with each $A$ brane that was removed are three of the
junctions necessary to form the fourth \textbf{8}$_v$ which is known
to be present in type IIB. The other $5$ junctions contributing to the
fourth \textbf{8}$_v$ are massive BPS states for $\beta > 0$ that
become massless when $\beta = 0$.

The appearance of additional light states from ones that are massive
at generic points in moduli space is common in string theory. In
F-theory it is simple to see in the junction picture, because
junctions between separated branes have finite length, but become
massless when the branes come together. The statements we have made
about massless and massive junctions upon movement in moduli space are
not limited to charged instanton zero modes, though. In particular,
though we were interested in attaching the ``free end" with asymptotic
charge $(1,0)$ to a $D3$ instanton transverse to the 7-branes, it
could also attach to an $A$ brane, or a $D3$ parallel to the
7-branes. In that case, one would have extra matter, rather than extra
instanton zero modes, as one moves in moduli space. In realistic
scenarios involving instantons one will have a combination of both
phenomena at play, so we now proceed to study situations in which both
light matter and extra zero modes appear at particular points in
moduli space.

\section{Instanton physics via heterotic / F-theory duality}
\label{sec:heterotic duality}

In section \ref{sec:string junctions} we demonstrated that string
junctions provide a useful tool for uncovering interesting
relationships between charged instanton zero modes in F-theory and
type IIB. In an example, we showed that to reproduce known type IIB
results smoothly from F-theory one has to take into account both the
Higgsing of massless junctions and massive junctions which become
massless at certain points in moduli space. We expect this behavior to
be fairly common.
Though the string junction picture gives a consistent and illuminating
picture of charged zero modes, it is still not clear how to use such
knowledge to actually compute superpotential corrections.  In this
section we utilize duality between the heterotic string and F-theory
to compute (via duality) instanton corrections in F-theory at points
in moduli space with exceptional gauge symmetry. On the heterotic
side, worldsheet instantons provide the superpotential corrections we
will study
\cite{Dine:1986zy,Dine:1987bq,Distler:1986wm,Distler:1987ee,Berglund:1995yu}.
Of particular interest to us is that
\cite{Buchbinder:2002ic,Buchbinder:2002pr} obtained the dependence of
the worldsheet instanton generated superpotential on the heterotic
vector bundle moduli space. This dualizes to ED3/M5 instanton
corrections in F-theory depending on the fourfold complex structure
moduli. Microscopically, the means that the structure of the
correction is dependent on the position of $7$-branes, due to the
appearance of extra zero modes when the positions and structure of
$7$-branes take a particular form. By reinterpreting the known answer
from the heterotic in the F-theory language we will obtain information
about the physics away from weakly coupled limits.

\medskip

For the convenience of the reader, since we introduce a good deal of
notation in what follows, we collect our naming and notational
conventions here. As will be discussed in section \ref{sec:het f},
heterotic / F-theory duality requires a number of fibration
structures, given by
\begin{align}
&T^2 \longhookrightarrow X \overset{\pi_H}{\longtwoheadedrightarrow} B_2 \qquad \qquad
T^2 \longhookrightarrow Y \overset{\pi_F}{\longtwoheadedrightarrow} B_3 \notag \\
&K3 \longhookrightarrow Y \overset{\pi_{K3}}{\longtwoheadedrightarrow} B_2 \qquad \qquad
\bP^1 \longhookrightarrow B_3 \overset{\pi_{\bP^1}}{\longtwoheadedrightarrow} B_2,
\end{align}
where $X$ is the heterotic elliptic $CY_3$ and $Y$ is the F-theory
elliptic $CY_4$ that is also $K3$ fibered. Our convention is to name
projection maps with a subscript denoting the fiber.  The only
ambiguity is for the two elliptic fibrations, in which case we name
the maps $\pi_H$ and $\pi_F$ for heterotic and F-theory,
respectively. We give a summary of the notation used in this section
in tables~\ref{table:geometry notation table} and \ref{table:bundle
  notation table}.
\begin{table}[htbp]
	\centering
	\begin{tabular}{l|c}
          Symbol & Definition and/or Comment \\ \hline
          $X$ & Elliptic $CY_3$ of the heterotic compactification. \\
          $\sigma_H$ & Section of $X$. \\
          $Y$ & Elliptic $CY_4$ of F-theory. Also $K3$ fibered. \\
          $\sigma_F$ & Section of $Y$. \\
          $B_2$ & Twofold base of $X$ and $Y$. \\
          $B_3$ & Threefold base of $Y$. It is a $\bP^1$ fibration over $B_2$.\\
          $\Sigma$ & Curve in $B_2$ wrapped by the instanton. \\
          $\cE$ & $\pi_H^{-1}(\Sigma)$.\\
          $\pi_{K3}^{-1}(\Sigma)$ & Divisor wrapped by $M5$
          instanton. \\
          $\chi$ & $\Sigma \cdot c_1(TB_2)$
	\end{tabular}
	\caption{Table of notation for geometric objects used in discussing heterotic / F-theory
	duality and instantons. Please see the text for context and discussion.}
	\label{table:geometry notation table}
\end{table}
\begin{table}[htbp]
	\centering
	\begin{tabular}{l|c}
          Symbol & Definition and/or Comment \\ \hline
          $V$ & Holomorphic vector bundle on $X$, $V = V_1 \oplus V_2$. \\
          $V_1$ & Bundle studied for WS instanton corrections. \\
          $(C,L)$ & Spectral pair equivalent to $V_1$ under Fourier-Mukai. \\ \hline
          $C$ & A divisor in $X$ of class $n\, \sigma_H + \pi_H^{-1} \eta$, \, with $\eta$ a curve in $B_2$ \\
          $f_C$ & Polynomial whose zero locus is $C$. \\
          $a_q$ & Sections of line bundles on $B_2$ appearing in $f_C$. \\ \hline
          $c$ & $c\equiv C\cdot \cE$. Divisor in $\cE$ of class $n\sigma_\cE + rF$ \\
          $f_c$ & Polynomial whose zero locus in $\cE$ is $c$. $f_c\equiv f_C|_\cE$ \\
          $\tilde a_q$ & Section of line bundles on $\Sigma$ appearing in $f_c$. \\ \hline
          $L$ & Line bundle on $X$, gives a line bundle on $C$ via restriction. \\
          $\tilde \cL$ & Line bundle on $\cE$, $\tilde \cL\equiv L|_\cE \otimes O_\cE(-F)$. \\
          $\cL$ & Line bundle on c, $\tilde \cL \equiv \tilde \cL|_c$. $\Pfaff_\Sigma = 0 \Leftrightarrow h^0(c,\cL) > 0$.
	\end{tabular}
	\caption{Table of notation for objects related to the rank $n$ holomorphic vector bundle
          $V$ on $X$.}
	\label{table:bundle notation table}
\end{table}

\begin{figure}[thbp]
\begin{center}
\includegraphics[scale=.8]{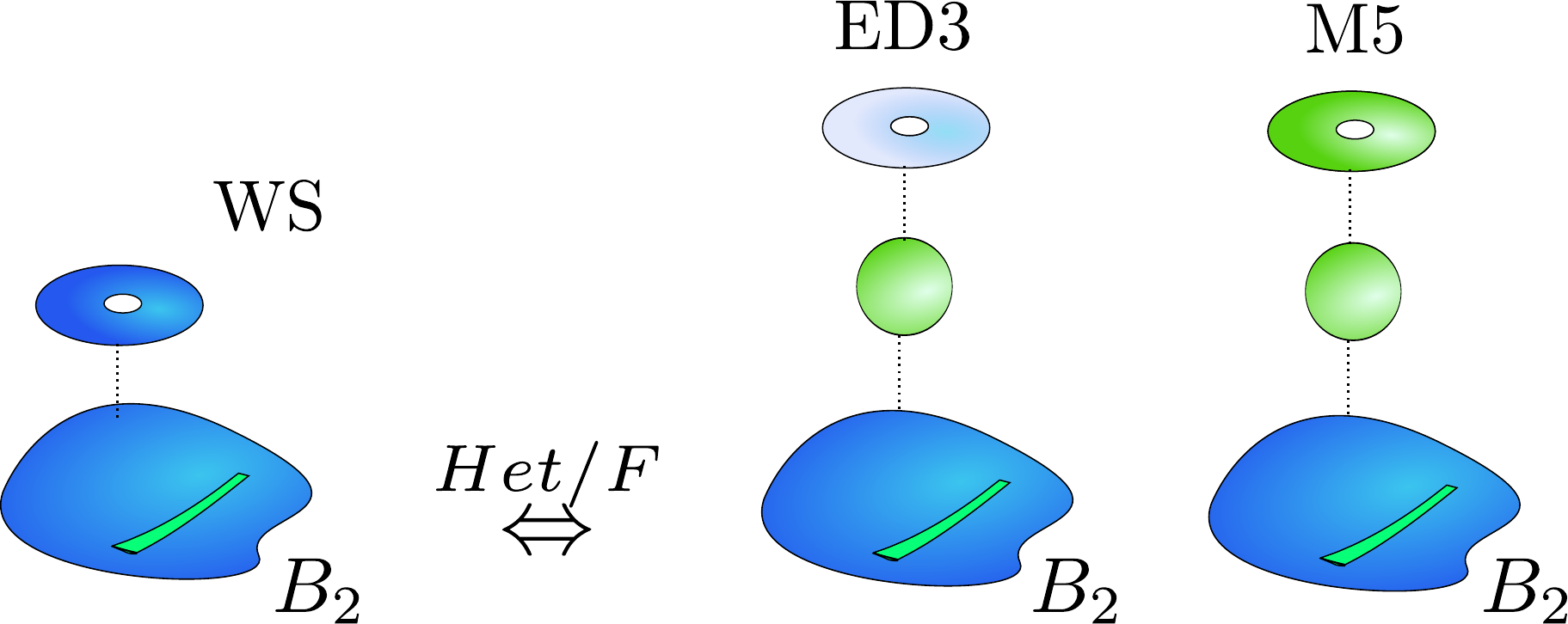}
\caption{A depiction of a heterotic worldsheet instanton wrapped on a
  curve $\Sigma$ in $B_2$ and the dual $ED3/M5$ instanton in
  F-theory. The heterotic $T^2$ becomes a $\bP^1$ base of the $K3$
  fiber in F-theory which is wrapped by the instanton. Green denotes
  the instanton wrapping that part of the geometry. On the right we
  show both the IIB and M-theory descriptions of the dual F-theory. In
  the first the elliptic fiber is an auxiliary description for the
  variation of the axio-dilaton $\tau$, and the physical object is an
  euclidean $D3$, while in the second the fiber is physical, and the
  $M5$ brane instanton wraps it.}
\label{fig:instanton comparison}
\end{center}
\end{figure}

\subsection{Heterotic / F-theory duality }
\label{sec:het f}
We begin by reviewing the basics of heterotic / F-theory duality,
focusing on the aspects most relevant for our discussion. For a more
detailed summary with an emphasis on F-theory GUTs, see for example
\cite{Weigand:2010wm}. The simplest example is that the $E_8\times E_8$
heterotic string on $T^2$ is dual to F-theory on $K3$. The heterotic
side is endowed with a rank $n$ holomorphic vector bundle over $T^2$
which encodes the breaking of $E_8\times E_8$. On the F-theory side, the
bundle moduli which specify the symmetry breaking are encoded purely
in terms of geometry. That is, F-theory gives a geometric depiction of
the moduli space of holomorphic vector bundles on $T^2$. By making
this precise, we will be able to map results from heterotic
compactifications to F-theory.

In six dimensions, the heterotic string on $K3$ is dual to F-theory
compactified on a Calabi-Yau threefold which is an elliptic fibration
over a Hirzebruch surface $\bF_p$
\cite{Morrison:1996na,Morrison:1996pp}. This can be understood by
fibering the compactification manifolds in the eight-dimensional case
over a $\bP^1$. Since the Hirzebruch surface is a $\bP^1$ fibration
over $\bP^1$, the F-theory $CY_3$ is also a $K3$-fibration over
$\bP^1$. That is, heterotic on an elliptic Calabi-Yau over $B$ is dual
to F-theory on a $K3$-fibered Calabi-Yau over $B$, with here
$B=\bP^1$.

As one might expect, a similar story holds in four dimensions. In this
paper, we will focus on cases where the heterotic string is
compactified to four dimensions on an elliptic Calabi-Yau $X$ with
section $\sigma_H$ over a two-fold base $B_2$, which is dual to
F-theory on a $K3$-fibration over $B_2$.  We take homogeneous fiber
coordinates $(x_H,y_H,z_H)\in \bP_{231}$ so that the $X$ is given by
the Weierstrass equation
\begin{equation}
y_H^2 = x_H^3 + \tilde f\,x_Hz_H^2 + \tilde g\,z_H^6,
\end{equation}
where $\tilde f$ and $\tilde g$ are sections of $K_{B_2}^{-4}$ and
$K_{B_2}^{-6}$, respectively.  The F-theory compactification manifold
is an elliptic Calabi-Yau over $B_3$, which is itself a $\bP^1$
fibration over $B_2$, and has Weierstrass equation
\begin{equation}
y_F^2 = x_F^3 + f\,x_F z_F^2 + g\,z_F^6,
\end{equation}
where $f$ and $g$ are sections of $K_{B_3}^{-4}$ and
$K_{B_3}^{-6}$. When the need to specify arises, we will take
$B_2=\bF_p$, so that $B_3$ is a generalized Hirzebruch variety
$\bF_{pmk}$. (A generalized Hirzebruch variety is a $\bP^1$ fibration
over a Hirzebruch surface. We include some background material on this
geometry in appendix~\ref{sec:gen hirz}.)

Heterotic / F-theory duality is made manifest by using the spectral
cover formalism \cite{Friedman:1997yq,Donagi:1997mg}, in which a
holomorphic vector bundle $V$ is determined by specifying a spectral
surface $C$ and a holomorphic line bundle $L$ on it. The pair $(C,L)$
determines $V$ by Fourier-Mukai transform. $V$ can be decomposed as $V
= V_1 \oplus V_2$, with $V_1$ associated to one $E_8$ factor and $V_2$
to the other. The geometric moduli which determine $C$ map directly to
complex structure moduli in F-theory. For concreteness and since it is
what we need to dualize the worldsheet instanton calculation, we focus
on a single bundle $V_1$.

As a divisor in $X$, the spectral
surface $C$ has homology class
\begin{equation}
	\label{eqn:C homology}
	C = n\sigma_H + \pi_H^* \eta
\end{equation}
for some curve $\eta$ in $B_2$ which is determined by $c_2(V_1)$ under
Fourier-Mukai transform.  Deformation moduli of the spectral surface
$C$ are bundle moduli which are counted by $h^0(X,\cO_X(C))$. One can
decompose sections of $h^0(X,\cO_X(C))$ appearing in the polynomial
$f_C$, whose zero locus determines $C$, in terms of fiber coordinates
and sections of the base as
\begin{equation}
	\label{eqn:fc}
	f_C = a_0 \,\, z_H^n + a_2 \,\, x_H z_H^{n-2} + a_3 \,\,  y_H z_H^{n-3} + a_4 \,\, x_H^2 z_H^{n-4}+ a_5 \,\, x_H y_H z_H^{n-5}+ \dots ,
\end{equation}
where a generic term $a_{q}x_H^{n_x} y_H^{n_y} z_H^{n-q}$ (subject the
constraint $q=2n_x+3n_y$) has
\begin{equation}
\label{eqn:aq}
a_q\in H^0(B_2,K_{B_2}^{\otimes q} \otimes \cO_{B_2}(\eta)) \cong
H^0(B_2,K_{B_2}^{\otimes q-6}\otimes \cO_{B_2}({\tilde \eta}))\, .
\end{equation}
As will become clear, it is convenient to write $\eta \equiv \tilde
\eta -6[K_{B_2}]$.  Intuitively, one might expect that as the rank $n$
of $V_1$ increases, the additional monomials allowed in \eqref{eqn:fc}
correspond to degrees of freedom that further break the $E_8$
factor. That is, $a_q$ degrees of freedom for increasingly large $q$
correspond to larger amounts of breaking. This fact can be seen
plainly on the heterotic side, but we will present it from the
F-theory point of view, where the spectral surface moduli\footnote{For
  the time being, we only consider those bundle moduli which
  correspond to spectral surface moduli.}  control the degenerations
of the elliptic fiber, and hence the structure of $7$-branes.

The spectral line bundle $L$ on $C$ is often presented as the
restriction of a line bundle $\un L$ on $X$.  To avoid introducing too
much notation, we will abuse notation and refer to $\un L$ simply as
$L$, as is common in the literature, so that $L$ is a line bundle on
$X$ which restricted to $C$ gives the spectral line bundle, also
called $L$. We will mostly deal with bundles of $SU(3)$ structure. The
constraint $c_1(V)=0$ then requires that
\begin{equation}
	\label{eqn:L chern}
	c_1(L) = \frac{n}{2} \sigma_H + \frac{1}{2} \pi_H^* \eta + \frac{1}{2} \pi_H^* c_1(TB_2) + \gamma,
\end{equation}
so that $L$ is determined up to a choice of a twist parameter
$\gamma$, of the form
\begin{equation}
	\gamma = \lambda (n\sigma_H -(\pi_H^*\eta - nc_1(TB_2)),
\end{equation}
with $\lambda$ integral if $n$ is even and $\lambda$ half integral if
$n$ is odd. This parameter maps into $G$-flux in F-theory, where the
parameter $\lambda$ controls the ``amount" of flux and the choice of
$\eta$ determines its structure.

\subsubsection{Mapping bundle moduli}
We now wish to identify complex structure moduli on the F-theory side
with vector bundle moduli on the heterotic side. The basics of the
discussion are presented in \cite{Bershadsky:1997zs}, and we fill in
some details relevant for our instanton computation, where the bundle
moduli are restricted to the curve in the base which the worldsheet
instanton wraps.

As discussed, the compactification manifold on the F-theory side is a
Calabi-Yau fourfold $Y$ that is an elliptic fibration over $B_3$,
which itself is a $\bP^1$ fibration over $B_2$. $Y$ is also a $K3$
fibration over $B_2$. The dual seven-branes wrap the base $B_2$ of the
$K3$-fibration. We decompose $f$ and $g$, which are sections of line
bundles on $B_3$, in terms of the homogeneous coordinates $y$ and $z$
of the $\bP^1$ fiber of $B_3$ and sections of $B_2$. This gives
\begin{equation}
	\label{eqn:f and g}
	f = \sum_{a=0}^8  y^{8-a} z^a \,\, f_a \qquad\qquad g = \sum_{b = 0}^{12} y^{12-b} z^b \,\, g_b,
\end{equation}
where the sections $f_a$ and $g_b$ are
\begin{align}
  \begin{split}
    f_a &\in H^0(B_2,K_{B_2}^{-4} \otimes \cO_{B_2}(\tilde \eta)^{\otimes (4-a)}) \\
    g_b &\in H^0(B_2,K_{B_2}^{-6} \otimes \cO_{B_2}(\tilde
    \eta)^{\otimes (6-b)}).
  \end{split}
\label{eqn:fa gb}
\end{align}
From this parameterization, it is clear that in much of the moduli
space the discriminant $\Delta = 4\,f^3 + 27\,g^2$ has zeros at $z=0$,
signaling the presence of a 7-brane at $z=0$ which wraps the base
$B_2$ of the $K3$ fibration. In this notation, $f_4$ and $g_6$ are
equivalent to $\tilde f$ and $\tilde g$, which determine the
Weierstrass equation for the heterotic elliptic Calabi-Yau $X$. From
\eqref{eqn:fa gb} it is easy to see that they do not depend on $\tilde
\eta$, which is data associated to the heterotic bundle $V$. Rather,
they depend only on geometry of the heterotic compactification.

For our purposes, we will often (although not exclusively) focus on
the gauge group along the 7-brane at $z=0$ at various points in
complex structure moduli space. It is known from the Kodaira
classification that an $E_8$ singularity has orders of vanishing
$ord(f)\ge 4$, $ord(g)=5$, $ord(\Delta) = 10$.  Thus one can take
$f=0$ \footnote{This might seem strange, but recall that in the case
  of $K3$ in section~\ref{sec:string junctions} we took $f=0$ to get
  an $E_6$ singularity, which has $ord(f) \ge 3$, $ord(f)=4$,
  $ord(\Delta) = 10$. That is, $f$ can vanish everywhere, as long as
  $g,\Delta$ vanish to the correct order.}  and
$g_0=g_1=g_2=g_3=g_4=0$. If $g_5$ does not vanish at a point in moduli
space, then $g$ vanishes to order $5$ and $\Delta$ vanishes to order
$10$ at $z=0$, ensuring an $E_8$ singularity there. That is, the moduli
specifying $g_5$ are the moduli which preserve an $E_8$ singularity.

It is known on the heterotic side that the moduli specifying $a_0$
preserve one $E_8$ if the higher $a_q$'s vanish. In
\cite{Bershadsky:1997zs} a careful counting showed that $a_0$ and
$g_5$ have the same number of moduli. The simple reason for this, as
seen from \eqref{eqn:aq} and \eqref{eqn:fa gb} is that $g_5$ and
$a_0$, are sections of the same bundle, and thus they should be
identified. One statement of heterotic / F-theory duality, then, is
that the mathematical object $a_0$ which specifies part of a vector
bundle on the heterotic side contributes purely to the geometry of a
$K3$ on the F-theory side. By similar arguments, one can show that
moduli in $f_3$ preserve an $E_7$ singularity, which is identified with
$a_2$ since they are sections of the same bundle. The section that
preserves an $E_6$ is $g_4$. From an initial glance at \eqref{eqn:aq},
it is easy to see that there is no single $a_q$ which can be
identified with $g_4$, due to the tensor power of the $\cO$-bundle in
$g_4$, but that $a_3^2$ and $g_4$ are sections of the same bundle. One
might be concerned that $g_4$ should not be identified with $a_3^2$,
since $g_4$ in its most general form does not have to be a perfect
square. If $g_4$ were not a perfect square, though, the singular fiber
would undergo non-trivial monodromy upon taking certain paths in $B_2$
and the singularity structure would be $F_4$ rather than $E_6$
\cite{Bershadsky:1996nh,Bershadsky:1997zs}. Thus, we have $g_4 \equiv
a_3^2$ if $E_6$ is to be preserved.

\subsection{Heterotic worldsheet instantons and their F-theory duals}
\label{sec:instanton duals}
Having reviewed the basics of heterotic F-theory duality necessary for
our discussion of instantons, we now discuss heterotic worldsheet
instantons and their ED3/M5 instanton duals in F-theory.

Non-perturbative worldsheet instanton effects in heterotic string
vacua \cite{Dine:1986zy,Dine:1987bq,Distler:1986wm,Distler:1987ee} can
give corrections to the superpotential that play important roles in
low energy physics, lifting vector bundle moduli (see
\cite{Moore:2000fs,Curio:2001qi,Buchbinder:2003pi} for early examples)
or giving corrections \cite{Berglund:1995yu} to couplings involving
charged matter fields. Consider the compactification of the heterotic
$E_8\times E_8$ theory on a smooth Calabi-Yau threefold $X$ endowed with
a holomorphic vector bundle $V$.  Then a superpotential correction due
to a heterotic worldsheet instanton wrapped on a holomorphic curve
$\Sigma$ in $X$ generically takes the form
\begin{equation}
	W_\Sigma \,\,\, \sim \,\,\, \Pfaff_\Sigma \,\,\,e^{i \int_\Sigma J}
\end{equation}
where $\Pfaff_\Sigma$ is the one-loop Pfaffian prefactor and $J$ is the
complexified K\"ahler form. The Pfaffian prefactor is a function of
vector bundle moduli and greatly influences the structure of the
superpotential correction, including its zeroes. This will be the main
object of our study via duality. It was computed directly using
algebro-geometric techniques in
\cite{Buchbinder:2002ic,Buchbinder:2002pr}, where in the examples
studied it took the form
\begin{equation}
	\Pfaff_\Sigma \,\,\, \sim \,\,\, p^k \,q \qquad \qquad k \in \bZ,
\end{equation}
with $p$ and $q$ being complicated polynomials in the vector bundle
moduli which define the spectral surface $C$. We will show that the
structure of $p$, $q$, and $k$ have interesting physics in terms of
intersecting branes in F-theory. We review the details of the
calculation of $\Pfaff_\Sigma$ since it is relevant for discussion of
the physics of ED3/M5 instantons via duality.

An important condition for the study of worldsheet instanton
corrections is that $\Pfaff_\Sigma$ (and therefore the superpotential
$W_\Sigma$) vanishes if and only if the sheaf cohomology
$H^0(\Sigma,V_1|_\Sigma\otimes \cO_\Sigma(-1))$ is
non-trivial\footnote{For brevity we omit the derivation of this
  condition, instead referring the reader to
  \cite{Buchbinder:2002ic}.}. Since we are interested in heterotic
compactifications with F-duals, $X$ is elliptically fibered and
therefore there must be an isomorphism between
$H^0(\Sigma,V_1|_\Sigma\otimes \cO_\Sigma(-1))$ and some sheaf
cohomology involving only spectral data. Before stating the
isomorphism, let us make a few definitions.

Taking $\pi_H$ the projection of $X$ onto $B_2$, we define $\cE \equiv
\pi_H^{-1} (\Sigma)$, $r\equiv \pi_H^* \eta \cdot \cE$ and $c\equiv
C|_\cE = n\sigma_\cE + rF$, where $r$ is an integer calculable in
specific examples.  The divisor $c$ in $\cE$ has a defining equation
given by\footnote{Here there are two departures from the notation of
  \cite{Curio:2009wn}, where we call the defining equation $f_c$
  instead of $s$ and the sections $\tilde a_q$ instead of $a_q$, as
  the $a_q$'s are typically reserved for sections appearing in $f_C$,
  not $f_c$.}
\begin{equation}
	\label{eqn:fc}
	f_c =  \tilde a_0 \,\,  z_H^n +  \tilde a_2 \,\,  x_H  z_H^{n-2} +  \tilde a_3 \,\,   y_H  z_H^{n-3} + \tilde a_4 \,\,  x_H^2  z_H^{n-4}+ \tilde a_5 \,\,  x_H  y_H  z_H^{n-5}+ \dots ,
\end{equation}
where $\tilde a_q = a_q|_\Sigma$. It was shown in \cite{Buchbinder:2002ic} that 
\begin{equation}
	H^0(\Sigma,V_1|_\Sigma\otimes \cO_\Sigma(-1)) \cong H^0(c,(L|_\cE \otimes O_\cE(-F))|_c)
\end{equation}
where $F$ is the elliptic fiber class and a straightforward
computation from equation \eqref{eqn:L chern} gives $L|_\cE =
\cO_\cE(n(\frac{1}{2} + \lambda)\, \sigma|_\cE +
(r(\frac{1}{2}-\lambda) + \chi(\frac{1}{2} + n\lambda))\,F)$.  Here we
have defined $\chi = c_1(TB_2) \cdot \Sigma$. This isomorphism
gives the sheaf cohomology to be calculated when using the spectral
cover formalism and implies that
\begin{equation}
\Pfaff_\Sigma = 0 \qquad \Leftrightarrow \qquad h^0(c,\cL) > 0,	
\end{equation}
where we have defined $\tilde \cL\equiv L|_\cE \otimes O_\cE(-F)$ and
$\cL \equiv \tilde \cL|_c$ for notational simplicity in what follows.
Determining the zeroes of the Pfaffian then amounts to calculating
sheaf cohomology.  One way of doing this is starting from the short
exact sequence
\begin{equation}
	\label{eqn:fc sequence}
	0 \rightarrow \tilde \cL\otimes \cO_\cE(-c) \xrightarrow{f_c} \tilde \cL \rightarrow \cL \rightarrow 0
\end{equation}
and calculating $h^0(c,\cL)$ via the corresponding long exact sequence in cohomology.

The non-trivial physics is determined by the fact that $h^0(c,\cL)$
can jump by integral values as one moves in the vector bundle moduli
space $\cM_X(V)$, signaling the presence of extra zero modes. This was
first done in \cite{Buchbinder:2002ic}, where the map $f_c$ determines
matrices $f_c^{(0)}$ and $f_c^{(1)}$ appearing in the long exact
sequence
\begin{align}
	\label{eqn:cohomology sequence}
	0 \rightarrow H^0(\cE,\tilde \cL \otimes \cO_\cE(-c)) &\xrightarrow{f_c^{(0)}} H^0(\cE,\tilde \cL) \rightarrow H^0(c,\cL) \rightarrow  H^1(\cE,\tilde \cL \otimes \cO_\cE(-c)) \notag \\
	& \xrightarrow{f_c^{(1)}} H^1(\cE,\tilde \cL) \rightarrow \dots
\end{align}
which gives $h^0(c,\cL) = h^0(\cE,\tilde \cL) + h^1(\cE,\tilde \cL
\otimes \cO_\cE(-c)) - rk(f_c^{(0)}) - rk(f_c^{(1)})$. In most cases
studied in the literature, one of the groups in \eqref{eqn:cohomology
  sequence} vanishes so that one must only compute the rank of one of
the maps. Jumps in $h^0$ occur when one of the maps loses rank, which
was determined in \cite{Buchbinder:2002ic} by calculating the
determinants $det(f_c^{(0)})$ and $det(f_c^{(1)})$. These determinants
are polynomial functions of moduli appearing in $\tilde a_q$ and thus
their zero loci determine algebraic subvarieties in the moduli space
of deformations of $c$ in $\cE$, henceforth called $\cM_\cE(c)$. The
Pfaffian prefactor, and hence the superpotential, vanishes on these
subvarieties of moduli space, which we call $(\Pfaff_\Sigma)$.  Our
goal is to understand the physics of $(\Pfaff_\Sigma)$ and further
subvarieties in it. In particular, we wish to relate those subvarieties
to new features in the low energy 4d theory.

It is illuminating to consider the physics of such a calculation in
the dual F-theory compactification to four dimensions. A worldsheet
instanton wrapping a curve $\Sigma$ in X is dual to a vertical $M5$
instanton wrapping $\pi_{K3}^{-1} (\Sigma)$, or alternatively a
euclidean D3 wrapping $\cE_F \equiv \pi_{K3}^{-1}
(\Sigma)|_{B_3}=\pi_{\bP^1}^{-1}(\Sigma)$, where we have made use of the
multiple fibration structures $Y\xrightarrow{\pi_{K3}}B$ and
$B_3\xrightarrow{\pi_{\bP^1}}B$. The two-folds $\cE$ and $\cE_F$ are
related via duality, where the elliptic fiber is mapped to a $\bP^1$
as usual.

The F-theory dual of the heterotic zero mode calculation makes
physical sense from an open string point of view. The subvariety
$(\Pfaff_\Sigma)$ in $\cM_\cE(c)$ is a polynomial in moduli of the
$\tilde a_q$'s. In F-theory these determine the structure of the
discriminant $\Delta$ over the instanton worldvolume $\cE_F$, and
therefore $(\Pfaff_\Sigma)$ is also a subvariety in
$\cM_{\cE_F}(\Delta|_{\cE_F})$. Physically, this means that the
structure of $7$-branes inside the instanton worldvolume (roughly, the
structure of 3-7 zero modes) governs whether or not the superpotential
vanishes, as is expected from the type II perspective. For this
reason, in \cite{Donagi:2010pd} the curve $c$ was suggestively called
$\Sigma_{37}$, since the moduli determining $c$ determine the locus of
$7$-brane intersections with the $3$-brane instanton,
$\Delta|_{\cE_F}$. The Pfaffians calculated on the heterotic side
therefore map directly to the F-theory side, where each point in
$\cM_\cE(c)$ determines a brane configuration in the worldvolume of
the instanton which governs (along with $G$-flux) whether or not the
superpotential vanishes.\footnote{Note that, though only moduli
  describing deformations of $c$ appear in the Pfaffian, the structure
  of the polynomial is determined by the matrix structure of maps in
  \ref{eqn:cohomology sequence}.}

It is worth noting that $\Delta |_{\cE_F}$ is determined by the
restriction of $f_a$ and $g_b$ to the instanton, which in turn are
sections of line bundles determined by the parameter $r\equiv
\pi_H^{-1}(\Sigma) \cdot \cE$. There are typically multiple curves
$\eta$ which could give rise to the same $r$, each of which would
determine a different heterotic compactification, and therefore a
different F-dual. Physically, this means that there are different
compactifications which each typically have different structures of
7-branes away from the instanton. On the instanton, though, every
compactification with the same $r$ has the same discriminant $\Delta
|_{\cE_F}$. This is a strong hint that the structure of the
superpotential depends heavily on the structure of 7-branes and matter
curves within the instanton worldvolume, as one might expect. We will
see this in the examples we consider.

These interpretations of the physics make physical sense at a
qualitative level, but still leave a number of questions
unanswered. One important question is how the superpotentials of $ED3$
instantons in F-theory calculated as a function of F-theory complex
structure moduli via duality with heterotic relate to standard open
string considerations in type IIB. In that context, the action of a
$ED3$ instanton could contain a coupling of the form $\lambda X \ov
\lambda$, where $X$ is a matter field and $\lambda$ and $\ov \lambda$
are charged instanton zero modes.  An instanton with such a coupling
would --- assuming that the structure of uncharged zero modes
satisfies the appropriate conditions --- generate a superpotential
coupling of the form $W\sim X$.  At generic points in moduli space
where $X$ has a non-zero vev, this is an uncharged superpotential
correction, whereas at special points when $X$ has zero vev it is a
charged superpotential correction. $ED3$ instanton corrections in
F-theory should also have such a structure, so the duals of worldsheet
instanton corrections should be able to be interpreted in this way. We
will show in explicit examples how at certain points in
$(\Pfaff_\Sigma)$ there are indeed jumps in the amount of chiral matter
in the theory.

\subsubsection{Understanding the structure of $(\Pfaff_\Sigma)$}
\label{sec:structure of Pfaff}

From the type II perspective, where the vanishing of the
superpotential corresponds to some matter fields having zero vev, one
would expect that at points in $(\Pfaff_\Sigma)$ there is some special
gauge enhancement corresponding to non-trivial intersections of
$7$-branes. Recall that the $\tilde a_q$ are sections of a line bundle
on $\Sigma$, which is the $\bP^1$ that the instanton wraps on the
heterotic side and is the intersection of the instanton with the GUT
stack in F-theory. They determine the degenerations of the elliptic
fiber over the instanton, and thus the positions of 7-branes in the
instanton worldvolume. A natural conjecture is that gauge enhancement
will often occur when the sections have common zeroes, leading to the
study of resultants of these sections.  In \cite{Curio:2008cm,
  Curio:2009wn,Curio:2010hd}, Curio studied the structure of
$(\Pfaff_\Sigma)$ with a number of techniques, including that of
resultants. We restate some results here, and interpret in terms of
$ED3$ instantons in F-theory.

As we move in the moduli space of the vector bundle, the curve $c =
C\cdot \cE \subset \cE$ changes. We will denote the corresponding
moduli space of curves as $\cM_\cE(c)$, and a point in this moduli
space $t \in \cM_\cE(c)$. We will sometimes denote the corresponding
curve $c_t$, to emphasize the dependence of $c$ on the particular point in
moduli space $t$. 
Following
\cite{Curio:2008cm,Curio:2009wn,Curio:2010hd}, we introduce the
following sublocus in $\cM_\cE(c)$:\footnote{In order to avoid a clash
  of conventions, we have changed notation from
  \cite{Curio:2008cm,Curio:2009wn,Curio:2010hd}. The dictionary is
  $\nabla_{here} = \Sigma_{there}$, and $\Sigma_{here} = b_{there}$.}
\begin{align}
  \nabla = \{t\in \cM_\cE(c)\,\,\,|\,\,\, \Lambda|_{c_t} = \cO_{c_t}\},
\end{align}
where we have introduced
\begin{align}
  \Lambda = \cO_\cE(n \sigma|_\cE - (r-n\chi)F)\, .
\end{align}
In addition to $\Sigma$, we would like to define another sublocus in
moduli space:
\begin{align}
  \label{eq:R-locus}
  \cR = \{ t \in \cR \,\,\,|\,\,\, \tilde a_n | \tilde a_j
  \quad \textrm{for}\quad j=2,\ldots,n\}\, .
\end{align}
It was shown in \cite{Curio:2009wn} that $\cR \subseteq \nabla \subset
(\Pfaff_\Sigma)$. In fact, it is crucial in F-theory that $\tilde a_n$
does not divide $\tilde a_0$, as we will see that it would correspond to
an enhancement beyond $E_8$, which would give a singularity that cannot
be resolved while maintaining the Calabi-Yau condition. It is also worth
noting that for rank 3 bundles $\cR = \nabla$, which is the case for
all four examples in \cite{Buchbinder:2002pr}. (It has been further
argued in \cite{Curio:2010hd} that $\cR=\nabla$ in general, not just
for rank 3 bundles, but we will just need the weaker $SU(3)$ result.)

What does the F-theory dual look like at points $t \in \cR$?  Taking
$n=3$ for the sake of illustration, so that the GUT group is $E_6$,
the discriminant restricted to the instanton is
\begin{equation}
  \Delta|_{\cE_F} = z^8(27\tilde a_3^4 + z(4\tilde a_2^3 + 54\tilde a_0 \tilde a_3^2) + z^2(12\tilde a_2^2 f_{4|\cE_F} + 27 \tilde a_0^2 + 54 \tilde a_3^2 g_{6|\cE_F})+ ..).
\end{equation}
For $t \in\cR$ we have that $\tilde a_3$ divides $\tilde a_2$. We can
then factor out a factor of $\tilde a_3$ from the first two terms in
the expansion in $z$. Since in general $\tilde a_n$ is a section of a
line bundle of degree $r-n\chi$ in $\Sigma=\bP^1$, we have that at $r-n\chi$
points of $\Sigma$ the singularity is enhanced to $E_8$.

This result did not really require $n=3$, so we see that also in the
case of higher rank vector bundles, for $t\in \cR$ we find a $E_8$
singularity in the worldvolume of the instanton. Finally, the
correspondence between an $E_8$ singularity in the worldvolume of the
instanton and the vanishing of the superpotential also applies to
lower rank groups. For a rank $1$ bundle there is $E_8$ symmetry on
the entire $\bP^1$, and for a rank $2$ bundle there is generically a
point of $E_8$ enhancement in the instanton, since three divisors (the
instanton, the $E_7$-brane, and a $U(1)$ flavor brane) generically
intersect at a point inside a Calabi-Yau. For these ranks, there are
points of $E_8$ enhancement in the instanton worldvolume at all points
in moduli space, giving a physical reason for the fact
\cite{Curio:2009wn} that the superpotential is identically zero for
rank $1$ and rank $2$ bundles.

\medskip

As an somewhat speculative side remark, it is perhaps worth pointing
out that for F-theory GUTs with $G=SU(5)$, points of local $E_8$
enhancement have been advocated as giving rise to interesting
phenomenological properties \cite{Heckman:2009mn}. Nevertheless, such
a point is rather non-generic in moduli space, so one may wonder why
dynamics should prefer such a point (one could perhaps try to make
some kind of cosmological argument along the lines of
\cite{Kofman:2004yc}). The observation here that local enhancements to
$E_8$ symmetry are associated with vanishing non-perturbative
superpotentials could perhaps be useful to give a dynamical reason why
$E_8$ points may be preferred. It would be interesting to explore this
point further.

\subsection{Explicit examples}
\label{sec:het/f instanton examples}
In this section we discuss the F-theory duals of explicit heterotic
worldsheet instanton calculations presented in
\cite{Buchbinder:2002pr}. Those examples took the heterotic base to be
a Hirzebruch surface $\bF_p$, which is the toric variety specified in
table \ref{table:Fr}, and wrapped a worldsheet instanton on
$\Sigma=\{x=0\}$.
\begin{table}[ht]
	\centering
	\begin{tabular}{c|c|c|c|c|c|c|c}
		& $u$ & $v$ & $w$ & $x$\\ \hline
		$A$ & $1$ & $1$ & $p$ & $0$\\
		$\Sigma$ & $0$ & $0$ & $1$ & $1$
	\end{tabular}
	\caption{Homogeneous coordinates, divisor classes, and GLSM charges for the Hirzebruch surface $\bF_r$.}
	\label{table:Fr}
\end{table}
The Stanley-Reisner ideal is given by $SRI = \langle uv,wx\rangle$ and
the intersection numbers are given by $A^2 = 0$, $A\cdot\Sigma=1$,
$\Sigma^2 = -p$. The canonical bundle is $K_{\bF_p} = \cO_{\bF_p}(-
(p+2) A - 2 \Sigma)\equiv
\cO_{\bF_p}(-(r+2),-2)$.\footnote{\cite{Buchbinder:2002pr} uses the
  same coefficients, but $\cS$ instead of $\Sigma$ and $\cE$ instead
  of $A$. We deviate from this because we have defined $\cE$
  elsewhere. }

The curve $\eta$ in $B_2=\bF_p$ which determines the spectral cover $C$
is given by
\begin{equation}
	\label{eqn:eta in examples}
	\eta = b A + (a+1) \, \Sigma  \qquad \qquad a,b \in \bZ.
\end{equation}
From \eqref{eqn:aq} it can be seen that the $a_q$ coefficients
appearing in the defining equation $f_C$ for the spectral cover are
sections of
\begin{equation}
	a_q \in  H^0(\bF_p,\cO_{\bF_p}(-q(p+2)+b,a+1-2q))
\end{equation}
and a simple calculation, noting that $\cO_{\bF_p}(\Sigma)|_\Sigma =
\cO_\Sigma(-p)$ and $\cO_{\bF_p}(A)|_\Sigma=\cO_\Sigma(1)$, gives
\begin{equation}
	\tilde a_q \equiv a_q|_\Sigma\in H^0(\bP^1,\cO_{\bP^1}(b-pa-p+q(p-2))),
\end{equation}
where we have also used the fact that $\Sigma$ is a $\bP^1$, as it
must be if a worldsheet instanton on it gives a superpotential
correction. We write the section $a_q$ as
\begin{equation}
  \label{eq:aq-example}
  a_q = \sum_{i=0}^{a+1-2q} x^i w^{a+1-2q-i}\sum_{j=0}^{-q(p+2)+b-p(a+1-2q-i)} a_{q;ij}\,\,u^jv^{-q(p+2)+b-p(a+1-2q-i)-j}.
\end{equation}
The coefficients of $\tilde a_q$ are the vector bundle moduli that
appear in $\Pfaff_\Sigma$, and are given by the terms in
\eqref{eq:aq-example} that do not vanish when we set $x=0$:
\begin{equation}
  \tilde a_q = \sum_{i=0}^{b-pa-p+q(p-2)} a_{q;0i} \,\, u^i
  v^{b-pa-p+q(p-2) - i}\, .
\end{equation}
Interestingly, in the case of $p=2$ the $q$ dependence of $\tilde a_q$
drops out and therefore the $\tilde a_q$ are all sections of the same
bundle.  Also, it is important to note that the sections $\tilde a_q$
depend only on the combination $b-pa$, rather than $b$ and $a$
individually. This means that two heterotic (or F-theory, via duality)
compactifications which differ in $a$ and/or $b$ but not $b-pa$ have
the same Pfaffian, $\Pfaff_\Sigma$. We will explain this phenomenon
from an F-theory point of view.

\medskip

We wish to discuss in particular F-theory duals of the examples in
\cite{Buchbinder:2002pr} which took $V$ to be a rank $3$ holomorphic
vector bundle, so $n$ = 3, and therefore from \eqref{eqn:f and g} the
F-theory fourfold $Y$ in Weierstrass form has
\begin{align}
  \begin{split}
    f = z^3\,\,f_3 + z^4\,\,f_4 \qquad g = z^4 \,\, g_4 + z^5 \,\, g_5
    + z^6 g_{6} \\
    \Delta = z^8\,\big(27\,g_4^2 + z\,\, (4\,f_3^3 + 54\,g_4g_5) +
    z^{2}\,\,(27\, g_5^2+ 12\,f_3^2 f_4 + 54\,g_4 g_6) + \dots\big)
  \end{split}
\end{align}
which can be written, via duality, in terms of the spectral surface sections
$a_q$ as 
\begin{equation}
\Delta =  z^8\,\big(27\,a_3^4 + z\,\, (4\,a_2^3 + 54\,a_3^2a_0) +
z^{2}\,\,(27\, a_0^2+ 12\,a_2^2 f_4 + 54\,a_3^2 g_6)+\dots\big)\, .
\end{equation}
The dots represent higher order terms in $z$ that are not relevant for the
discussion at hand. 

From the structure of the discriminant, it is easy to see that there
is a GUT stack with $E_6$ singularity at $z=0$. The rest of the
discriminant gives a divisor in $B_3$ which has $I_1$ singularity for
generic moduli. The lowest order terms in $z$ appearing
in the discriminant will be the most interesting for us. For example,
the intersection of $a_3=0$ and the GUT stack at $z=0$ corresponds to
an $E_7$ enhancement over a curve, and thus the possibility of having
\textbf{27}/$\overline{\textbf{27}}$ matter localized there. We
include the next order in $z$ by defining
\begin{equation}
\Gamma \equiv 27\,a_3^4 + z\,\, (4\,a_2^3 + 54\,a_3^2a_0) \qquad\qquad  \tilde\Gamma \equiv \Gamma|_{x=0} = 27\,\tilde a_3^4 + z\,\, (4\,\tilde a_2^3 + 54\,\tilde a_3^2\tilde a_0).
\end{equation}
If $\Gamma$ is identically zero everywhere in $B_2$, then the $I_1$
part of the discriminant enhances to $I_2$ (giving $A_1\cong SU(2)$)
at $z=0$, which (including the $z^8$ term) enhances to $E_8$ symmetry
on the GUT stack. $\Gamma$ does not have to be identically zero, of
course, but may become zero on divisors or curves in $B_3$. The
appearance of such divisors and/or curves at certain loci in moduli
space will be important for the structure of instanton corrections.

For euclidean D-instanton corrections in weakly coupled type II, the
presence of additional abelian symmetries plays an important role in
determining charged superpotential corrections.
In F-theory, it was suggested in \cite{Grimm:2010ez} that the existence of curves
of $SU(2)$ enhancement in the $I_1$ part of the discriminant is an
indication of the existence of unhiggsed abelian symmetries in the
compactification. The argument is essentially that chiral matter under
the abelian symmetries is localized at the curve of $SU(2)$
enhancement. As one varies the moduli away from regions of $SU(2)$
enhancement the branes recombine, giving a vev to the chiral matter
and higgsing the symmetry. Though the curves of $SU(2)$ enhancement
that we study will often be contained inside the GUT stack, such
curves generically occur away from the GUT stack and are difficult to
study. This is one of the major disadvantages of the spectral cover
formalism in F-theory that necessitates a more global point of view:
GUT singlet matter that is chiral under abelian symmetries is
generically localized away from the GUT stack. We will see that this
matter can be of crucial importance for $ED3/M5$ instanton
corrections.

\subsubsection{A very symmetric example}

We discuss a simple non-trivial example, which is example 4 of
\cite{Buchbinder:2002pr}. The parameters specifying the
compactification are given by
\begin{equation}
n = 3, \qquad p=2, \qquad b-pa = 4, \qquad \lambda = \frac{3}{2} \qquad a>5
\end{equation}
where the inequality for $a$ ensures a positive spectral cover and the other
parameters are as discussed previously.
We emphasize again that this example includes many heterotic compactifications, since $a$ and $b$ are  not specified, but instead only satisfy some constraints
$b-pa=4$ and $a>5$.  Thus, in each compactification, the $a_q$'s may be sections of different bundles based on the particular values of $a$ and $b$.
It is only the $\tilde a_q$'s whose bundles are explicitly determined, which are
\begin{equation}
 \tilde a_0, \tilde a_2, \tilde a_3 \in H^0(\bP^1,\cO_{\bP^1}(2)).
\end{equation}
They are all sections of the same bundle because $p=2$, which eliminates
the $q$ dependence of the bundles.

Due to the one-parameter ambiguity and $a$ and $b$, there are also
multiple F-theory compactifications corresponding to this example,
each dual to a particular heterotic compactification specified by a
choice of $a$ and $b$. Each different F-theory compactification that
gives rise to the same $\Pfaff_\Sigma$ can have
a different structure of 7-branes away from the instanton. The 7-brane
structure common to all such compactifications is the structure of 7-brane
over the worldvolume of the instanton, as the discriminant restricted
to the instanton is the same in all of those examples, and therefore one
might expect $\Delta|_{\cE_F}$ to be crucial for the structure of $\Pfaff_\Sigma$.

Using the techniques reviewed in section \ref{sec:instanton duals}, the
Pfaffian prefactor of a superpotential correction due to a worldsheet
instanton on $\Sigma$ can be calculated, giving \cite{Buchbinder:2002pr}
\begin{equation}
\label{eqn:pfaff symmetric example}
	\Pfaff_\Sigma = \epsilon_{ijk} \,\,  a_{0;0i}\,\,  a_{2;0j} \,\, a_{3;0k},
\end{equation}
which is nothing but the determinant of the matrix
\begin{equation*}
	\begin{pmatrix} 
		 a_{0;00} & a_{0;01} & a_{0;02} \\
		 a_{1;00} & a_{1;01} & a_{1;02} \\
		 a_{2;00} & a_{2;01} & a_{2;02}
	\end{pmatrix}.
\end{equation*}
We have $\Pfaff_\Sigma$ as an explicit function of the complex
structure moduli space of F-theory, and an important question is how
the structure of its vanishing locus, $(\Pfaff_\Sigma)$ relates to 4d
matter fields. At a naive glance, this connection seems like it might
be possible to make because $(\Pfaff_\Sigma)$ is a function of complex
structure moduli, which determine the structure of 7-branes and thus
also of matter fields. Ultimately, we wish to relate the vanishing of
$\Pfaff_\Sigma$ to the appearance of additional matter at certain point
in moduli space.

\begin{figure}
\centering
\includegraphics[scale=.5]{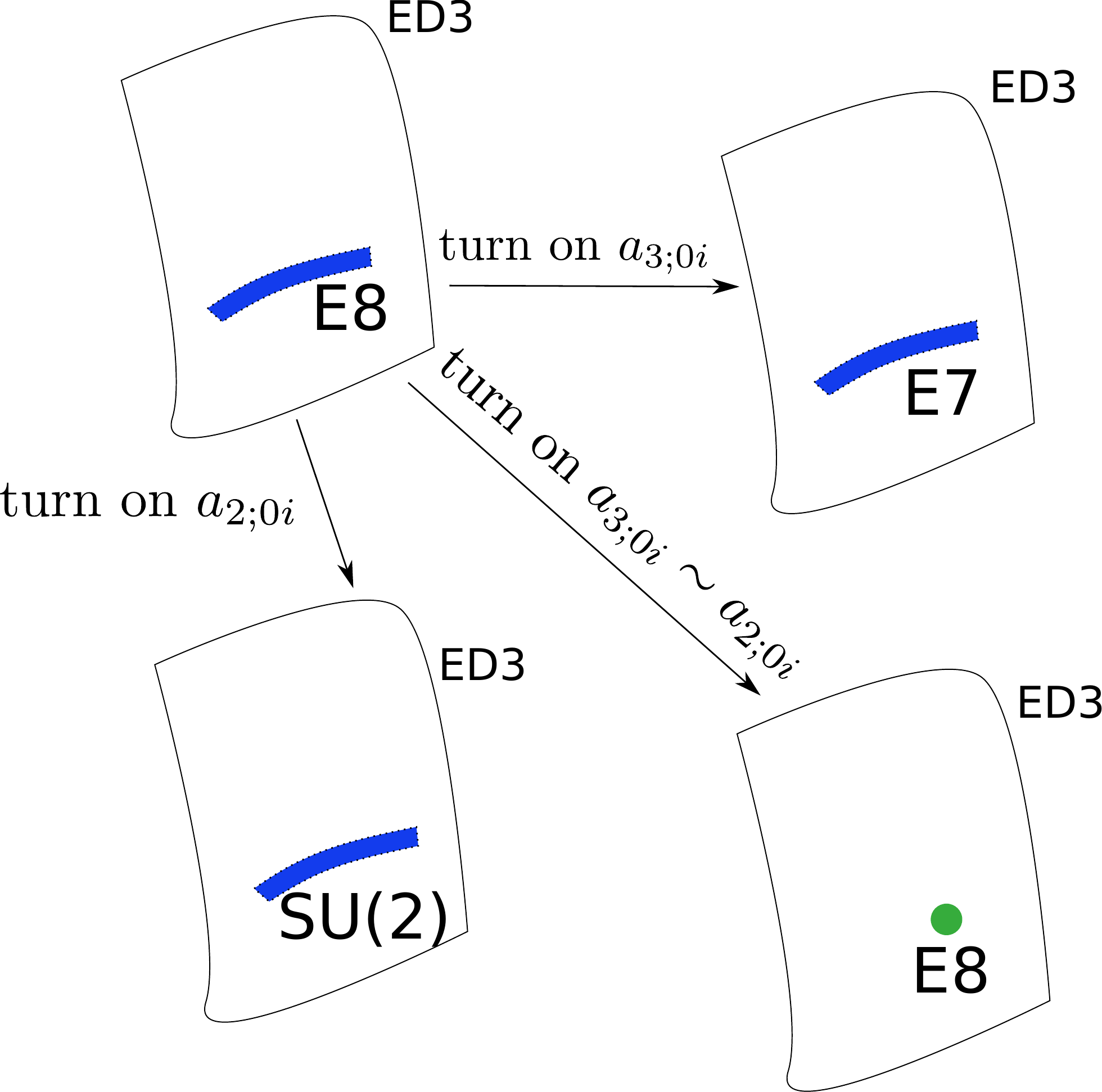}
\caption{Depicted are three deformations from the region in moduli
  space with a curve of $E_8$ enhancement in the worldvolume of the
  instanton. Two correspond to deformations which leave a matter curve
  present in the instanton worldvolume, represented by curves of
  $SU(2)$ and $E_7$ enhancement, while a third deformation maintains a
  point of $E_8$ enhancement (generically multiple) but no matter curve
  is present. All four instanton configuration represent regions in
  moduli space where the Pfaffian \eqref{eqn:pfaff symmetric example}
  is zero.}
\label{fig:e8 curve deform}
\end{figure}

We do this by considering particular subloci of $(\Pfaff_\Sigma)$. Four
of the subloci and the deformations interpolating between them are
displayed in figure~\ref{fig:e8 curve deform}. First consider the
region in moduli space where $a_{0;0i}$ are turned on, but $a_{2;0i}$
and $a_{3;0i}$ are identically zero for all $i$. That is, $\tilde a_2$
and $\tilde a_3$ are zero. From the structure of the discriminant
restricted to the instanton,
\begin{equation}
\label{eqn:instanton disc example}
  \Delta|_{\cE_F} = z^8(27\tilde a_3^4 + z(4\tilde a_2^3 + 54\tilde
  a_0 \tilde a_3^2) + z^2(12\tilde a_2^2 f_{4|\cE_F} + 27 \tilde a_0^2
  + 54 \tilde a_3^2 g_{6|\cE_F})+ ..)\, ,
\end{equation}
it is clear that such a region in moduli space corresponds to matter
curves of $E_8$ enhancement in the instanton worldvolume. As we have
said nothing about the moduli $a_{q;ij}$ for $i>0$, which only
influence the discriminant away from the instanton, this is only a
curve of $E_8$ enhancement in $Y$, rather than the restoration of $E_8$
symmetry along the entire locus $z=0$. The curve of $E_8$ involves the
multiple intersections: it is located in both the instanton and the
GUT stack, and also is in the locus where the $I_1$ part of the
discriminant self intersects. It is certainly non-generic that such a
curve would be present in the instanton worldvolume. From
\eqref{eqn:pfaff symmetric example} it is straightforward to see that
the Pfaffian vanishes at this locus in moduli space.

From this region with a curve of $E_8$ present in the instanton
worldvolume, consider three simple deformations. First, turning on
$a_{2;0i}$, it is clear from \ref{eqn:instanton disc example} that the
curve of $E_8$ in the instanton worldvolume $\cE_F$ becomes a curve of
$E_7$ in $\cE_F$. This matter curve is generically the intersection of
the $E_6$ GUT stack with the $I_1$ part of the discriminant, so that
$\textbf{27}$'s or $\overline{\textbf{27}}$'s of $E_6$ are present at
the curve of $E_7$. An explicit calculation using the techniques of
\cite{Donagi:2004ia} shows that this curve corresponds to a
\emph{jump} in the number of $\overline{\textbf{27}}$'s. There is new
light matter in the spectrum, which makes it not particularly
surprising that the Pfaffian \eqref{eqn:pfaff symmetric example}
vanishes at this point in moduli space.

Going back to the region of moduli space with a curve of $E_8$,
consider instead turning on $a_{3;0i}$.  In \cite{Hayashi:2010zp} it
was argued that in F-theory compactifications with an $E_6$ GUT stack,
the $I_1$ part of the discriminant enhances to an $I_2=SU(2)$
singularity when $a_2=0$.\footnote{The conventions of
  \cite{Hayashi:2010zp} are slightly different than those used here,
  including a rescaling of one of the sections by $\frac{1}{4}$. For
  brevity, we do not delve into details here.} Therefore, there is a
curve of $SU(2)$ enhancement in the instanton worldvolume when $\tilde
a_2=0$, as is the case here. There is new light matter in the
spectrum, which again makes it not particularly surprising that the
Pfaffian \eqref{eqn:pfaff symmetric example} vanishes at this region
in moduli space.

Going back to the region with a curve of $E_8$, turn on both $a_{2;0i}$
and $a_{3;0i}$, but so that they are proportional to one another. In
this case, we are on the locus $\cR$ in moduli space (see the
discussion around \eqref{eq:R-locus}), and $\tilde a_3$ divides
$\tilde a_2$. There is therefore a point of $E_8$ enhancement at every
zero of $\tilde a_3$, so that (in this case) there are two points of
$E_8$ enhancement in the worldvolume of the instanton. In this region,
there is no matter curve obviously present in the instanton
worldvolume. Nevertheless, it is clear from \eqref{eqn:pfaff symmetric
  example} that the Pfaffian vanishes, as expected from being on
$\cR$.

Finally, we wish to mention one other locus in $(\Pfaff_\Sigma)$ that
is puzzling. If one turns off $\tilde a_{0,i}$ but turns on $\tilde
a_{2,i}$ and $\tilde a_{3,i}$, the superpotential vanishes. However,
the moduli responsible for preserving $E_8$ in the worldvolume of the
instanton have been turned off, so that if $\tilde a_{2,i}$ and
$\tilde a_{3,i}$ were also turned off, the singularity would enhance
past $E_8$. The implications of turning off $a_0$ were studied in
\cite{Hayashi:2010zp}, with a focus on the influence of $U(1)$'s. In
doing so, one leaves the regime where an $8d$ gauge theory localized
at $z=0$ is a good approximation. We leave the study of instantons in
this region of moduli space for future work.

\section{Conclusions}

In this paper we have studied various aspects of the behavior of
D-brane instantons in regions of strong coupling. Due to the fact that
worldsheet CFT techniques and intuitions are no longer available to
us, we have taken a number of complementary approaches in order to
illuminate the physics.

\medskip

In section~\ref{sec:anomaly inflow} we gave a description of neutral
zero modes valid at strong coupling. The description is based on the
observation of \cite{Harvey:2007ab} that there is anomaly inflow
towards the orientifold, and it is the $\theta$ modes that cancel the
anomaly on the orientifold. We reinterpreted this observation from the
point of view of string junctions, which gave us a clear picture of
how to describe the $\theta$ mode at strong coupling. One particularly
interesting result of this analysis is that it explains how can the
neutral $\theta$ mode arise out of the light, mutually non-local,
charged degrees of freedom localizing in each of the components of the
$O7^-$ at strong coupling.

\medskip

In section~\ref{sec:string junctions} we turned our attention to
describing the related issue of charged modes at strong coupling. In
particular, we dealt with the behavior of zero modes as we move in
moduli space from perturbative points ($SO(8)^4$ in our analysis) to
very non-perturbative points ($E_6^3$). One surprising feature is that
the string junctions that in one corner of moduli space give light
degrees of freedom localized at each of the 7-brane stacks, in the
opposite corner become very massive states, transforming in
non-trivial representations of the gauge groups on the 7-brane stacks.

\medskip

Finally, having obtained some understanding of how to compute the
spectrum of neutral and charged zero modes at strong coupling, in
section~\ref{sec:heterotic duality} we proceed to study our main
object of interest: non-perturbative superpotentials in strongly
coupled regimes. In particular, we give a physical analysis of the
results in  \cite{Buchbinder:2002ic,Buchbinder:2002pr}, who give an
exact expression for the vector bundle moduli dependence of the
superpotential generated by a particular worldsheet instanton in a
heterotic compactification preserving $E_6$ symmetry. By dualizing the
configuration to F-theory, we observe that zeroes of the
superpotential are related to local enhancements of the singularity of
the elliptic fiber over the worldvolume of the instanton. The superpotential
correction is therefore fairly independent
of 7-brane behavior away from the instanton, so that the same correction
could arise from many compactifications. We show that regions in moduli
space where the superpotential vanishes are often accompanied by the presence
of matter curves
in the worldvolume of the instanton, corresponding to the presence of
additional light matter in the low energy spectrum. This agrees with what
is expected from the type II perspective. We also discuss a region in moduli space where
zeroes of the superpotential are accompanied by points of $E8$ enhancement
in the instanton worldvolume.

\medskip

There are various questions which, despite being clearly suggested by
our line of attack, we have not answered in this paper, and to which
we hope to come back shortly. A first question is generalizing the
anomaly inflow analysis to configurations with strongly coupled
stacks, such as $E_6$. This should provide a better understanding of
the dynamics of instanton zero modes living on the stack, and clarify
the mixing of charged/neutral zero modes as we move to strong
coupling.

\medskip

When we studied the behavior of the non-perturbative potential in
section~\ref{sec:heterotic duality}, we observed a clear pattern of
enhancement associated with vanishing of the superpotential, at least
for some loci of moduli space. Nevertheless, our analysis was not
general --- we do not identify the behavior of the dual F-theory
background for all moduli making $W$ vanish --- and it does not give a
clear F-theoretic explanation of what precisely makes $W$ vanish. A
full description of the physics will doubtlessly require some
understanding of the F-theory background beyond pure geometry, in
particular issues of massive $U(1)$s and G-flux will have to be
addressed (see \cite{Grimm:2010ez,Marsano:2010ix,Marsano:2011nn} for
recent papers clarifying some of the aspects that we believe will be
involved in a complete analysis).

\acknowledgments We would like to thank Lara Anderson, Pablo Camara,
Jeff Harvey, Hirotaka Hayashi, Joe Marsano, Denis Klevers and Sakura
Sch\" afer-Nameki for useful conversations during the course of this
work. We are particularly grateful to Ron Donagi for extensive and
illuminating conversations. I.G.-E. would like to thank the Hong-Kong
IAS for hospitality while parts of this work were being completed, and
N. Hasegawa for kind encouragement and support. This research was
supported in part by the National Science Foundation under Grant
No. NSF PHY05-51164, DOE under grant DE-FG05-95ER40893-A020, NSF RTG
grant DMS-0636606, the Fay R. and Eugene L. Langberg Chair, and the
Slovenian Research Agency (ARRS).

\appendix

\section{Generalized Hirzebruch variety $\bF_{pmk}$}
\label{sec:gen hirz}
\label{sec:gen hirz}
The worldsheet instanton examples discussed in section \ref{sec:het/f
  instanton examples} took $B_2=\bF_p$ as a base. Generically, the
F-theory elliptic fibration base $B_3$ is a $\bP^1$ fibration over
$B_2$. In this case, $B_2$ is a Hirzebruch surface, which is itself a
$\bP^1$ fibration over $\bP^1$. Therefore, $B_3$ is a $\bP^1$
fibration over a $\bP^1$ fibration over $\bP^1$. This geometry is
known as a \emph{generalized Hirzebruch variety} $\bF_{pmk}$
\cite{Bershadsky:1997zs}. Parametrizing the curve $\tilde \eta \equiv
\eta + 6\, [K_{\bF_p}]$ as $\tilde \eta = m \, A + n \, \Sigma$, the
GLSM charges of $\bF_{pmk}$ are given in table~\ref{table:gen Hirz}.
\begin{table}[htbp]
	\centering
	\begin{tabular}{c|c|c|c|c|c|c}
		& $u$ & $v$ & $w$ & $x$ & $y$ & $z$ \\ \hline
		$A$ & $1$ & $1$ & $p$ & $0$ & $0$ & $m$ \\
		$\Sigma$ & $0$ & $0$ & $1$ & $1$ & $0$ & $k$ \\
		$P$ & $0$ & $0$ & $0$ & $0$ & $1$ & $1$
	\end{tabular}
	\caption{The charge matrix for the generalized Hirzebruch variety $\bF_{pmk}$
	with homogeneous coordinates $u,v,w,x,y,z$.}
	\label{table:gen Hirz}
\end{table}

We abuse notation and call two of the divisor classes in $\bF_{pmk}$
$A$ and $\Sigma$, which are merely pullbacks from the corresponding
divisor classes in $\bF_p$. The parameters $m$ and $k$, which are
related to the parameters $a$ and $b$ in \eqref{eqn:eta in examples}
by $a+1 = 12+6p+m$ and $b=12+k$ determine the Chern classes of $V$ on
the heterotic side and are purely geometric on the F-theory side,
describing how the $\bP^1$ fiber of $B_3$ is fibered over the base
$B_2$.

The Stanley-Reisner ideal of $\bF_{pmk}$ is given by:
\begin{align}
  SRI_{\bF_{pmk}} = \left\langle uv, wx, yz \right\rangle
\end{align}
and the corresponding intersection ring is given by:
\begin{align}
  I_{\bF_{pmk}} = P\left(A\Sigma - k AP - p EE + (kp-m) EP +
    (2km-k^2p)PP\right)\, .
\end{align}

The anticanonical class of $\bF_{pmk}$ is given by the sum of its
divisors:
\begin{align}
  K_{\bF_{pmk}}^{-1} = \sum D_i = (p+m+2) A + (k+2) \Sigma + 2 P\, .
\end{align}

Since $\bF_{pmk}$ is a toric variety, itself, we can construct the
F-theory Calabi-Yau fourfold $Y$as a hypersurface in a toric variety
$\Sigma_F$. This is done by taking homogeneous coordinates $(\tilde
x_F, \tilde y_F, \tilde z_F)\in \bP_{231}$ and choosing the GLSM
charges of $\tilde x_F$, $\tilde y_F$ and $\tilde z_F$ so that the
Weierstrass equation defines a Calabi-Yau hypersurface in
$\Sigma_F$. The details of the toric variety $\Sigma_F$ are given in
table~\ref{table:cy f examples}. Per usual, $f$ and $g$ appearing in
the defining equation for $Y$ are sections of $4\,K_{\bF_{pmk}}^{-1}$
and $6\, K_{\bF_{pmk}}^{-1}$, respectively. Writing the global
sections $f$ and $g$ as a power series in $z$ gives the
parametrization of $f$ and $g$ discussed in section \ref{sec:het/f
  instanton examples}.

\begin{table}[tbp]
	\centering
	\begin{tabular}{c|c|c|c|c|c|c|c|c|c}
		& $u$ & $v$ & $w$ & $x$ & $y$ & $z$ & $\tilde x_F$ & $\tilde y_F$ & $\tilde z_F$  \\ \hline
	    $A$ & $1$ & $1$ & $p$ & $0$ & $0$ & $m$ & $2(p+m+2)$ & $3(p+m+2)$ & $0$\\
		$\Sigma$ & $0$ & $0$ & $1$ & $1$ & $0$ & $k$ & $2(k+2)$ & $3(k+2)$ & $0$\\
		$P$ & $0$ & $0$ & $0$ & $0$ & $1$ & $1$ & $4$ & $6$ & $0$ \\
		$\sigma_F$ & $0$ & $0$ & $0$ & $0$ & $0$ & $0$ & $2$ & $3$ & $1$
	\end{tabular}
	\caption{The GLSM charges for a toric variety whose Calabi-Yau hypersurface is 
		a fourfold elliptic fibration over a generalized Hirzebruch variety.}
	\label{table:cy f examples}
\end{table}

\newpage

\section{$\textbf{8}_v$ untangling under $A^4BC \mapsto ACACAC$}
\label{sec:untangle}

\begin{figure}[ht]
\begin{minipage}[b]{0.3\linewidth}
(1)\includegraphics[scale=.5]{1AAAABC.pdf}
\end{minipage}
\hspace{3.5cm}
\begin{minipage}[b]{0.3\linewidth}
(2)\includegraphics[scale=.5]{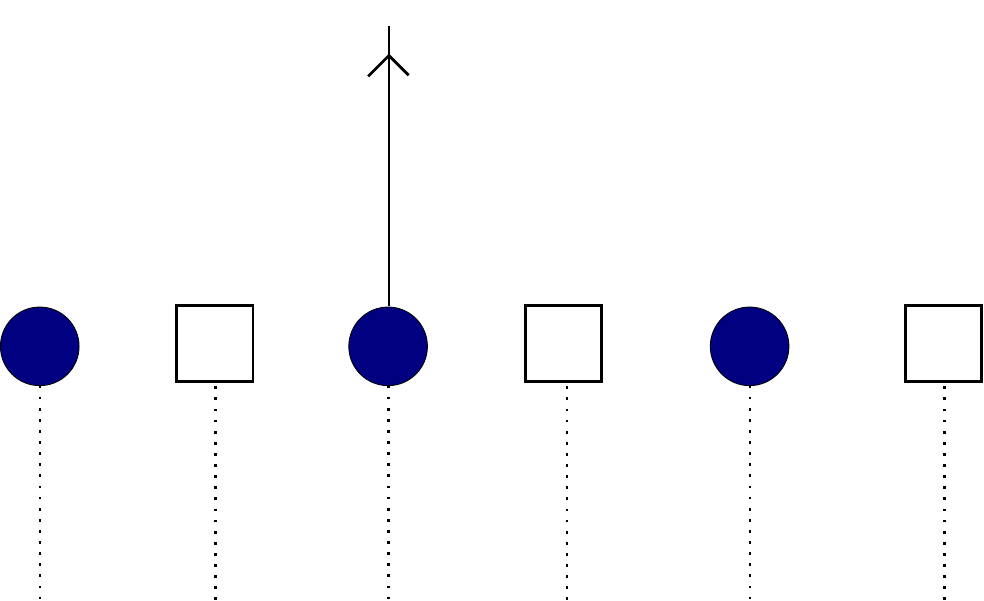}
\end{minipage}
\vspace{1cm}\\
\begin{minipage}[b]{0.3\linewidth}
(3)\includegraphics[scale=.5]{2AAAABC.pdf}
\end{minipage}
\hspace{3.5cm}
\begin{minipage}[b]{0.3\linewidth}
(4)\includegraphics[scale=.5]{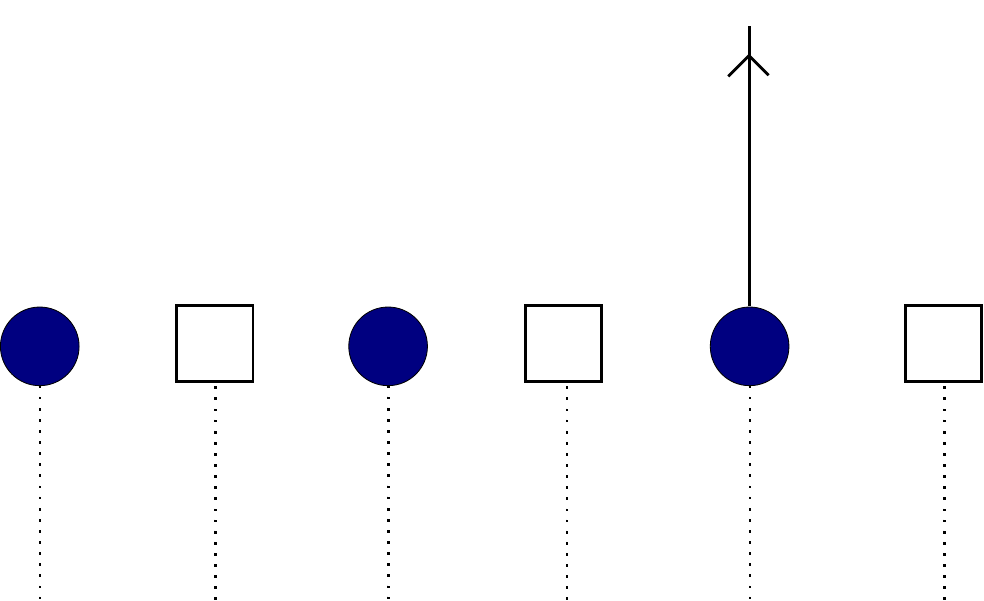}
\end{minipage}
\vspace{1cm}\\
\begin{minipage}[b]{0.3\linewidth}
(5)\includegraphics[scale=.5]{3AAAABC.pdf}
\end{minipage}
\hspace{3.5cm}
\begin{minipage}[b]{0.3\linewidth}
(6)\includegraphics[scale=.5]{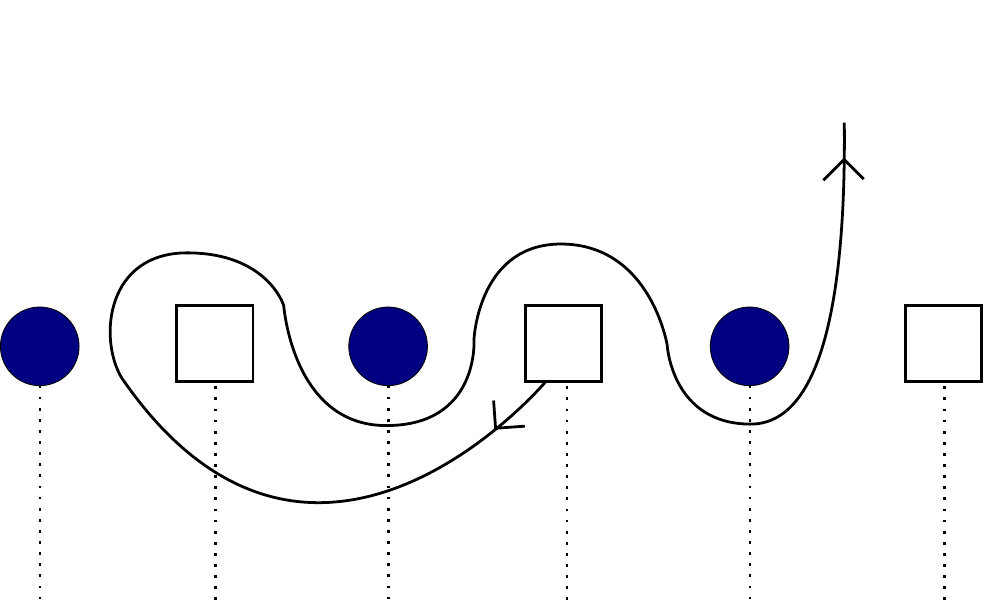}
\end{minipage}
\vspace{1cm}\\
\begin{minipage}[b]{0.3\linewidth}
(7)\includegraphics[scale=.5]{4AAAABC.pdf}
\end{minipage}
\hspace{3.5cm}
\begin{minipage}[b]{0.3\linewidth}
(8)\includegraphics[scale=.5]{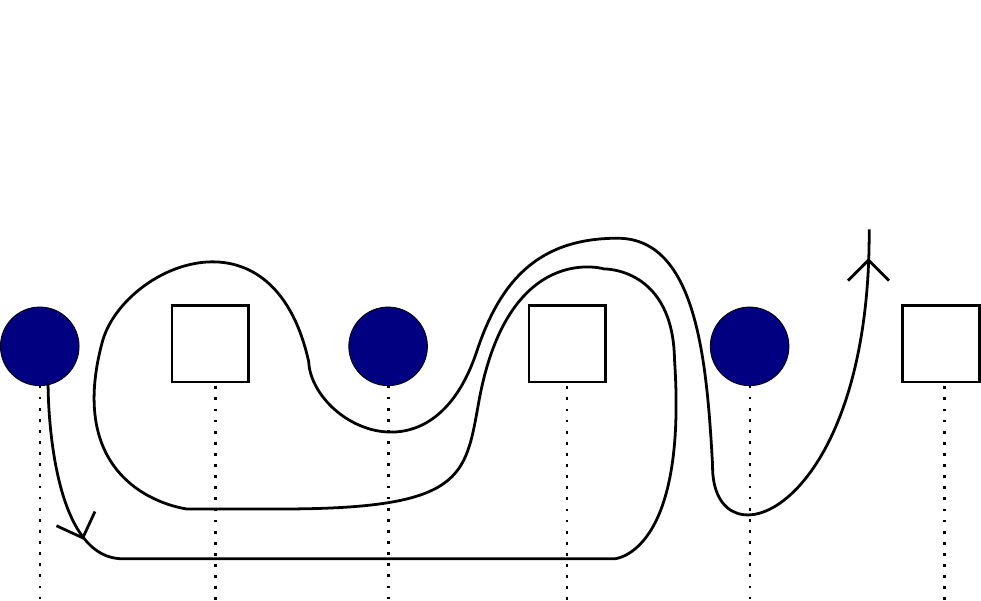}
\end{minipage}
\caption{Depicted on the left are four strings of asymptotic charge $(1,0)$
which can be thought of as standard $ED3-D7$ $\lambda$ modes which form half
of the $\textbf{8}_v$ at points of $SO(8)$ enhancement. To right of each
string configuration is a representation of the string after performing
the brane motion $A^4BC \mapsto ACACAC$.}
\label{fig:untangle1}
\end{figure}

\begin{figure}[ht]
\begin{minipage}[b]{0.3\linewidth}
(1)\includegraphics[scale=.5]{5AAAABC.pdf}
\end{minipage}
\hspace{3.5cm}
\begin{minipage}[b]{0.3\linewidth}
(2)\includegraphics[scale=.5]{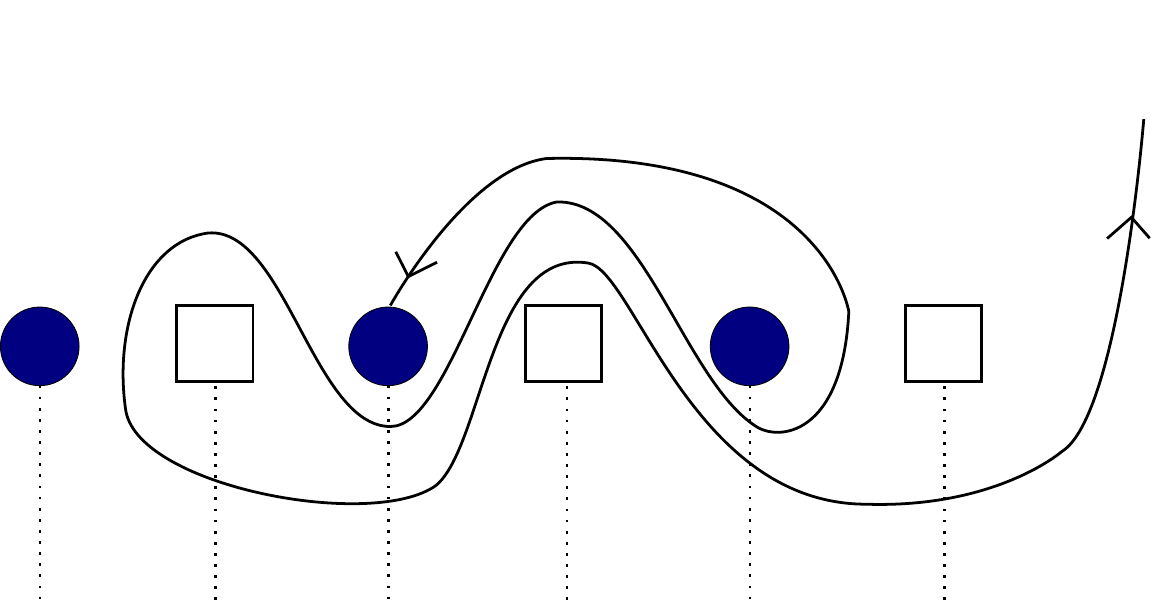}
\end{minipage}
\vspace{1cm}\\
\begin{minipage}[b]{0.3\linewidth}
(3)\includegraphics[scale=.5]{6AAAABC.pdf}
\end{minipage}
\hspace{3.5cm}
\begin{minipage}[b]{0.3\linewidth}
(4)\includegraphics[scale=.5]{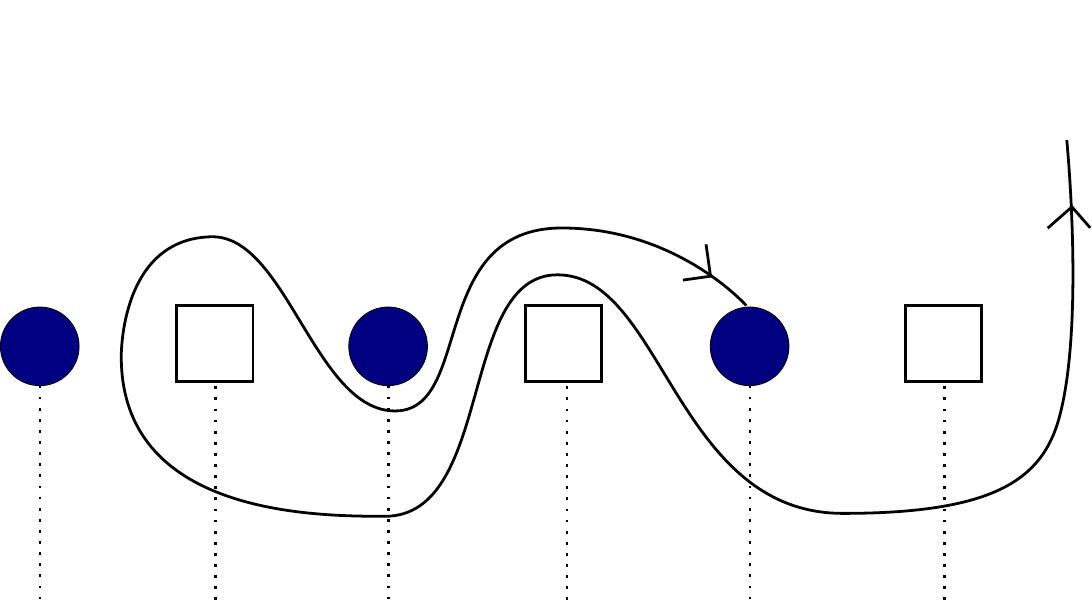}
\end{minipage}
\vspace{1cm}\\
\begin{minipage}[b]{0.3\linewidth}
(5)\includegraphics[scale=.5]{7AAAABC.pdf}
\end{minipage}
\hspace{3.5cm}
\begin{minipage}[b]{0.3\linewidth}
(6)\includegraphics[scale=.5]{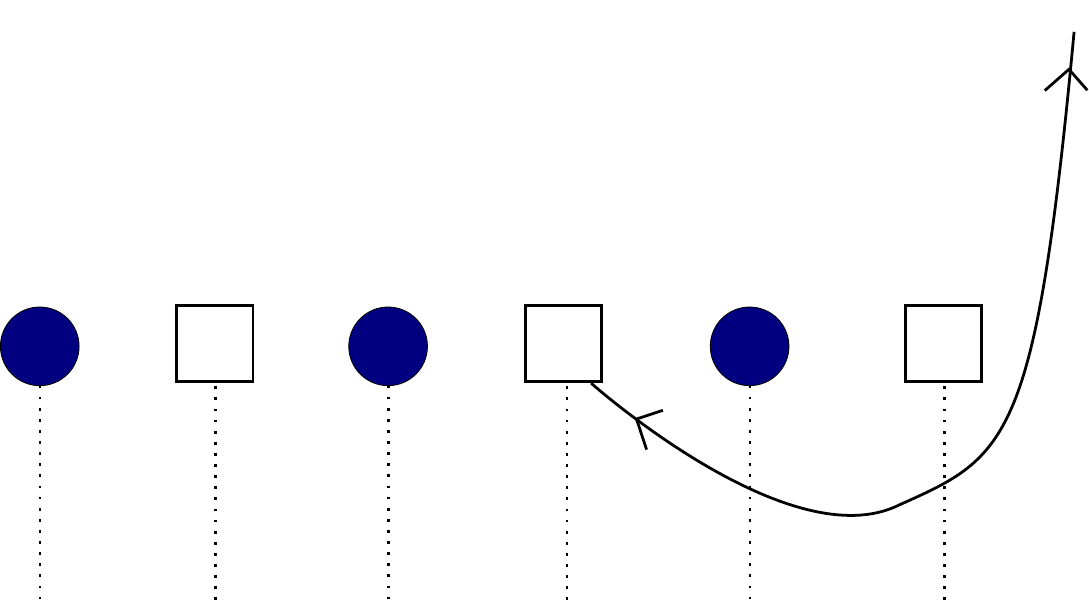}
\end{minipage}
\vspace{1cm}\\
\begin{minipage}[b]{0.3\linewidth}
(7)\includegraphics[scale=.5]{8AAAABC.pdf}
\end{minipage}
\hspace{3.5cm}
\begin{minipage}[b]{0.3\linewidth}
(8)\includegraphics[scale=.5]{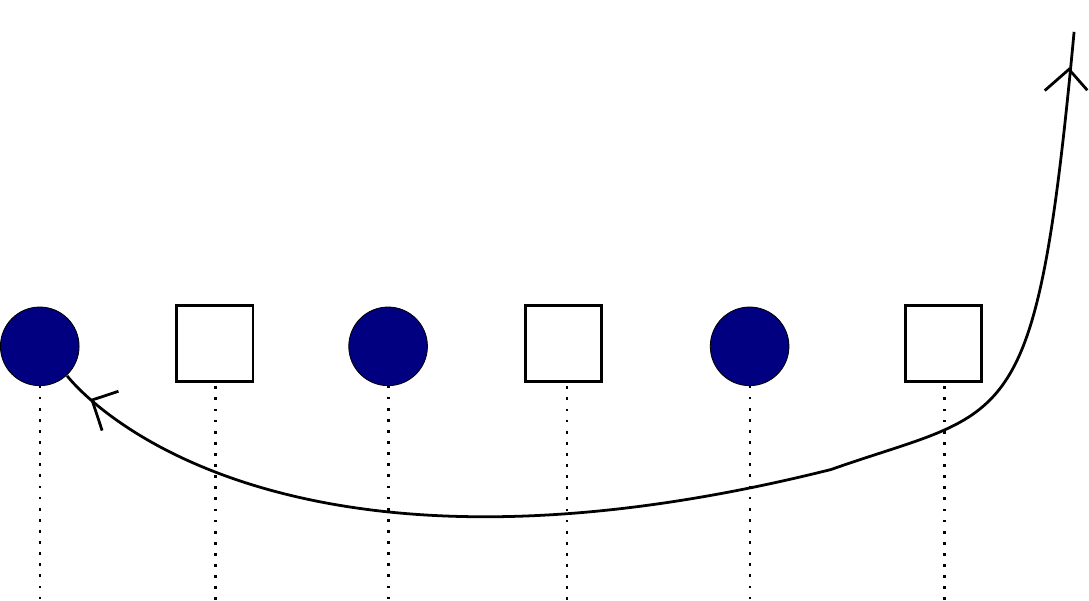}
\end{minipage}
\caption{Depicted on the left are four $\lambda$ modes whose orientations are
flipped by the action of $B$ and $C$, which together become an orientifold
at weak coupling. They fill out the other half
of the $\textbf{8}_v$ at points of $SO(8)$ enhancement. To right of each
string configuration is a representation of the string after performing
the brane motion $A^4BC \mapsto ACACAC$.}
\label{fig:untangle2}
\end{figure}

\newpage

\bibliographystyle{JHEP}
\bibliography{refs}

\providecommand{\href}[2]{#2}\begingroup\raggedright\begin{thebibliography}{10}

\bibitem{Blumenhagen:2009qh}
R.~Blumenhagen, M.~Cveti{\v c}, S.~Kachru, and T.~Weigand, {\it {D-Brane
  Instantons in Type {II} Orientifolds}},  {\em Ann. Rev. Nucl. Part. Sci.}
  {\bf 59} (2009) 269--296, [\href{http://xxx.lanl.gov/abs/0902.3251}{{\tt
  arXiv:0902.3251}}].

\bibitem{Blumenhagen:2005mu}
R.~Blumenhagen, M.~Cveti{\v c}, P.~Langacker, and G.~Shiu, {\it {Toward
  Realistic Intersecting D-Brane Models}},  {\em Ann. Rev. Nucl. Part. Sci.}
  {\bf 55} (2005) 71--139, [\href{http://xxx.lanl.gov/abs/hep-th/0502005}{{\tt
  hep-th/0502005}}].

\bibitem{Blumenhagen:2006ci}
R.~Blumenhagen, B.~Kors, D.~L{\" u}st, and S.~Stieberger, {\it
  {Four-Dimensional String Compactifications with D-Branes, Orientifolds and
  Fluxes}},  {\em Phys. Rept.} {\bf 445} (2007) 1--193,
  [\href{http://xxx.lanl.gov/abs/hep-th/0610327}{{\tt hep-th/0610327}}].

\bibitem{Cvetic:2011vz}
M.~Cveti{\v c} and J.~Halverson, {\it {Tasi Lectures: Particle Physics from
  Perturbative and Non- Perturbative Effects in D-Braneworlds}},
  \href{http://xxx.lanl.gov/abs/1101.2907}{{\tt arXiv:1101.2907}}.

\bibitem{Blumenhagen:2006xt}
R.~Blumenhagen, M.~Cveti{\v c}, and T.~Weigand, {\it {Spacetime Instanton
  Corrections in 4D String Vacua - the Seesaw Mechanism for D-Brane Models}},
  {\em Nucl. Phys.} {\bf B771} (2007) 113--142,
  [\href{http://xxx.lanl.gov/abs/hep-th/0609191}{{\tt hep-th/0609191}}].

\bibitem{Ibanez:2006da}
L.~E. Ibanez and A.~M. Uranga, {\it {Neutrino Majorana Masses from String
  Theory Instanton Effects}},  {\em JHEP} {\bf 03} (2007) 052,
  [\href{http://xxx.lanl.gov/abs/hep-th/0609213}{{\tt hep-th/0609213}}].

\bibitem{Florea:2006si}
B.~Florea, S.~Kachru, J.~McGreevy, and N.~Saulina, {\it {Stringy Instantons and
  Quiver Gauge Theories}},  {\em JHEP} {\bf 05} (2007) 024,
  [\href{http://xxx.lanl.gov/abs/hep-th/0610003}{{\tt hep-th/0610003}}].

\bibitem{Blumenhagen:2007zk}
R.~Blumenhagen, M.~Cveti{\v c}, D.~L{\" u}st, .~Richter, Robert, and
  T.~Weigand, {\it {Non-Perturbative Yukawa Couplings from String Instantons}},
   {\em Phys. Rev. Lett.} {\bf 100} (2008) 061602,
  [\href{http://xxx.lanl.gov/abs/0707.1871}{{\tt arXiv:0707.1871}}].

\bibitem{Kachru:2003aw}
S.~Kachru, R.~Kallosh, A.~D. Linde, and S.~P. Trivedi, {\it {De Sitter Vacua in
  String Theory}},  {\em Phys. Rev.} {\bf D68} (2003) 046005,
  [\href{http://xxx.lanl.gov/abs/hep-th/0301240}{{\tt hep-th/0301240}}].

\bibitem{Vafa:1996xn}
C.~Vafa, {\it {Evidence for F-Theory}},  {\em Nucl. Phys.} {\bf B469} (1996)
  403--418, [\href{http://xxx.lanl.gov/abs/hep-th/9602022}{{\tt
  hep-th/9602022}}].

\bibitem{Sen:1997gv}
A.~Sen, {\it {Orientifold limit of F theory vacua}},  {\em Phys.Rev.} {\bf D55}
  (1997) 7345--7349, [\href{http://xxx.lanl.gov/abs/hep-th/9702165}{{\tt
  hep-th/9702165}}].

\bibitem{Donagi:2008ca}
R.~Donagi and M.~Wijnholt, {\it {Model Building with F-Theory}},
  \href{http://xxx.lanl.gov/abs/0802.2969}{{\tt arXiv:0802.2969}}.

\bibitem{Beasley:2008dc}
C.~Beasley, J.~J. Heckman, and C.~Vafa, {\it {Guts and Exceptional Branes in
  F-Theory - I}},  {\em JHEP} {\bf 01} (2009) 058,
  [\href{http://xxx.lanl.gov/abs/0802.3391}{{\tt arXiv:0802.3391}}].

\bibitem{Beasley:2008kw}
C.~Beasley, J.~J. Heckman, and C.~Vafa, {\it {Guts and Exceptional Branes in
  F-Theory - Ii: Experimental Predictions}},  {\em JHEP} {\bf 01} (2009) 059,
  [\href{http://xxx.lanl.gov/abs/0806.0102}{{\tt arXiv:0806.0102}}].

\bibitem{Donagi:2008kj}
R.~Donagi and M.~Wijnholt, {\it {Breaking GUT Groups in F-Theory}},
  \href{http://xxx.lanl.gov/abs/0808.2223}{{\tt arXiv:0808.2223}}.

\bibitem{Collinucci:2008zs}
A.~Collinucci, {\it {New F-theory lifts}},  {\em JHEP} {\bf 0908} (2009) 076,
  [\href{http://xxx.lanl.gov/abs/0812.0175}{{\tt arXiv:0812.0175}}].

\bibitem{Donagi:2009ra}
R.~Donagi and M.~Wijnholt, {\it {Higgs Bundles and UV Completion in F-Theory}},
   \href{http://xxx.lanl.gov/abs/0904.1218}{{\tt arXiv:0904.1218}}.

\bibitem{Marsano:2009gv}
J.~Marsano, N.~Saulina, and S.~Sch{\" a}fer-Nameki, {\it {Monodromies, Fluxes,
  and Compact Three-Generation F-Theory Guts}},  {\em JHEP} {\bf 08} (2009)
  046, [\href{http://xxx.lanl.gov/abs/0906.4672}{{\tt arXiv:0906.4672}}].

\bibitem{Marsano:2009wr}
J.~Marsano, N.~Saulina, and S.~Sch{\" a}fer-Nameki, {\it Compact f-theory guts
  with $u(1)_{PQ}$},  {\em JHEP} {\bf 04} (2010) 095,
  [\href{http://xxx.lanl.gov/abs/0912.0272}{{\tt arXiv:0912.0272}}].

\bibitem{Collinucci:2009uh}
A.~Collinucci, {\it {New F-theory lifts. II. Permutation orientifolds and
  enhanced singularities}},  {\em JHEP} {\bf 1004} (2010) 076,
  [\href{http://xxx.lanl.gov/abs/0906.0003}{{\tt arXiv:0906.0003}}].

\bibitem{Blumenhagen:2009up}
R.~Blumenhagen, T.~W. Grimm, B.~Jurke, and T.~Weigand, {\it {F-Theory Uplifts
  and Guts}},  {\em JHEP} {\bf 09} (2009) 053,
  [\href{http://xxx.lanl.gov/abs/0906.0013}{{\tt arXiv:0906.0013}}].

\bibitem{Blumenhagen:2009yv}
R.~Blumenhagen, T.~W. Grimm, B.~Jurke, and T.~Weigand, {\it {Global F-Theory
  Guts}},  {\em Nucl. Phys.} {\bf B829} (2010) 325--369,
  [\href{http://xxx.lanl.gov/abs/0908.1784}{{\tt arXiv:0908.1784}}].

\bibitem{Hayashi:2008ba}
H.~Hayashi, R.~Tatar, Y.~Toda, T.~Watari, and M.~Yamazaki, {\it {New Aspects of
  Heterotic--F Theory Duality}},  {\em Nucl. Phys.} {\bf B806} (2009) 224--299,
  [\href{http://xxx.lanl.gov/abs/0805.1057}{{\tt arXiv:0805.1057}}].

\bibitem{Grimm:2009yu}
T.~W. Grimm, S.~Krause, and T.~Weigand, {\it {F-Theory GUT Vacua on Compact
  Calabi-Yau Fourfolds}},  {\em JHEP} {\bf 07} (2010) 037,
  [\href{http://xxx.lanl.gov/abs/0912.3524}{{\tt arXiv:0912.3524}}].

\bibitem{Cvetic:2010rq}
M.~Cveti{\v c}, I.~Garcia-Etxebarria, and J.~Halverson, {\it {Global F-theory
  Models: Instantons and Gauge Dynamics}},
  \href{http://xxx.lanl.gov/abs/1003.5337}{{\tt arXiv:1003.5337}}.

\bibitem{Chen:2010tp}
C.-M. Chen and Y.-C. Chung, {\it {Flipped SU(5) GUTs from $E_8$ Singularities
  in F-theory}},  {\em JHEP} {\bf 1103} (2011) 049,
  [\href{http://xxx.lanl.gov/abs/1005.5728}{{\tt arXiv:1005.5728}}].

\bibitem{Chen:2010ts}
C.-M. Chen, J.~Knapp, M.~Kreuzer, and C.~Mayrhofer, {\it {Global SO(10)
  F-Theory Guts}},  {\em JHEP} {\bf 10} (2010) 057,
  [\href{http://xxx.lanl.gov/abs/1005.5735}{{\tt arXiv:1005.5735}}].

\bibitem{Grimm:2010ez}
T.~W. Grimm and T.~Weigand, {\it {On Abelian Gauge Symmetries and Proton Decay
  in Global F- Theory Guts}},  {\em Phys. Rev.} {\bf D82} (2010) 086009,
  [\href{http://xxx.lanl.gov/abs/1006.0226}{{\tt arXiv:1006.0226}}].

\bibitem{Chung:2010bn}
Y.-C. Chung, {\it {On Global Flipped SU(5) GUTs in F-theory}},  {\em JHEP} {\bf
  1103} (2011) 126, [\href{http://xxx.lanl.gov/abs/1008.2506}{{\tt
  arXiv:1008.2506}}].

\bibitem{Chen:2010tg}
C.-M. Chen and Y.-C. Chung, {\it {On F-theory $E_6$ GUTs}},  {\em JHEP} {\bf
  1103} (2011) 129, [\href{http://xxx.lanl.gov/abs/1010.5536}{{\tt
  arXiv:1010.5536}}].

\bibitem{Knapp:2011wk}
J.~Knapp, M.~Kreuzer, C.~Mayrhofer, and N.-O. Walliser, {\it {Toric
  Construction of Global F-Theory Guts}},  {\em JHEP} {\bf 03} (2011) 138,
  [\href{http://xxx.lanl.gov/abs/1101.4908}{{\tt arXiv:1101.4908}}].

\bibitem{Witten:1996bn}
E.~Witten, {\it {Non-Perturbative Superpotentials in String Theory}},  {\em
  Nucl. Phys.} {\bf B474} (1996) 343--360,
  [\href{http://xxx.lanl.gov/abs/hep-th/9604030}{{\tt hep-th/9604030}}].

\bibitem{Blumenhagen:2010ja}
R.~Blumenhagen, A.~Collinucci, and B.~Jurke, {\it {On Instanton Effects in
  F-theory}},  \href{http://xxx.lanl.gov/abs/1002.1894}{{\tt arXiv:1002.1894}}.

\bibitem{Donagi:2010pd}
R.~Donagi and M.~Wijnholt, {\it {Msw Instantons}},
  \href{http://xxx.lanl.gov/abs/1005.5391}{{\tt arXiv:1005.5391}}.

\bibitem{Heckman:2008es}
J.~J. Heckman, J.~Marsano, N.~Saulina, S.~Sch{\" a}fer-Nameki, and C.~Vafa,
  {\it {Instantons and SUSY Breaking in F-Theory}},
  \href{http://xxx.lanl.gov/abs/0808.1286}{{\tt arXiv:0808.1286}}.

\bibitem{Marsano:2008py}
J.~Marsano, N.~Saulina, and S.~Sch{\" a}fer-Nameki, {\it {An Instanton Toolbox
  for F-Theory Model Building}},  {\em JHEP} {\bf 01} (2010) 128,
  [\href{http://xxx.lanl.gov/abs/0808.2450}{{\tt arXiv:0808.2450}}].

\bibitem{Grimm:2011dj}
T.~W. Grimm, M.~Kerstan, E.~Palti, and T.~Weigand, {\it {On Fluxed Instantons
  and Moduli Stabilisation in IIB Orientifolds and F-Theory}},
  \href{http://xxx.lanl.gov/abs/1105.3193}{{\tt arXiv:1105.3193}}.

\bibitem{Blumenhagen:2007sm}
R.~Blumenhagen, S.~Moster, and E.~Plauschinn, {\it {Moduli Stabilisation Versus
  Chirality for MSSM Like Type IIB Orientifolds}},  {\em JHEP} {\bf 01} (2008)
  058, [\href{http://xxx.lanl.gov/abs/0711.3389}{{\tt arXiv:0711.3389}}].

\bibitem{Marsano:2010ix}
J.~Marsano, N.~Saulina, and S.~Sch{\" a}fer-Nameki, {\it {A Note on G-Fluxes
  for F-Theory Model Building}},  {\em JHEP} {\bf 11} (2010) 088,
  [\href{http://xxx.lanl.gov/abs/1006.0483}{{\tt arXiv:1006.0483}}].

\bibitem{Marsano:2011nn}
J.~Marsano, N.~Saulina, and S.~Sch{\" a}fer-Nameki, {\it {On G-flux, M5
  instantons, and U(1)s in F-theory}},
  \href{http://xxx.lanl.gov/abs/1107.1718}{{\tt arXiv:1107.1718}}.

\bibitem{Harvey:2007ab}
J.~A. Harvey and A.~B. Royston, {\it {Localized Modes at a D-Brane--O-Plane
  Intersection and Heterotic Alice Strings}},  {\em JHEP} {\bf 04} (2008) 018,
  [\href{http://xxx.lanl.gov/abs/0709.1482}{{\tt arXiv:0709.1482}}].

\bibitem{Buchbinder:2002ic}
E.~I. Buchbinder, R.~Donagi, and B.~A. Ovrut, {\it {Superpotentials for Vector
  Bundle Moduli}},  {\em Nucl. Phys.} {\bf B653} (2003) 400--420,
  [\href{http://xxx.lanl.gov/abs/hep-th/0205190}{{\tt hep-th/0205190}}].

\bibitem{Buchbinder:2002pr}
E.~I. Buchbinder, R.~Donagi, and B.~A. Ovrut, {\it {Vector Bundle Moduli
  Superpotentials in Heterotic Superstrings and M-Theory}},  {\em JHEP} {\bf
  07} (2002) 066, [\href{http://xxx.lanl.gov/abs/hep-th/0206203}{{\tt
  hep-th/0206203}}].

\bibitem{Cvetic:2009ah}
M.~Cveti{\v c}, I.~Garcia-Etxebarria, and R.~Richter, {\it {Branes and
  instantons at angles and the F-theory lift of O(1) instantons}},  {\em AIP
  Conf.Proc.} {\bf 1200} (2010) 246--260,
  [\href{http://xxx.lanl.gov/abs/0911.0012}{{\tt arXiv:0911.0012}}].

\bibitem{Dasgupta:1997cd}
K.~Dasgupta, D.~P. Jatkar, and S.~Mukhi, {\it {Gravitational couplings and Z(2)
  orientifolds}},  {\em Nucl.Phys.} {\bf B523} (1998) 465--484,
  [\href{http://xxx.lanl.gov/abs/hep-th/9707224}{{\tt hep-th/9707224}}].

\bibitem{Dasgupta:1997wd}
K.~Dasgupta and S.~Mukhi, {\it {Anomaly inflow on orientifold planes}},  {\em
  JHEP} {\bf 9803} (1998) 004,
  [\href{http://xxx.lanl.gov/abs/hep-th/9709219}{{\tt hep-th/9709219}}].

\bibitem{Morales:1998ux}
J.~F. Morales, C.~A. Scrucca, and M.~Serone, {\it {Anomalous couplings for
  D-branes and O-planes}},  {\em Nucl.Phys.} {\bf B552} (1999) 291--315,
  [\href{http://xxx.lanl.gov/abs/hep-th/9812071}{{\tt hep-th/9812071}}].

\bibitem{Stefanski:1998yx}
J.~Stefanski, Bogdan, {\it {Gravitational couplings of D-branes and O-planes}},
   {\em Nucl.Phys.} {\bf B548} (1999) 275--290,
  [\href{http://xxx.lanl.gov/abs/hep-th/9812088}{{\tt hep-th/9812088}}].

\bibitem{Scrucca:1999uz}
C.~A. Scrucca and M.~Serone, {\it {Anomalies and inflow on D-branes and O -
  planes}},  {\em Nucl.Phys.} {\bf B556} (1999) 197--221,
  [\href{http://xxx.lanl.gov/abs/hep-th/9903145}{{\tt hep-th/9903145}}].

\bibitem{Harvey:2005it}
J.~A. Harvey, {\it {TASI 2003 lectures on anomalies}},
  \href{http://xxx.lanl.gov/abs/hep-th/0509097}{{\tt hep-th/0509097}}.

\bibitem{Sen:1996vd}
A.~Sen, {\it {F-Theory and Orientifolds}},  {\em Nucl. Phys.} {\bf B475} (1996)
  562--578, [\href{http://xxx.lanl.gov/abs/hep-th/9605150}{{\tt
  hep-th/9605150}}].

\bibitem{Kapustin:2006pk}
A.~Kapustin and E.~Witten, {\it {Electric-magnetic duality and the geometric
  Langlands program}},  \href{http://xxx.lanl.gov/abs/hep-th/0604151}{{\tt
  hep-th/0604151}}.

\bibitem{Seiberg:1994rs}
N.~Seiberg and E.~Witten, {\it {Electric - magnetic duality, monopole
  condensation, and confinement in N=2 supersymmetric Yang-Mills theory}},
  {\em Nucl.Phys.} {\bf B426} (1994) 19--52,
  [\href{http://xxx.lanl.gov/abs/hep-th/9407087}{{\tt hep-th/9407087}}].

\bibitem{Seiberg:1994aj}
N.~Seiberg and E.~Witten, {\it {Monopoles, duality and chiral symmetry breaking
  in N=2 supersymmetric QCD}},  {\em Nucl.Phys.} {\bf B431} (1994) 484--550,
  [\href{http://xxx.lanl.gov/abs/hep-th/9408099}{{\tt hep-th/9408099}}].

\bibitem{Naculich:1987ci}
S.~G. Naculich, {\it {AXIONIC STRINGS: COVARIANT ANOMALIES AND BOSONIZATION OF
  CHIRAL ZERO MODES}},  {\em Nucl.Phys.} {\bf B296} (1988) 837.

\bibitem{Fayyazuddin:1995pk}
A.~Fayyazuddin, {\it {Some comments on N=2 supersymmetric Yang-Mills}},  {\em
  Mod.Phys.Lett.} {\bf A10} (1995) 2703--2708,
  [\href{http://xxx.lanl.gov/abs/hep-th/9504120}{{\tt hep-th/9504120}}].

\bibitem{Ferrari:1996sv}
F.~Ferrari and A.~Bilal, {\it {The Strong coupling spectrum of the
  Seiberg-Witten theory}},  {\em Nucl.Phys.} {\bf B469} (1996) 387--402,
  [\href{http://xxx.lanl.gov/abs/hep-th/9602082}{{\tt hep-th/9602082}}].

\bibitem{Bilal:1996sk}
A.~Bilal and F.~Ferrari, {\it {Curves of marginal stability, and weak and
  strong coupling BPS spectra in N=2 supersymmetric QCD}},  {\em Nucl.Phys.}
  {\bf B480} (1996) 589--622,
  [\href{http://xxx.lanl.gov/abs/hep-th/9605101}{{\tt hep-th/9605101}}].

\bibitem{Bergman:1998br}
O.~Bergman and A.~Fayyazuddin, {\it {String junctions and BPS states in
  Seiberg-Witten theory}},  {\em Nucl.Phys.} {\bf B531} (1998) 108--124,
  [\href{http://xxx.lanl.gov/abs/hep-th/9802033}{{\tt hep-th/9802033}}].

\bibitem{Mikhailov:1998bx}
A.~Mikhailov, N.~Nekrasov, and S.~Sethi, {\it {Geometric realizations of BPS
  states in N=2 theories}},  {\em Nucl.Phys.} {\bf B531} (1998) 345--362,
  [\href{http://xxx.lanl.gov/abs/hep-th/9803142}{{\tt hep-th/9803142}}].

\bibitem{DeWolfe:1998zf}
O.~DeWolfe and B.~Zwiebach, {\it {String junctions for arbitrary Lie algebra
  representations}},  {\em Nucl.Phys.} {\bf B541} (1999) 509--565,
  [\href{http://xxx.lanl.gov/abs/hep-th/9804210}{{\tt hep-th/9804210}}].

\bibitem{Gaberdiel:1998mv}
M.~R. Gaberdiel, T.~Hauer, and B.~Zwiebach, {\it {Open String-String Junction
  Transitions}},  {\em Nucl. Phys.} {\bf B525} (1998) 117--145,
  [\href{http://xxx.lanl.gov/abs/hep-th/9801205}{{\tt hep-th/9801205}}].

\bibitem{Dasgupta:1996ij}
K.~Dasgupta and S.~Mukhi, {\it {F-Theory at Constant Coupling}},  {\em Phys.
  Lett.} {\bf B385} (1996) 125--131,
  [\href{http://xxx.lanl.gov/abs/hep-th/9606044}{{\tt hep-th/9606044}}].

\bibitem{Gaberdiel:1997ud}
M.~R. Gaberdiel and B.~Zwiebach, {\it {Exceptional Groups from Open Strings}},
  {\em Nucl. Phys.} {\bf B518} (1998) 151--172,
  [\href{http://xxx.lanl.gov/abs/hep-th/9709013}{{\tt hep-th/9709013}}].

\bibitem{Hanany:1996ie}
A.~Hanany and E.~Witten, {\it {Type IIB Superstrings, BPS Monopoles, and Three-
  Dimensional Gauge Dynamics}},  {\em Nucl. Phys.} {\bf B492} (1997) 152--190,
  [\href{http://xxx.lanl.gov/abs/hep-th/9611230}{{\tt hep-th/9611230}}].

\bibitem{DeWolfe:1998bi}
O.~DeWolfe, T.~Hauer, A.~Iqbal, and B.~Zwiebach, {\it {Constraints on the BPS
  Spectrum of ${\mathcal{N}}\!=2$, D = 4 Theories with A-D-E Flavor Symmetry}},
   {\em Nucl. Phys.} {\bf B534} (1998) 261--274,
  [\href{http://xxx.lanl.gov/abs/hep-th/9805220}{{\tt hep-th/9805220}}].

\bibitem{Dine:1986zy}
M.~Dine, N.~Seiberg, X.~G. Wen, and E.~Witten, {\it {Nonperturbative Effects on
  the String World Sheet}},  {\em Nucl. Phys.} {\bf B278} (1986) 769.

\bibitem{Dine:1987bq}
M.~Dine, N.~Seiberg, X.~G. Wen, and E.~Witten, {\it {Nonperturbative Effects on
  the String World Sheet. 2}},  {\em Nucl. Phys.} {\bf B289} (1987) 319.

\bibitem{Distler:1986wm}
J.~Distler, {\it {Resurrecting (2,0) Compactifications}},  {\em Phys. Lett.}
  {\bf B188} (1987) 431--436.

\bibitem{Distler:1987ee}
J.~Distler and B.~R. Greene, {\it {Aspects of (2,0) String Compactifications}},
   {\em Nucl. Phys.} {\bf B304} (1988) 1.

\bibitem{Berglund:1995yu}
P.~Berglund {\em et.~al.}, {\it {On the Instanton Contributions to the Masses
  and Couplings of $E_{6}$ Singlets}},  {\em Nucl. Phys.} {\bf B454} (1995)
  127--163, [\href{http://xxx.lanl.gov/abs/hep-th/9505164}{{\tt
  hep-th/9505164}}].

\bibitem{Weigand:2010wm}
T.~Weigand, {\it {Lectures on F-Theory Compactifications and Model Building}},
  {\em Class. Quant. Grav.} {\bf 27} (2010) 214004,
  [\href{http://xxx.lanl.gov/abs/1009.3497}{{\tt arXiv:1009.3497}}].

\bibitem{Morrison:1996na}
D.~R. Morrison and C.~Vafa, {\it {Compactifications of F-Theory on Calabi--Yau
  Threefolds -- I}},  {\em Nucl. Phys.} {\bf B473} (1996) 74--92,
  [\href{http://xxx.lanl.gov/abs/hep-th/9602114}{{\tt hep-th/9602114}}].

\bibitem{Morrison:1996pp}
D.~R. Morrison and C.~Vafa, {\it {Compactifications of F-Theory on Calabi--Yau
  Threefolds -- II}},  {\em Nucl. Phys.} {\bf B476} (1996) 437--469,
  [\href{http://xxx.lanl.gov/abs/hep-th/9603161}{{\tt hep-th/9603161}}].

\bibitem{Friedman:1997yq}
R.~Friedman, J.~Morgan, and E.~Witten, {\it {Vector Bundles and F Theory}},
  {\em Commun. Math. Phys.} {\bf 187} (1997) 679--743,
  [\href{http://xxx.lanl.gov/abs/hep-th/9701162}{{\tt hep-th/9701162}}].

\bibitem{Donagi:1997mg}
R.~Y. Donagi, {\it {Principal Bundles on Elliptic Fibrations}},  {\em Asian J.
  Math.} {\bf 1} (1997) 214--223.

\bibitem{Bershadsky:1997zs}
M.~Bershadsky, A.~Johansen, T.~Pantev, and V.~Sadov, {\it {On Four-Dimensional
  Compactifications of F-Theory}},  {\em Nucl. Phys.} {\bf B505} (1997)
  165--201, [\href{http://xxx.lanl.gov/abs/hep-th/9701165}{{\tt
  hep-th/9701165}}].

\bibitem{Bershadsky:1996nh}
M.~Bershadsky {\em et.~al.}, {\it {Geometric Singularities and Enhanced Gauge
  Symmetries}},  {\em Nucl. Phys.} {\bf B481} (1996) 215--252,
  [\href{http://xxx.lanl.gov/abs/hep-th/9605200}{{\tt hep-th/9605200}}].

\bibitem{Moore:2000fs}
G.~W. Moore, G.~Peradze, and N.~Saulina, {\it {Instabilities in heterotic M
  theory induced by open membrane instantons}},  {\em Nucl.Phys.} {\bf B607}
  (2001) 117--154, [\href{http://xxx.lanl.gov/abs/hep-th/0012104}{{\tt
  hep-th/0012104}}].

\bibitem{Curio:2001qi}
G.~Curio and A.~Krause, {\it {G fluxes and nonperturbative stabilization of
  heterotic M theory}},  {\em Nucl.Phys.} {\bf B643} (2002) 131--156,
  [\href{http://xxx.lanl.gov/abs/hep-th/0108220}{{\tt hep-th/0108220}}].

\bibitem{Buchbinder:2003pi}
E.~I. Buchbinder and B.~A. Ovrut, {\it {Vacuum stability in heterotic M
  theory}},  {\em Phys.Rev.} {\bf D69} (2004) 086010,
  [\href{http://xxx.lanl.gov/abs/hep-th/0310112}{{\tt hep-th/0310112}}].

\bibitem{Curio:2009wn}
G.~Curio, {\it {Perspectives on Pfaffians of Heterotic World-Sheet
  Instantons}},  {\em JHEP} {\bf 09} (2009) 131,
  [\href{http://xxx.lanl.gov/abs/0904.2738}{{\tt arXiv:0904.2738}}].

\bibitem{Curio:2008cm}
G.~Curio, {\it {World-Sheet Instanton Superpotentials in Heterotic String
  Theory and Their Moduli Dependence}},  {\em JHEP} {\bf 09} (2009) 125,
  [\href{http://xxx.lanl.gov/abs/0810.3087}{{\tt arXiv:0810.3087}}].

\bibitem{Curio:2010hd}
G.~Curio, {\it {On the Heterotic World-Sheet Instanton Superpotential and Its
  Individual Contributions}},  {\em JHEP} {\bf 08} (2010) 092,
  [\href{http://xxx.lanl.gov/abs/1006.5568}{{\tt arXiv:1006.5568}}].

\bibitem{Heckman:2009mn}
J.~J. Heckman, A.~Tavanfar, and C.~Vafa, {\it {The Point of $E_{8}$ in F-Theory
  Guts}},  {\em JHEP} {\bf 08} (2010) 040,
  [\href{http://xxx.lanl.gov/abs/0906.0581}{{\tt arXiv:0906.0581}}].

\bibitem{Kofman:2004yc}
L.~Kofman, A.~D. Linde, X.~Liu, A.~Maloney, L.~McAllister, {\em et.~al.}, {\it
  {Beauty is attractive: Moduli trapping at enhanced symmetry points}},  {\em
  JHEP} {\bf 0405} (2004) 030,
  [\href{http://xxx.lanl.gov/abs/hep-th/0403001}{{\tt hep-th/0403001}}].

\bibitem{Donagi:2004ia}
R.~Donagi, Y.-H. He, B.~A. Ovrut, and R.~Reinbacher, {\it {The Particle
  Spectrum of Heterotic Compactifications}},  {\em JHEP} {\bf 12} (2004) 054,
  [\href{http://xxx.lanl.gov/abs/hep-th/0405014}{{\tt hep-th/0405014}}].

\bibitem{Hayashi:2010zp}
H.~Hayashi, T.~Kawano, Y.~Tsuchiya, and T.~Watari, {\it {More on Dimension-4
  Proton Decay Problem in F-theory -- Spectral Surface, Discriminant Locus and
  Monodromy}},  {\em Nucl. Phys.} {\bf B840} (2010) 304--348,
  [\href{http://xxx.lanl.gov/abs/1004.3870}{{\tt arXiv:1004.3870}}].

\end{thebibliography}\endgroup

\end{document}